\documentclass[11pt,a4paper]{article}

\pdfoutput=1
\usepackage{amsmath,amssymb,authblk,multirow,collref}
\usepackage{graphicx}
\usepackage{pgfplots}
\usepackage{xparse}
\usepackage{subfigure}

\usepackage{tikz}
\usetikzlibrary{positioning,arrows,patterns}
\usetikzlibrary{decorations.pathmorphing}
\usetikzlibrary{decorations.markings}
\usetikzlibrary{calc}
\tikzset{
    photon/.style={decorate, decoration={snake}, draw=black, thick},
    fermionnoarrow/.style={draw=black, postaction={decorate}, thick},
    scalar/.style={draw=black, postaction={decorate}, thick, dashed},
    fermion/.style={draw=black, postaction={decorate},decoration={markings,mark=at position .55 with {\arrow{>}}}, thick},
    gluon/.style={decorate, draw=black, decoration={coil,amplitude=4pt, segment length=5pt}, thick},
    vertex/.style={draw,shape=circle,fill=black,minimum size=3pt,inner sep=0pt} 
}

\graphicspath{{./}{./Figures/}}

\usepackage{geometry}
\newcommand{\lag}{\mathcal{L}}
\newcommand{\sigv}{\langle \sigma v \rangle}
\newcommand{\lhp}{\lambda_{h\phi}}
\newcommand{\spand}{\quad\text{and}\quad}
\newcommand{\vew}{v_{EW}}

\newcommand{\set}[1]{\mathbb{#1}}

\newcommand{\modeqref}[1]{Eq.~\eqref{#1}}
\newcommand{\twoeqref}[2]{Eqs.~\eqref{#1} and~\eqref{#2}}
\newcommand{\tabref}[1]{Table~\ref{#1}}
\newcommand{\secref}[1]{Section~\ref{#1}}
\newcommand{\secsref}[1]{Sections~\ref{#1}}
\newcommand{\threesecsref}[3]{Sections~\ref{#1}, \ref{#2} and~\ref{#3}}

\newcommand{\refcite}[1]{Ref.~\cite{#1}}
\newcommand{\refscite}[1]{Refs.~\cite{#1}}
\newcommand{\figref}[1]{Figure~\ref{#1}}
\newcommand{\figsref}[2]{Figures~\ref{#1} and~\ref{#2}}

\usepackage[sort&compress,numbers,colon]{natbib}
\usepackage{hyperref}

\begin{document}

\title{Fermionic Semi-Annihilating Dark Matter}
\author[1]{Yi Cai\thanks{\texttt{yi.cai@unimelb.edu.au}}}
\author[1,2]{Andrew Spray\thanks{\texttt{andrew.spray@coepp.org.au}}}
\affil[1]{ARC Centre of Excellence for Particle Physics at the Terascale, School of Physics, 
The University of Melbourne, Victoria 3010, Australia}
\affil[2]{Center for Theoretical Physics of the Universe, Institute for Basic Science (IBS), Daejeon, 34051, Korea}

\maketitle

\begin{abstract}
Semi-annihilation is a generic feature of dark matter theories with symmetries larger than $\set{Z}_2$.  We investigate two examples with multi-component dark sectors comprised of an $SU(2)_L$ singlet or triplet fermion besides a scalar singlet.  These are respectively the minimal fermionic semi-annihilating model, and the minimal case for a gauge-charged fermion.  We study the relevant dark matter phenomenology, including the interplay of semi-annihilation and the Sommerfeld effect.  We demonstrate that semi-annihilation in the singlet model can explain the gamma ray excess from the galactic center.  For the triplet model we scan the parameter space, and explore how signals and constraints are modified by semi-annihilation.  We find that the entire region where the model comprises all the observed dark matter is accessible to current and planned direct and indirect searches.  
\end{abstract}

\section{Introduction}

The dark matter (DM) problem remains one of the most important questions in contemporary particle physics.  Measurements across multiple scales all point to the existence of a cold non-baryonic component of matter in the Universe, from galaxy rotation curves to fluctuations in the cosmic microwave background.  Over the last few decades, enormous experimental efforts have been made to uncover its true identity.  However, no unambiguous non-gravitational signal has been found and the microscopic properties of DM remain unknown.

The quest to explore DM phenomenology has involved several different approaches.  One can construct complete models of UV physics which, in addition to solving the DM problem, address other issues within the Standard Model (SM), such as naturalness or the flavour puzzle.  The most well-known example of this approach is the neutralino of supersymmetry.  This direction has the benefit of completeness, but the need to address multiple problems at once may be too constraining.  Additionally, it can be hard to construct a top-down model that reproduces a given DM phenomenology.  For this reason, the use of effective theories has also been very popular.  These models allow one to focus on only the DM degrees of freedom, connecting to the SM through higher-dimensional operators.  This has the weakness that the cut-off scale that can be probed at high-energy machines such as the LHC tends to be too low for the results to hold any validity.  Additionally, even if the cut-off scale is sufficiently large the sensitivity to low scales remains uncertain.

In light of this, a third approach has risen based on constructing simple but complete DM models~\cite{Abdallah:2015ter}.  They combine the strengths of effective theories with a greater range of validity in the results.  In particular, it is relatively easy to construct models with a particular dark sector phenomenology.  These considerations motivate us to adopt this direction here.

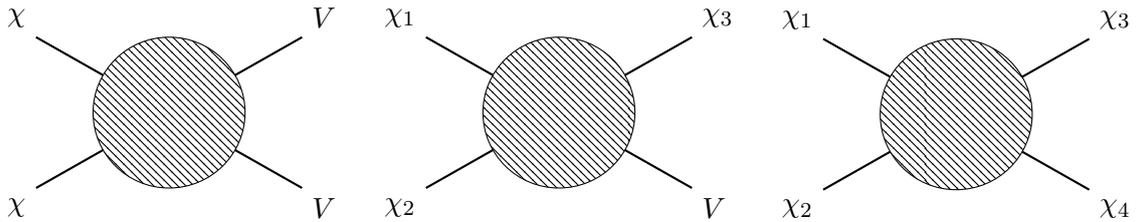
\begin{figure}
  \centering
  \begin{tikzpicture}[node distance=1cm and 1.75cm]
    \coordinate (v1);
    \coordinate[above left = of v1, label=above left:$\chi$] (i1);
    \coordinate[below left = of v1, label=below left:$\chi$] (i2);
    \coordinate[above right = of v1, label=above right:{$V$}] (o1);
    \coordinate[below right = of v1, label=below right:{$V$}] (o2);
    \draw[fermionnoarrow] (i1) -- (v1);
    \draw[fermionnoarrow] (v1) -- (i2);
    \draw[fermionnoarrow] (o2) -- (v1);
    \draw[fermionnoarrow] (v1) -- (o1);
    \draw[fill = white] (v1) circle (1);
    \fill[pattern = north west lines] (v1) circle (1);
  \end{tikzpicture}\quad
  \begin{tikzpicture}[node distance=1cm and 1.75cm]
    \coordinate (v1);
    \coordinate[above left = of v1, label=above left:$\chi_1$] (i1);
    \coordinate[below left = of v1, label=below left:$\chi_2$] (i2);
    \coordinate[above right = of v1, label=above right:{$\chi_3$}] (o1);
    \coordinate[below right = of v1, label=below right:{$V$}] (o2);
    \draw[fermionnoarrow] (i1) -- (v1);
    \draw[fermionnoarrow] (v1) -- (i2);
    \draw[fermionnoarrow] (o2) -- (v1);
    \draw[fermionnoarrow] (v1) -- (o1);
    \draw[fill = white] (v1) circle (1);
    \fill[pattern = north west lines] (v1) circle (1);
  \end{tikzpicture}\quad
  \begin{tikzpicture}[node distance=1cm and 1.75cm]
    \coordinate (v1);
    \coordinate[above left = of v1, label=above left:$\chi_1$] (i1);
    \coordinate[below left = of v1, label=below left:$\chi_2$] (i2);
    \coordinate[above right = of v1, label=above right:{$\chi_3$}] (o1);
    \coordinate[below right = of v1, label=below right:{$\chi_4$}] (o2);
    \draw[fermionnoarrow] (i1) -- (v1);
    \draw[fermionnoarrow] (v1) -- (i2);
    \draw[fermionnoarrow] (o2) -- (v1);
    \draw[fermionnoarrow] (v1) -- (o1);
    \draw[fill = white] (v1) circle (1);
    \fill[pattern = north west lines] (v1) circle (1);
  \end{tikzpicture}
  \caption{Three types of dark sector processes, where $\chi$ ($V$) is a dark (visible) sector field.  (Left): DM annihilation to/from the SM; this is the only process possible when the dark matter is stabilised by a $\set{Z}_2$ symmetry.  (Centre): Semi-annihilation, a non-decay process with an odd number of external visible particles.  (Right): DM exchange, only possible when the dark sector is multicomponent.}\label{fig:SAandDE}
\end{figure}

Semi-annihilation (SA) is a generic feature of dark sector phenomenology that occurs whenever the symmetry that stabilizes DM is larger than $\set{Z}_2$~\cite{D'Eramo:2010ep}.  It is shown in \figref{fig:SAandDE}.  For the usually-considered case, the only allowed $2\to 2$ diagram is that on the left: DM annihilation to/from or scattering off the SM.  SA is shown by the central diagram, and is characterised by a non-decay process with an odd number of external dark sector particles.  Finally, many models of semi-annihilating dark matter (SADM) involve multicomponent dark sectors, in which case dark matter exchange (DME), the process shown in the right diagram, can be relevant.

Previous studies of SADM have mostly focused on scalar DM candidates~\cite{D'Eramo:2012rr,Belanger:2012vp,Belanger:2012zr,Ivanov:2012hc,Lovrekovic:2012bz,Belanger:2014bga} (see \refscite{Aoki:2012ub,Aoki:2014cja} for exceptions).  This is natural, as renormalisable quartic scalar couplings involving one visible sector particle and three dark sector ones can be easily realized with a $\set{Z}_3$ symmetry.  For fermions, such couplings are non-renormalisable. Further, the Higgs portal means that scalars can always couple renormalisably to the visible sector; this is not true for gauge-singlet fermions.  Nonetheless, it would be interesting to explore beyond the simplest scenarios and examine what possible SA models for fermionic DM could exist and what special phenomenology they might have.

There are two immediate conclusions we can draw about fermionic SADM.  First, the lack of renormalisable couplings \emph{demands} multi-component dark sectors.  Multi-component DM generically leads to richer phenomenology~\cite{Khlopov:1995pa}, but when simply imposed by hand it lacks motivation.  Additionally, directly probing fermionic SA leads to considering non-singlet fermions.  The simplest such example is a fermion $SU(2)_L$ triplet, which is subject the Sommerfeld effect (SE), a non-perturbative enhancement of cross sections at low velocities.  In particular, the SE can be relevant both for determining the relic density and for indirect signals in the present day. The intersection of SA and the SE has not previously been considered.

The outline of the paper is as follows.  In \secref{sec:model}, we introduce two models of SADM, one with a fermion singlet and one with a fermion triplet. We discuss the complications in the calculation of the DM relic density, including semi-annihilation and the Sommerfeld effect in \secref{sec:RD}.  We then analyze constraints on the model parameter space from colliders, direct detection and indirect detection in \secsref{sec:LHC}, \ref{sec:DD} and \ref{sec:ID} respectively.  We also probe the possibility to explain the galactic centre excess with these models in \secref{sec:FXS}.  We present the combined results for the fermion triplet model in \secref{sec:triplet} and conclude with \secref{sec:conc}.

\section{Models of Fermion Semi-Annihilating Dark Matter}\label{sec:model}
Fermionic examples of SADM must include bosonic degrees of freedom in the dark sector.  This is a consequence of renormalisable operators involving at most two fermion fields.  An interaction term with a scalar $\phi$ and two fermions $\psi_{1,2}$
\begin{equation}
  \lag \supset \phi \, \psi^\dagger_1 \psi_2 + h.c.
\label{eq:sdmint}\end{equation}
will only generate SA involving the fermions if the product of the fermion spinors is charged under the dark sector symmetry, otherwise the action of the symmetry on the fermions is indistinguishable from a $\set{Z}_2$.  Hence $\phi$ must also have dark sector charge.  Non-renormalisable theories can avoid this constraint, but necessitate integrating in bosonic degrees of freedom at a high scale.  It follows that while the minimal models of scalar SADM are single-component, the minimal \emph{fermionic} constructions are necessarily two-component, a fermion and a scalar.

If we restrict ourselves to fully minimal models with only two dark particles, then the dark sector interaction \modeqref{eq:sdmint} must take the form
\begin{equation}
  \lag \supset \phi \, \bar{\psi}^c \psi + h.c. \; ,
\label{eq:darkyuk}\end{equation}
where $\psi^c$ is the charge conjugate fermion.  Before going on to discuss this case in more detail, we briefly note an interesting non-minimal scenario, where instead \modeqref{eq:sdmint} couples a dark and SM fermion:
\begin{equation}
  \lag \supset \phi \, \bar{f} \psi + h.c. \; .
\label{eq:fermportal}\end{equation}
Such a model can be thought of as a generalisation of fermion portal DM~\cite{Bai:2013iqa,DiFranzo:2013vra,Bai:2014osa}.  However, in order for such a model to lead to SA, we must include at least one more dark sector state.  We thus defer such an interesting possibility for future study\footnote{It is possible to have a model with one dark fermion and one dark scalar with both couplings \twoeqref{eq:darkyuk}{eq:fermportal}.  However, both dark sector states would have non-zero hypercharge, and would thus be excluded by direct detection searches.}.

An interesting feature of minimal models is that in large regions of parameter space both dark states are stable.  Absent a coupling of the form \modeqref{eq:fermportal}, the fermion is always stable thanks to an accidental $\set{Z}_2$.  The scalar will also be stable unless the decay $\phi \to \psi\psi$ is open.  In this case, the large hierarchy between the scalar and fermion masses would make the scalar irrelevant to freeze out or dark matter searches, and the phenomenology would reduce to one without semi-annihilation.  For this reason, we will restrict our attention to $m_\phi < 2 m_\psi$.

\begin{table}
  \centering
  \begin{tabular}{|c|c|c|c|}
    \hline
    & Field & $G_{SM}$ & $\set{Z}_4$ \\
    \hline
    \multirow{2}{*}{Singlet Model} &
    $\phi$ & (1, 1, 0) & 2 \\
    & $\psi$ & (1, 1, 0) & 1 \\
    \hline
    \hline
    \multirow{2}{*}{Triplet Model} &
    $\phi$ & (1, 1, 0) & 2 \\
    & $\psi$ & (1, 3, 0) & 1 \\
    \hline
  \end{tabular}
  \caption{New particle content for the two models we consider in this paper.}\label{tab:parts}
\end{table}

The simplest possible model with the interaction \modeqref{eq:sdmint} has a Dirac fermion singlet with charge $q$ under the dark global symmetry, and a scalar singlet with charge $-2q$.  If the global symmetry is $\set{Z}_3$, then there will be a scalar cubic term that can lead to scalar SA.  Since we want to focus on fermion SA, we will instead consider a $\set{Z}_4$ global symmetry.  This acts on the scalar as a $\set{Z}_2$, so that \modeqref{eq:sdmint} is the only source of SA.  In this model, the only connection between the dark sector and the SM is through a Higgs portal coupling.  In particular, the fermion has no direct couplings to any SM states.  This means that, phenomenologically, this model is equivalent to the well-studied scalar singlet model~\cite{Cline:2013gha,Duerr:2015aka,Bishara:2015cha} with only two modifications: it is natural for the scalar to be only a fraction of the DM density; and we have the SA process $\phi \psi \to \bar{\psi} h$ that can contribute to indirect detection signals.

In order to more directly observe on fermion SA, we also consider a non-minimal model where the fermion has SM gauge charges.  We expect that if the fermion has non-zero hypercharge, it will be severely constrained by direct detection measurements due to its unsuppressed coupling to the $Z$ boson.  The next smallest $SU(2)_L$ representation with zero hypercharge and a neutral state is the triplet.  This model thus has some similarities to a supersymmetric wino.  There are three physical fermion states, two charged ($\psi^+,\ \psi^-$) and one neutral ($\psi^0$).  Note that because our fermion is Dirac, each of these three states is likewise Dirac and in particular, $\psi^+$ and $\psi^-$ are not anti-particles of one another.  Radiative corrections split these three states, with the charged states being slightly heavier.  In the limit of heavy fermions, $m_\psi \gg m_W$, the mass splitting is~\cite{Cirelli:2005uq}
\begin{equation}
  \delta m_\psi \equiv m_{\psi^\pm} - m_{\psi^0} \approx 167\text{ MeV.}
\end{equation}
The Dirac nature of the fermion triplet modifies its phenomenology slightly from that of a pure wino, even without SA.  In particular, it weakens constraints involving indirect detection,  strengthens collider limits, and leads to the observed relic density occurring for cross sections below the Sommerfeld resonance instead of above it.

We summarise the new particle content in our two models in \tabref{tab:parts}.  The Lagrangians for both theories may be written as 
\begin{align}
  \lag & = \lag_{SM} + \bar{\psi} (i D\!\!\!\!/ - m_\psi) \psi + \frac{1}{2} (\partial_\mu \phi)^2 + \frac{1}{2} (m_\phi^2 - \lhp v^2) \phi^2 \notag \\
  & \quad + (y \phi \, \bar{\psi}^c \psi + h.c. ) + \frac{1}{2} \, \lhp \, H^\dagger H \, \phi^2 + \frac{1}{4} \, \lambda_{4\phi} \phi^4 \,.
\end{align}
There are five new parameters compared to the SM: the masses of the two dark sector particles $m_\phi$ and $m_\psi$, the Higgs portal coupling $\lhp$, the semi-annihilation coupling $y$ and the new scalar quartic $\lambda_{4\phi}$.  Of these, the last is phenomenologically unimportant, so we effectively have a four-dimensional parameter space.  We may take $y$ real and positive without loss of generality.

\section{Relic Density}\label{sec:RD}
We begin by reviewing the calculation of the thermal relic density, and highlighting some issues that arise in our specific models.  The evolution of species density with time is given by the Boltzmann equation, which for stable particles is
\begin{equation}
  \frac{dY_a}{dx} = - \frac{sZ}{Hx} \sum_{b, i, j, \ldots} \frac{\Delta^a_{ab\to ij\ldots}}{S_{ab}} \biggl( Y_a Y_b - Y_a^{eq} Y_b^{eq} \frac{Y_i Y_j \ldots}{Y_i^{eq} Y_j^{eq} \ldots} \biggr) \sigv (ab \to ij\ldots) \,.
\label{eq:Boltz}\end{equation}
Here, $Y_a = n_a/s$, with $n_a$ the number density of species $a$ and $s$ the entropy density; the superscript $eq$ denotes thermal equilibrium values; $x = T_0/T$ is the inverse temperature, normalised to any convenient scale $T_0$; $H$ is the Hubble expansion rate; $S_{ab}$ is a symmetry factor, equal to 2 if $a = b$ and 1 otherwise; $\Delta^a_{ab\to ij\ldots}$ is the change in the number of particle $a$ in the process $ab \to ij\ldots$; $\sigv$ is the thermally averaged cross section; and
\begin{equation}
  Z = 1 - \frac{x}{3 g_{\ast S}} \frac{d g_{\ast S}}{d x} \,, \quad \text{with} \quad s = \frac{2\pi^2}{45} \, g_{\ast S} T^3 \, ,
\end{equation}
where $g_{\ast S}$ is the effective degrees of freedom in entropy.  Solutions to \modeqref{eq:Boltz} can generically be divided into two domains.  At high temperatures the dark sector states are in thermal equilibrium; the right-hand side vanishes and so $Y_a'(x) =0$.  However, as the temperature drops below the mass the number density begins to decrease exponentially.  Eventually annihilations are too slow to maintain thermal equilibrium, leading to $Y_a \gg Y_a^{eq}$. The process $ab \to ij\ldots$ is said to freeze out, the total number of species $a$ becomes fixed and the number density $n_a$ only changes due to the expansion of space.  For weakly-interacting particles, the temperature at which freeze-out occurs is typically $T_f \sim m/20$.
 
In both our models, we have two species densities, the fermion and the scalar.  We need not separately track the fermion components for two reasons.  First, because our dark sector preserves CP, the DM particle and anti-particle have the same number density, $n_\psi = n_{\psi^c} = \frac{1}{2} n_{\Psi}$, where $n_\Psi$ denotes the total DM number density.  We use $\Psi$ for any fermion or anti-fermion in the dark sector, including all component fields and their charged conjugate. Second, for the fermion triplet, scattering of $\psi$ off the SM thermal bath is much faster than (semi-)annihilations.  Both cross sections are comparable, but the former is proportional to the much larger number density of visible sector particles throughout freeze-out.  This ensures that the component fields of the fermion triplet are in chemical equilibrium and $n_{\psi^\pm} = n_{\psi^0} = \frac{1}{3} n_\Psi$ to high precision for $T \gg \delta m_\psi$.  Since $T_f \gg \delta m_\psi$, we neglect the mass splitting in almost all our relic density calculations except for the computation of the Sommerfeld enhancement, which is relevant at late times $T \sim \delta m_\psi$.  At such times, fermion scattering off the SM bath still maintains thermal equilibrium among the fermion components, but now suppresses $n_{\psi^\pm}$ compared to $n_{\psi^0}$.  After freeze out but before BBN, any remaining charged fermions will decay to the neutral states so $n_\Psi$ will give the correct fermionic DM relic density.

Our models have two complications over the standard single thermal relic.  Both models feature semi-annihilation by design, and we discuss the effects that has on the relic density calculation in \secref{sec:SA}.  The fermion triplet model also features the Sommerfeld enhancement, a non-perturbative increase in fermionic cross sections caused by long-distance interactions and bound state formation.  We discuss this in \secref{sec:SE}. 

\subsection{Semi-Annihilation}\label{sec:SA}

\begin{figure}
  \centering
  \parbox{0.3\textwidth}
  {
    \centering
    \begin{tikzpicture}[node distance=1cm and 0.75cm]
      \coordinate[vertex] (v1);
      \coordinate[right = of v1] (v3);
      \coordinate[vertex, right = of v3 ] (v2);
      \coordinate[above left = of v1, label=above left:$\psi$] (i1);
      \coordinate[below left = of v1, label=below left:$\bar{\psi}$] (i2);
      \coordinate[above right = of v2, label=above right:{$SM$}] (o1);
      \coordinate[below right = of v2, label=below right:{$SM$}] (o2);
      \draw[fermion] (i1) -- (v1);
      \draw[fermion] (v1) -- (i2);
      \draw[photon] (v1) -- (v2) node[midway, above=0.2cm] {$W,Z,\gamma$};
      \draw[fermionnoarrow] (o2) -- (v2);
      \draw[fermionnoarrow] (v2) -- (o1);
    \end{tikzpicture}\\
    \begin{tikzpicture}[node distance=0.5cm and 1.25cm]
      \coordinate[vertex] (v1);
      \coordinate[below = of v1] (v3);
      \coordinate[vertex, below = of v3 ] (v2);
      \coordinate[below = of v2, label=below:(a)] (v4);
      \coordinate[above left = of v1, label=above left:$\psi$] (i1);
      \coordinate[below left = of v2, label=below left:$\bar{\psi}$] (i2);
      \coordinate[above right = of v1, label=above:{$W,Z,\gamma$}] (o1);
      \coordinate[below right = of v2, label=below:{$W,Z,\gamma$}] (o2);
      \draw[fermion] (i1) -- (v1);
      \draw[fermion] (v1) -- (v2) node[midway, left=0.2cm] {$\psi$};
      \draw[fermion] (v2) -- (i2);
      \draw[photon] (o2) -- (v2);
      \draw[photon] (v1) -- (o1);
    \end{tikzpicture}
  }\qquad
  \parbox{0.3\textwidth}
  {
    \centering
    \begin{tikzpicture}[node distance=1cm and 0.75cm]
      \coordinate[vertex] (v1);
      \coordinate[right = of v1] (v3);
      \coordinate[vertex, right = of v3 ] (v2);
      \coordinate[above left = of v1, label=above left:$\psi$] (i1);
      \coordinate[below left = of v1, label=below left:$\psi$] (i2);
      \coordinate[above right = of v2, label=above right:{$h$}] (o1);
      \coordinate[below right = of v2, label=below right:{$\phi$}] (o2);
      \draw[fermion] (i1) -- (v1);
      \draw[fermion] (i2) -- (v1);
      \draw[scalar] (v1) -- (v2) node[midway, above=0.2cm] {$\phi$};
      \draw[scalar] (o2) -- (v2);
      \draw[scalar] (v2) -- (o1);
    \end{tikzpicture}\\
    \begin{tikzpicture}[node distance=0.5cm and 1.25cm]
      \coordinate[vertex] (v1);
      \coordinate[below = of v1] (v3);
      \coordinate[vertex, below = of v3 ] (v2);
      \coordinate[below = of v2, label=below:(b)] (v4);
      \coordinate[above left = of v1, label=above left:$\psi$] (i1);
      \coordinate[below left = of v2, label=below left:$\psi$] (i2);
      \coordinate[above right = of v1, label=above:{$W,Z,\gamma$}] (o1);
      \coordinate[below right = of v2, label=below right:{$\phi$}] (o2);
      \draw[fermion] (i1) -- (v1);
      \draw[fermion] (v2) -- (v1) node[midway, left=0.2cm] {$\psi$};
      \draw[fermion] (i2) -- (v2);
      \draw[scalar] (o2) -- (v2);
      \draw[photon] (v1) -- (o1);
    \end{tikzpicture}
  }\qquad
  \begin{tikzpicture}[node distance=0.5cm and 1.25cm]
    \coordinate[vertex] (v1);
    \coordinate[below = of v1 ] (v2);
    \coordinate[vertex, below = of v2 ] (v3);
    \coordinate[below = of v3, label=below:(c)] (v4);
    \coordinate[above right = of v1, label=above right:$\phi$] (i1);
    \coordinate[below right = of v3, label=below right:$\phi$] (i2);
    \coordinate[above left = of v1, label=above left:{$\bar{\psi}$}] (o1);
    \coordinate[below left = of v3, label=below left:{$\psi$}] (o2);
    \draw[scalar] (i1) -- (v1);
    \draw[scalar] (v3) -- (i2);
    \draw[fermion] (o1) -- (v1);
    \draw[fermion] (v3) -- (v1) node[midway, left=0.2cm] {$\psi$};;
    \draw[fermion] (v3) -- (o2);
  \end{tikzpicture}
  \caption{Processes relevant for the computation of the fermion relic density: (a) Fermion annihilation to the SM; (b) Fermion SA; (c) Dark matter exchange.  The diagrams of (a) and the second of (b) are only present for the fermion triplet model.}\label{fig:fer_FD}
\end{figure}
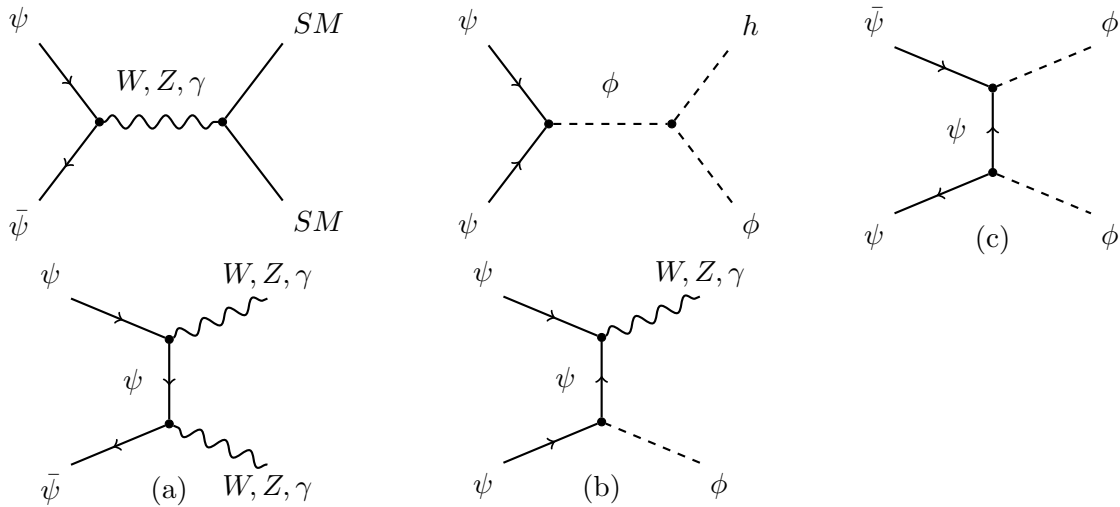

The general formalism for two-component DM considering only two-body final states can be found in \emph{e.g.} \refcite{micromegas4.1}.  Restricting our focus to only those processes which are non-vanishing in either of our models, we may write the coupled Boltzmann equations for $Y_\phi$ and $Y_\Psi$ as
\begin{align}
  \frac{d Y_\Psi}{d x} & = \frac{s Z}{H x} \, \biggl[ \bigl(Y_\Psi^2 - (Y_\Psi^{eq})^2 \bigr) \sigv (\Psi\Psi \to SM) + \biggl( Y_\Psi^2 - Y_\phi \frac{(Y_\Psi^{eq})^2}{Y_\phi^{eq}} \biggr) \sigv (\Psi\Psi \to\phi SM) \notag \\
  & \quad + \biggl( Y_\Psi^2 - Y_\phi^2 \frac{(Y_\Psi^{eq})^2}{(Y_\phi^{eq})^2} \biggr) \sigv (\Psi\Psi\to\phi\phi) \biggr] \,, \label{eq:fermdY} \displaybreak[0]\\
  \frac{d Y_\phi}{d x} & = \frac{s Z}{H x} \, \biggl[ \bigl(Y_\phi^2 - (Y_\phi^{eq})^2 \bigr) \sigv (\phi\phi \to SM) + Y_\Psi \bigl( Y_\phi - Y_\phi^{eq} \bigr) \sigv (\Psi\phi \to \Psi SM) \notag \\
  & \quad + \frac{1}{2} \biggl( Y_\phi \frac{(Y_\Psi^{eq})^2}{Y_\phi^{eq}} - Y_\Psi^2 \biggr) \sigv (\Psi\Psi \to \phi SM) + \biggl( Y_\phi^2 \frac{(Y_\Psi^{eq})^2}{(Y_\phi^{eq})^2} - Y_\Psi^2 \biggr)\sigv (\Psi\Psi \to \phi\phi) \biggr] \,. \label{eq:scadY}
\end{align}
We show relevant Feynman diagrams in \figsref{fig:fer_FD}{fig:sca_FD}.  There are two comments to make here.  First, cross sections for processes with two fermions in the initial state have an additional factor of one-half for averaging over both fermions and anti-fermions.  Second, we have written both expressions in terms of the DME process $\Psi\Psi \to \phi\phi$.  This is the most convenient form when $m_\psi > m_\phi$, as then $Y_\Psi^{eq} < Y_\phi^{eq}$ and we avoid numerically large ratios.  If instead $m_\psi < m_\phi$, we should make use of the identity
\begin{equation}
  \biggl( Y_\Psi^2 - Y_\phi^2 \frac{(Y_\Psi^{eq})^2}{(Y_\phi^{eq})^2} \biggr) \sigv (\phi\phi \to \Psi\Psi) = \biggl( Y_\Psi^2 \frac{(Y_\phi^{eq})^2}{(Y_\Psi^{eq})^2} - Y_\phi^2 \biggr) \sigv (\Psi\Psi\to\phi\phi) \,.
\end{equation}

\begin{figure}
  \centering
  \parbox{0.3\textwidth}
  {
    \centering
    \begin{tikzpicture}[node distance=1cm and 0.75cm]
      \coordinate[vertex] (v1);
      \coordinate[right = of v1] (v3);
      \coordinate[vertex, right = of v3 ] (v2);
      \coordinate[above left = of v1, label=above left:$\phi$] (i1);
      \coordinate[below left = of v1, label=below left:$\phi$] (i2);
      \coordinate[above right = of v2, label=above right:{$SM$}] (o1);
      \coordinate[below right = of v2, label=below right:{$SM$}] (o2);
      \draw[scalar] (i1) -- (v1);
      \draw[scalar] (v1) -- (i2);
      \draw[scalar] (v1) -- (v2) node[midway, above=0.2cm] {$h$};
      \draw[fermionnoarrow] (o2) -- (v2);
      \draw[fermionnoarrow] (v2) -- (o1);
    \end{tikzpicture}\\
    \begin{tikzpicture}[node distance=0.5cm and 1cm]
      \coordinate[vertex] (v1);
      \coordinate[below = of v1] (v3);
      \coordinate[vertex, below = of v3 ] (v2);
      \coordinate[below = of v2, label=below:(a)] (v4);
      \coordinate[above left = of v1, label=above left:$\phi$] (i1);
      \coordinate[below left = of v2, label=below left:$\phi$] (i2);
      \coordinate[above right = of v1, label=above right:{$h$}] (o1);
      \coordinate[below right = of v2, label=below right:{$h$}] (o2);
      \draw[scalar] (i1) -- (v1);
      \draw[scalar] (v2) -- (i2);
      \draw[scalar] (v1) -- (v2) node[midway, left=0.2cm] {$\phi$};
      \draw[scalar] (o2) -- (v2);
      \draw[scalar] (v1) -- (o1);
    \end{tikzpicture}
  }\qquad
  \parbox{0.3\textwidth}
  {
    \centering
    \begin{tikzpicture}[node distance=0.5cm and 1.25cm]
      \coordinate[vertex] (v1);
      \coordinate[below = of v1 ] (v2);
      \coordinate[vertex, below = of v2 ] (v3);
      \coordinate[above left = of v1, label=above left:$\psi$] (i1);
      \coordinate[below left = of v3, label=below left:$\phi$] (i2);
      \coordinate[above right = of v1, label=above right:{$\bar{\psi}$}] (o1);
      \coordinate[below right = of v3, label=below right:{$h$}] (o2);
      \draw[scalar] (i2) -- (v3);
      \draw[scalar] (v1) -- (v3) node[midway, left=0.2cm] {$\phi$};
      \draw[scalar] (v3) -- (o2);
      \draw[fermion] (i1) -- (v1);
      \draw[fermion] (o1) -- (v1);
    \end{tikzpicture}\\
    \begin{tikzpicture}[node distance=1cm and 0.75cm]
      \coordinate[vertex] (v1);
      \coordinate[right = of v1] (v3);
      \coordinate[vertex, right = of v3 ] (v2);
      \coordinate[below = of v3, label=below:(b)] (v4);
      \coordinate[above left = of v1, label=above left:$\psi$] (i1);
      \coordinate[below left = of v1, label=below left:$\phi$] (i2);
      \coordinate[above right = of v2, label=above right:{$\bar{\psi}$}] (o1);
      \coordinate[below right = of v2, label=below :{$W,Z,\gamma$}] (o2);
      \draw[fermion] (i1) -- (v1);
      \draw[scalar] (v1) -- (i2);
      \draw[fermion] (v2) -- (v1) node[midway, above=0.2cm] {$\bar{\psi}$};
      \draw[photon] (o2) -- (v2);
      \draw[fermion] (o1) -- (v2);
    \end{tikzpicture}
  }\qquad
  \begin{tikzpicture}[node distance=0.5cm and 1.25cm]
    \coordinate[vertex] (v1);
    \coordinate[below = of v1 ] (v2);
    \coordinate[vertex, below = of v2 ] (v3);
    \coordinate[below = of v3, label=below:(c)] (v4);
    \coordinate[above left = of v1, label=above left:$\phi$] (i1);
    \coordinate[below left = of v3, label=below left:$\phi$] (i2);
    \coordinate[above right = of v1, label=above right:{$\bar{\psi}$}] (o1);
    \coordinate[below right = of v3, label=below right:{$\psi$}] (o2);
    \draw[scalar] (i1) -- (v1);
    \draw[scalar] (v3) -- (i2);
    \draw[fermion] (o1) -- (v1);
    \draw[fermion] (v3) -- (v1) node[midway, left=0.2cm] {$\psi$};;
    \draw[fermion] (v3) -- (o2);
  \end{tikzpicture}
  \caption{Processes relevant for the computation of the scalar relic density: (a) Scalar annihilation to the SM; (b) Scalar SA; (c) DME.    The second diagram of (b) is only present in the fermion triplet model.  The diagrams of \figref{fig:fer_FD} (b) are also relevant.}\label{fig:sca_FD}
\end{figure}
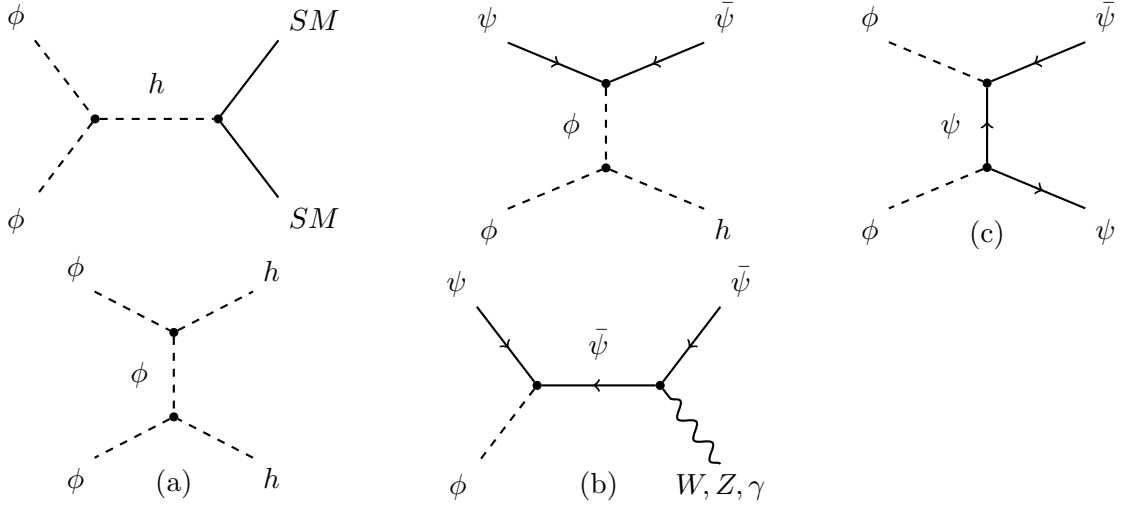

\begin{figure}
  \centering
  \includegraphics[width=0.48\textwidth]{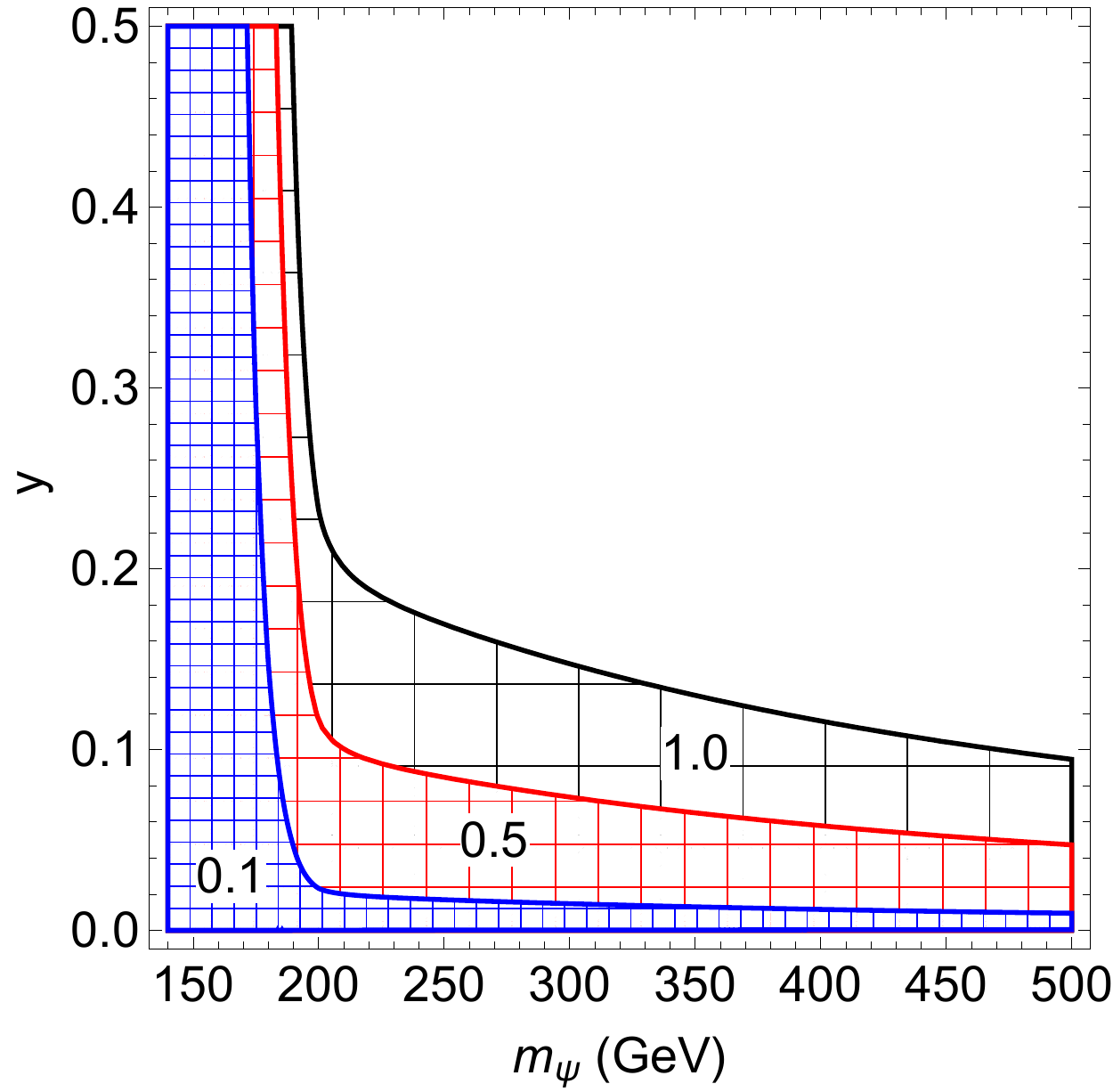}
  \includegraphics[width=0.48\textwidth]{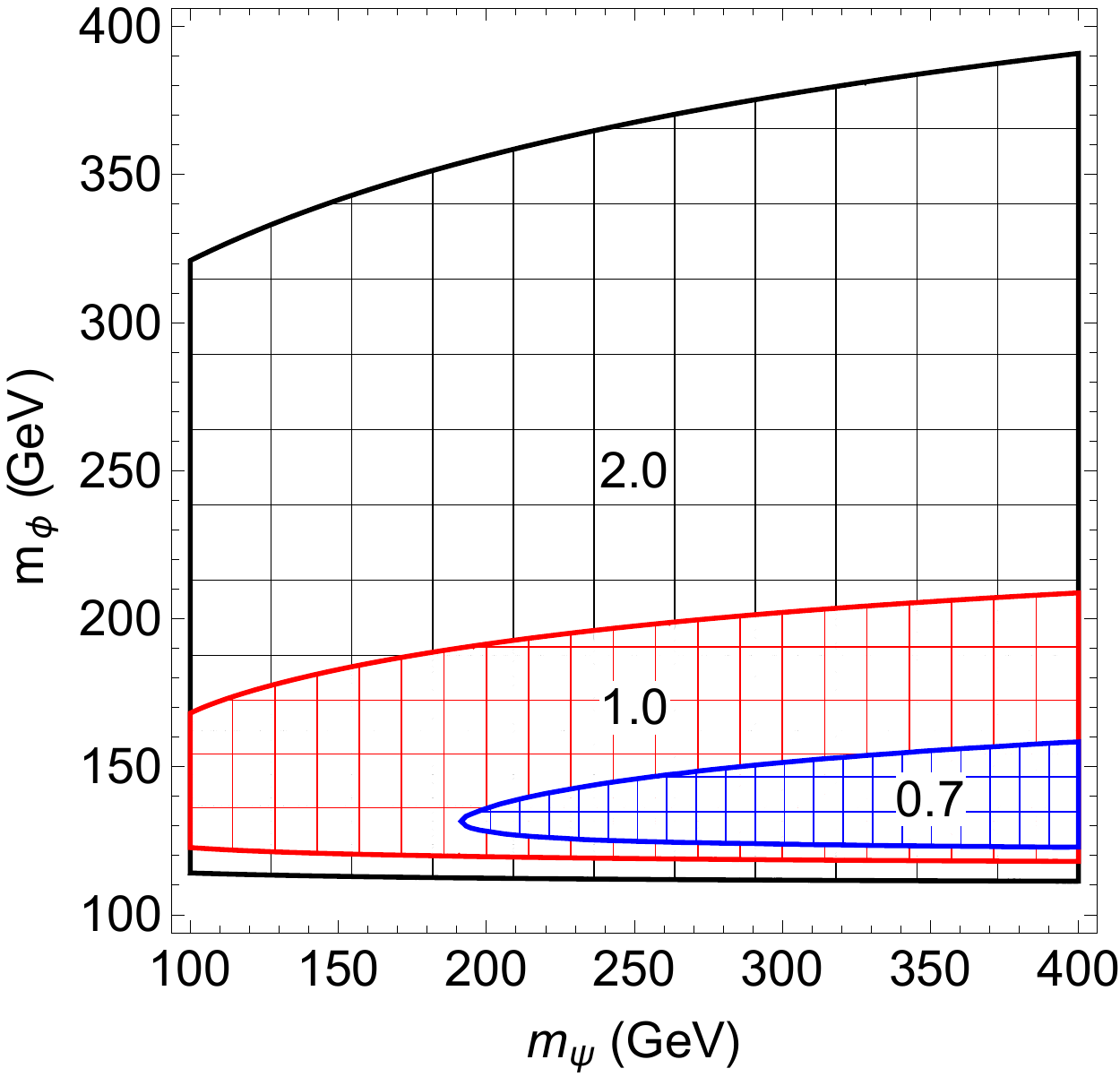}
  \caption{(Left:) Regions where fermion SA dominates over DME in the fermion singlet model, for $m_\phi = 200$~GeV and Higgs portal coupling $\lhp = 0.1$ (0.5, 1.0) in the blue (red, black) region. (Right:) Regions where scalar SA cross section is larger than that for annihilation to the SM, for fermion-scalar coupling $y = 0.7$ (1.0, 2.0) in the blue (red, black) region.}\label{fig:ferm1xsec}
\end{figure}

The three terms in \modeqref{eq:fermdY} are precisely the annihilation, SA and DME terms in order.  For \modeqref{eq:scadY}, there are four terms, of which the middle two are SA.  We can estimate when we expect the different terms to be important by comparing the parametric cross sections.  For the fermion singlet model, there is no fermion annihilation term while the SA and DME channels have no $s$-wave piece.  The leading terms at freeze-out are
\begin{equation}
  \sigma v (\psi\psi \to\phi h) \sim \frac{\lhp^2y^2}{1024\pi m_\psi^2} \frac{\vew^2}{m_\psi^2} \, v^2 \quad \text{and}\quad \sigma v (\psi\bar{\psi} \to\phi\phi) \sim \frac{3y^4}{128\pi m_\psi^2} v^2 \,,
\label{eq:f1bterms}\end{equation}
where $\vew = 246$~GeV is the SM Higgs VEV and $v$ is the relative annihilation velocity.  From this we see that DME will dominate unless either
\begin{equation}
  \lhp \gtrsim \frac{5 m_\psi}{\vew} \, y \,,
\label{eq:f1lhpbound}\end{equation}
or $m_\phi > m_\psi > \frac{1}{2} (m_\phi + m_h)$ so that DME is kinematically forbidden but SA is not.  To illustrate this, in the left-hand side of \figref{fig:ferm1xsec} we show regions where SA is the dominant channel for $m_\phi = 200$~GeV, $T = m_\psi/25$, and for differing values of $\lhp$.  This figure uses the full thermally-averaged cross section, rather than the approximations of Eqs.~\eqref{eq:f1bterms} or \eqref{eq:f1lhpbound}.

For the scalar in the fermion singlet model, we expect the $\phi\phi\to SM$ and $\phi\Psi \to \Psi h$ channels to dominate.  The other two channels, $\psi\psi\to \phi h$ and $\psi\bar{\psi}\to \phi\phi$, are precisely those that allow the fermion number to decrease, and so one or both must lead to an increase in the scalar number density.  To avoid too large a scalar relic density leads us to focus on the annihilation and SA channels mentioned.  Neglecting SM masses,
\begin{equation}
  \sigv (\phi\phi \to SM) \sim \frac{\lhp^2}{16\pi m_\phi^2} \quad\text{and}\quad \sigv (\Psi\phi \to \Psi SM) \sim \frac{\lhp^2y^2}{8\pi} \frac{\vew^2 m_\psi (2 m_\psi + m_\phi)}{m_\phi^3 (m_\psi + m_\phi)^3} \,.
\end{equation}
From this, including a factor of 2 for the different change in scalar number of the two processes, we expect the SA channel to dominate for
\begin{equation}
  y^2 \gtrsim \frac{m_\phi (m_\psi + m_\phi)^3}{\vew^2 m_\psi (2 m_\psi + m_\phi)} \sim 
  \begin{cases}
    m_\phi^3/(\vew^2 m_\psi) & \text{if } m_\phi \gg m_\psi \,,\\
    8 M^2/(3\vew^2) & \text{if } m_\psi \sim m_\phi \equiv M \,.
  \end{cases}
\end{equation}
Interestingly, note that SA for the scalar can dominate for \emph{any} value of $\lhp$.  We do not consider the case $m_\psi \gg m_\phi$, as in that case we would expect $n_\Psi \ll n_\phi$ during scalar freeze-out, suppressing SA.  In the right-hand side of \figref{fig:ferm1xsec},  we show regions where the scalar SA cross section is larger for different values of $y$ at $T = m_\phi/25$.

For the fermion triplet model, the fermion annihilation channel is set by SM parameters only, and has non-zero $s$-wave piece.  Additional SA channels to $\phi V$, with $V$ a SM gauge boson, also have non-vanishing $s$-wave terms, while DME remains pure $p$-wave.  As such, we expect either annihilation or SA to dominate.  At tree-level,
\begin{equation}
  \sigv (\psi\bar{\psi} \to SM) \sim \frac{37}{6} \, \pi \alpha_2^2 \quad \text{and}\quad \sigv (\psi\psi \to\phi SM) \sim \frac{4}{3} \, \alpha_2 y^2 \,,
\end{equation}
with $\alpha_2 = g_2^2/4\pi$.  From this, we expect the two processes to be comparable when
\begin{equation}
  y \approx \sqrt{\frac{37}{8} \, \pi \alpha_2} \approx 0.7 \,.
\end{equation}
For the scalar, we expect similar conclusions to in the fermion singlet case when $m_\phi < m_\psi$.  For heavy scalars, we generally expect the DME process to dominate, due to a large numerical enhancement:
\begin{equation}
  \sigv (\phi\phi \to \Psi\Psi) \sim \frac{12 y^4 m_\psi^2}{\pi m_\phi^4} \,.
\end{equation}
This will dominate the annihilation to the SM unless
\begin{equation}
  y^2 < \frac{\lhp}{192} \, \frac{m_\phi}{m_\psi}\,.
\end{equation}

For tree-level processes, including some 3- and 4-body final states from off-shell gauge bosons, the relic density for general models can be computed in micrOMEGAs~4.1~\cite{micromegas4.1}.  For our fermion singlet model, this is sufficient and hence the approach we take.  For our fermion triplet model, however, we must include the Sommerfeld effect as discussed in the following section.

\subsection{Sommerfeld Effect}\label{sec:SE}

The Sommerfeld effect (SE) is a non-perturbative modification of cross sections caused by the presence of a long-range interaction~\cite{Gamow,Sommerfeld,Hisano:2004ds,Hisano:2006nn,Iengo:2009ni,2010PhRvD..81h3502Z,Feng:2009hw}.  It can be understood as a modification of the incoming two-particle wavefunction, either enhancing or suppressing it at the origin (interaction point).  Generically, for a force mediated by a boson of mass $m_V$ and with coupling strength $\alpha_V$, the SE is important if the annihilating particles have mass $M \sim m_V/\alpha_V$.  For weak interactions, $m_W/\alpha_2 \approx 2.5$~TeV.  It is well-known (see \emph{e.g.}~\refscite{Cirelli:2005uq,Hisano:2004ds,Hisano:2006nn,Hryczuk:2011vi,Cirelli:2007xd}) that electroweak triplet dark matter has the correct relic density for masses of this order, and so this correction must be included.

As an aside, in our calculations we will include forces mediated by the dark scalar $\phi$.  One might ask why we do not then include the SE for our fermion singlet model.  The reason is that the effect would only be relevant for $m_\psi \gg m_\phi$.  In this region of parameter space, the SA channel $\psi\phi \to \bar{\psi}h$ would be irrelevant and the phenomenology of the model would reduce to that of the scalar singlet.

We follow the formalism of \refcite{Cirelli:2007xd}.  We split the annihilation channels into subgroups for the unbroken quantum numbers: charge $Q$, angular momentum $J$ and $\set{Z}_4$ charge $q$.  The Sommerfeld effect is relevant at late times, so we focus on $s$-wave annihilation $L = 0$; hence $J = S$, the total spin.  Because the scalar is a singlet, the only channels we need consider are those with two fermions in the initial state.  The possible quantum numbers are $Q\in \{0, \pm 1, \pm 2\}$, $S \in \{0, 1\}$ and $q \in \{0, 2\}$.  Because the dark sector respects $CP$, the SE is identical for initial states with opposite charge.

For each subspace, we solve a Schr\"odinger equation for a generally matrix-valued two-particle wavefunction $\Phi_{ij}$:
\begin{equation}
  - \frac{1}{M} \, \Phi_{ij}'' (r) + \sum_k V_{ik} (r) \Phi_{kj} (r) = K \Phi_{ij} (r) \,,
\label{eq:SomEffSchEq}\end{equation}
with $M$ the mass,  $K$ the centre of momentum frame kinetic energy at large separation, $r$ the separation and $V_{ij} (r)$ the long-range potential.  The indices label different two-particle states.  The wavefunction satisfies the boundary conditions
\begin{equation}
  \Phi_{ij} (0) = \delta_{ij} \quad\text{and}\quad \lim_{r\to\infty} \frac{\Phi_{ij}'(r)}{\Phi_{ij} (r)} = i \sqrt{M (K - V_{ii} (\infty)) } \text{ (no sum)} \,.
\label{eq:SEBC}\end{equation}
The enhancement matrix $A_{ij}$ is given by
\begin{equation}
  A_{ij} = \lim_{r\to\infty} \frac{\Phi_{ij} (r)}{\exp(i \Re \sqrt{M (K - V_{ii} (\infty))} r)} \,,
\end{equation}
such that $A_{ij} = \delta_{ij}$ in the absence of the SE.  The final cross section is given by
\begin{equation}
  \sigma_i = c_i (A \cdot \Gamma \cdot A^\dagger)_{ii} \,,
\end{equation}
where $\Gamma$ are annihilation matrices given below and $c_i = 1$ (2) if the state $i$ contains distinct (identical) particles.  This formalism automatically includes the fact that some states can only exist at large separation if $K$ is sufficiently large, and thus the thermal suppression in $n_{\psi^+}$ compared to $n_{\psi^0}$ discussed previously.

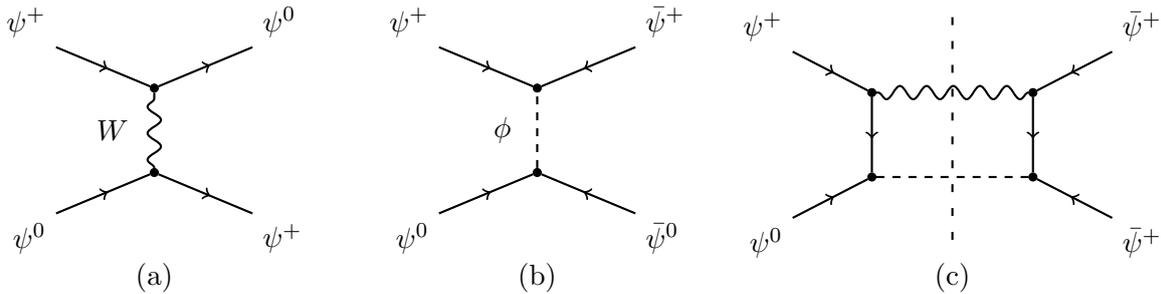
\begin{figure}
  \centering
  \begin{tikzpicture}[node distance=0.5cm and 1.25cm]
    \coordinate[vertex] (v1);
    \coordinate[below = of v1 ] (v2);
    \coordinate[vertex, below = of v2 ] (v3);
    \coordinate[below = of v3] (v5);
    \coordinate[below = of v5, label=below:(a)] (v4);
    \coordinate[above left = of v1, label=above left:$\psi^+$] (i1);
    \coordinate[below left = of v3, label=below left:$\psi^0$] (i2);
    \coordinate[above right = of v1, label=above right:{$\psi^0$}] (o1);
    \coordinate[below right = of v3, label=below right:{$\psi^+$}] (o2);
    \draw[fermion] (i1) -- (v1);
    \draw[fermion] (i2) -- (v3);
    \draw[fermion] (v1) -- (o1);
    \draw[photon] (v3) -- (v1) node[midway, left=0.2cm] {$W$};;
    \draw[fermion] (v3) -- (o2);
  \end{tikzpicture} \qquad
  \begin{tikzpicture}[node distance=0.5cm and 1.25cm]
    \coordinate[vertex] (v1);
    \coordinate[below = of v1 ] (v2);
    \coordinate[vertex, below = of v2 ] (v3);
    \coordinate[below = of v3] (v5);
    \coordinate[below = of v5, label=below:(b)] (v4);
    \coordinate[above left = of v1, label=above left:$\psi^+$] (i1);
    \coordinate[below left = of v3, label=below left:$\psi^0$] (i2);
    \coordinate[above right = of v1, label=above right:{$\bar{\psi}^+$}] (o1);
    \coordinate[below right = of v3, label=below right:{$\bar{\psi}^0$}] (o2);
    \draw[fermion] (i1) -- (v1);
    \draw[fermion] (i2) -- (v3);
    \draw[fermion] (o1) -- (v1);
    \draw[scalar] (v3) -- (v1) node[midway, left=0.2cm] {$\phi$};;
    \draw[fermion] (o2) -- (v3);
  \end{tikzpicture}  \quad
  \begin{tikzpicture}[node distance=0.5cm and 1cm]
    \coordinate[vertex] (v1);
    \coordinate[below = of v1 ] (u1);
    \coordinate[vertex, below = of u1 ] (v2);
    \coordinate[right = of v1] (u2);
    \coordinate[vertex, right = of u2] (v3);
    \coordinate[right = of v2] (u3);
    \coordinate[vertex, right = of u3] (v4);
    \coordinate[below = of u3] (u7);
    \coordinate[below = of u7, label=below:(c)] (u4);
    \coordinate[above = of u2] (u6);
    \coordinate[above = of u6] (u5);
    \coordinate[above left = of v1, label=above left:$\psi^+$] (i1);
    \coordinate[below left = of v2, label=below left:$\psi^0$] (i2);
    \coordinate[above right = of v3, label=above right:{$\bar{\psi}^+$}] (o1);
    \coordinate[below right = of v4, label=below right:{$\bar{\psi}^+$}] (o2);
    \draw[fermion] (i1) -- (v1);
    \draw[fermion] (v1) -- (v2);
    \draw[fermion] (i2) -- (v2);
    \draw[fermion] (o1) -- (v3);
    \draw[fermion] (v3) -- (v4);
    \draw[fermion] (o2) -- (v4);
    \draw[photon] (v3) -- (v1);
    \draw[scalar] (v2) -- (v4);
    \draw[style={draw=black, postaction={decorate}, thick, loosely dashed}] (u5) -- (u4);
  \end{tikzpicture} \quad
  \caption{Example diagrams contributing to the fermion potential and interaction matrices.  (a) Leading term for the $\psi^+ \psi^0$ diagonal potential. (b) Leading term for the off-diagonal potential term between $\psi^+ \psi^0$ and $\bar{\psi}^+ \bar{\psi}^0$. (c) Leading term for the off-diagonal ``annihilation'' term between $\psi^+ \psi^0$ and $\bar{\psi}^+ \bar{\psi}^0$.  The dashed lines denote Cutovsky cuts.}\label{fig:VandG}
\end{figure}

It is clear that the main model-dependent elements of this calculation are the matrices $V$ and $\Gamma$.  The diagonal entries of these matrices have obvious physical interpretations as the potential energies and annihilation cross sections of the associated two-body state.  The off-diagonal elements are less transparent; the matrices are formally defined in terms of the real ($V$) and imaginary ($\Gamma$) parts of the generalised two-body propagator $ij\to kl$.  We show some relevant Feynman diagrams in \figref{fig:VandG}.

The $q = 2$ subspace corresponds to the SA process $\Psi\Psi \to\phi SM$.  As such the SE in this sector has not previously been considered in the literature.  The $Q = \pm 2$ subspace is negligible, as it has no two-body final states into which it can annihilate.  The $Q = \pm 1$ subspaces are two-dimensional, $\{\psi^+\psi^0, \bar{\psi}^+\bar{\psi}^0\}$ where $\bar{\psi}^+$ is the antiparticle of $\psi^-$.  Note that the fermion and anti-fermion states have the same $\set{Z}_4$ charge $q$, and indeed can be converted into one another by $\phi$ exchange.  In the limit where we can neglect terms of order $m_{W}/m_\psi$, the spin-0 state has vanishing cross section while the spin-1 potential and annihilation matrices are
\begin{equation}
  V(r) = \delta m_\psi \, \delta_{ij} - \begin{pmatrix} W(r) & Y(r) \\ Y(r) & W(r) \end{pmatrix} ,\quad\text{and}\quad \Gamma = \frac{\alpha_2 y^2}{8m_\psi^2} \, \biggl(4 - \frac{m_\phi^2}{m_\psi^2} \biggr) \begin{pmatrix} 1 & 1 \\ 1 & 1 \end{pmatrix} ,
\end{equation}
where for later convenience we have introduced the functions
\begin{equation}
  W(r) = \frac{\alpha_2}{r}\, e^{-m_W r} \quad\text{and}\quad Y(r) = \frac{y^2}{4\pi r}\, e^{-m_\phi r} \,.
\end{equation}
Note that if we include the spin-0 channel, it has the same potential matrix and an annihilation matrix with the same structure (all entries equal).

There are four possible two-fermion neutral states: $\{\psi^+\psi^-,\psi^0\psi^0,\bar{\psi}^0\bar{\psi}^0,\bar{\psi}^+\bar{\psi}^-\}$.  However, two of these involve identical fermions and so can only exist in the anti-symmetric $S=0$ state.  For the two-dimensional $S=1$ space neglecting terms of order $m_Z/m_\psi$, we have
\begin{equation}
  V(r) = 2 \delta m_\psi \, \delta_{ij} - \begin{pmatrix} Z(r) & Y(r) \\ Y(r) & Z(r) \end{pmatrix} ,\quad\text{and}\quad \Gamma = \frac{\alpha_2 y^2}{8m_\psi^2} \, \biggl(4 - \frac{m_\phi^2}{m_\psi^2} \biggr) \begin{pmatrix} 1 & 1 \\ 1 & 1 \end{pmatrix} ,
\end{equation}
where we have introduce an additional function
\begin{equation}
  Z(r) = \frac{\alpha}{r} + \frac{\alpha_2 c^2_W}{r} \, e^{-m_Z r} .
\end{equation}
The $S=0$ cross section vanishes in the limit $m_Z \to 0$.  Keeping the leading term in the expansion in terms of $m_Z/m_\psi$, the potential and annihilation matrices are
\begin{equation}
  V(r) = - \begin{pmatrix} Z - 2 \delta m_\psi & \sqrt{2} W & 0 & Y \\ \sqrt{2}W & 0 & 2Y & 0\\0 & 2Y & 0 & \sqrt{2}W\\Y & 0 & \sqrt{2}W & Z - 2 \delta m_\psi \end{pmatrix} ,\ \text{and}\ \Gamma = \frac{\alpha_2 c_W^2 y^2 m_Z^2}{m_\psi^4} \begin{pmatrix} 1 & 0 & 0 & 1 \\ 0 & 0 & 0 & 0\\0 & 0 & 0 & 0\\1 & 0 & 0 & 1\end{pmatrix} .
\label{eq:S0Q0SA}\end{equation}
This is the only SA channel in this class after the charged fermions have decayed, and is suppressed by both $m_Z/m_\psi$ and the mass splitting.  This has the effect of suppressing indirect detection signals when $y$ is moderately large, as we discuss in \secref{sec:ID}.

The $q = 0$ subspace corresponds to the annihilation $\Psi\Psi \to SM$.  In principle, it would also modify the dark matter exchange process $\Psi\Psi \to \phi\phi$; however, that channel has   vanishing $s$-wave component, and so we neglect it.  This then is very similar to the well-studied case of Majorana triplets.  However, the Dirac nature of our fermions introduces some factors of 2, and enlarges the neutral $S=1$ subspace.  Additionally we keep contributions to the potential mediated by $\phi$.  We thus list all relevant terms below.

The $Q = \pm 2$ subspaces are one-dimensional, and the $s$-wave annihilation is pure spin-0.  Neglecting terms of order $m_W/m_\psi$, we have
\begin{equation}
  V(r) = 2 \delta m_\psi + Z(r) - Y(r) \spand \Gamma = \frac{\pi \alpha_2^2}{m_\psi^2} \,.
\end{equation}
The $Q = \pm 1$ subspaces are two-dimensional, with states $\{ \psi^+ \bar{\psi}^0, \bar{\psi}^+ \psi^0\}$.  There are non-trivial contributions in both the spin-0 and spin-1 channels.  For the $S=1$ case,
\begin{equation}
  V(r) = \begin{pmatrix} \delta m_\psi & W(r) - Y(r) \\ W(r) - Y(r) & \delta m_\psi \end{pmatrix} \spand \Gamma = \frac{25 \pi \alpha_2^2}{8 m_\psi^2} \begin{pmatrix} \phantom{-}1 & -1 \\ -1 &\phantom{-}1 \end{pmatrix} \,.
\end{equation}
For the $S=0$ case, the potential matrix is the same while the annihilation matrix is
\begin{equation}
  \Gamma = \frac{\pi \alpha_2^2}{2 m_\psi^2} \begin{pmatrix} 1 & 1 \\ 1 & 1 \end{pmatrix} \,.
\end{equation}
Finally, the neutral subspace is three-dimensional: $\{ \psi^+ \bar{\psi}^-, \psi^0 \bar{\psi}^0, \psi^- \bar{\psi}^+ \}$.  As noted, unlike for Majorana fermions, it remains three-dimensional in both the spin-0 and spin-1 cases.  For $S=1$, we have
\begin{equation}
  V(r) = - \begin{pmatrix} Z(r) - 2 \delta m_\psi & W(r) & Y(r) \\ W(r) & Y(r) & W(r) \\ Y(r) & W(r) & Z(r) - 2 \delta m_\psi \end{pmatrix} \text{ and } \Gamma = \frac{25\pi\alpha_2^2}{8m_\psi^2} \begin{pmatrix} \phantom{-}1 & 0 & -1\\ \phantom{-}0 & 0 & \phantom{-}0 \\ -1 & 0 & \phantom{-}1 \end{pmatrix} \,.
\label{eq:q0pot}\end{equation}
The $S=0$ subspace has the same potential matrix, and the annihilation matrix
\begin{equation}
  \Gamma = \frac{3\pi\alpha_2^2}{2m_\psi^2} \begin{pmatrix}1 & 0 & 1 \\ 0 & 0 & 0 \\ 1 & 0 & 1\end{pmatrix} + \frac{2\pi\alpha_2^2}{m_\psi^2} \begin{pmatrix} 0 & 1 & 0 \\ 1 & 1 & 1 \\ 0 & 1 & 0 \end{pmatrix} \,.
\label{eq:s0q0ann}\end{equation}

In computing the SE cross sections, we numerically solved the different Schr\"odinger equations~\eqref{eq:SomEffSchEq} for all the different subchannels.  We retained full dependence on SM masses in the annihilation matrices $\Gamma$.  We did not include thermal masses for the SM gauge bosons; this has been shown to not significantly affect the relic densities~\cite{Cirelli:2007xd}.  The SE equation of \modeqref{eq:SomEffSchEq} has boundary conditions at both ends of the interval $r \in [0, \infty)$; we first convert this to a problem with boundary conditions only at $r = \infty$.  If $\Phi$ is a solution to \modeqref{eq:SomEffSchEq} with boundary conditions~\eqref{eq:SEBC}, then for any constant matrix $C$, $\Xi \equiv \Phi \cdot C$ is \emph{also} a solution to \modeqref{eq:SomEffSchEq} with the same boundary conditions at infinity.  Further, $\Xi (0) = C$.  It follows that the SE matrix $A$ in terms of $\Xi$ is
\begin{equation}
  A_{ij} = \lim_{r\to\infty} \frac{\sum_j \Xi_{ik} (r) \Xi^{-1}_{kj} (0)}{\exp(i \Re \sqrt{M (K - V_{ii} (\infty))} r)} \,.
\end{equation}
This allows us to use (almost) any invertible matrix $\Xi^\infty$ as initial conditions at $r = \infty$ and still solve for $A$.\footnote{In practice, poor choices for $\Xi^\infty$ will enhance numerical errors.}  We used the Dormand-Prince (RKDP) algorithm~\cite{Dormand:1980ub} (an adaptive 4th-order Runge-Kutta method) to compute $A$ with a maximum step error of 0.1\%.  We compare a selection of points with the results for a maximum step error of 0.01\%, and find a numerical uncertainty of less than 1\% in the final cross sections.

\begin{figure}
  \centering
  \includegraphics[width=0.49\textwidth]{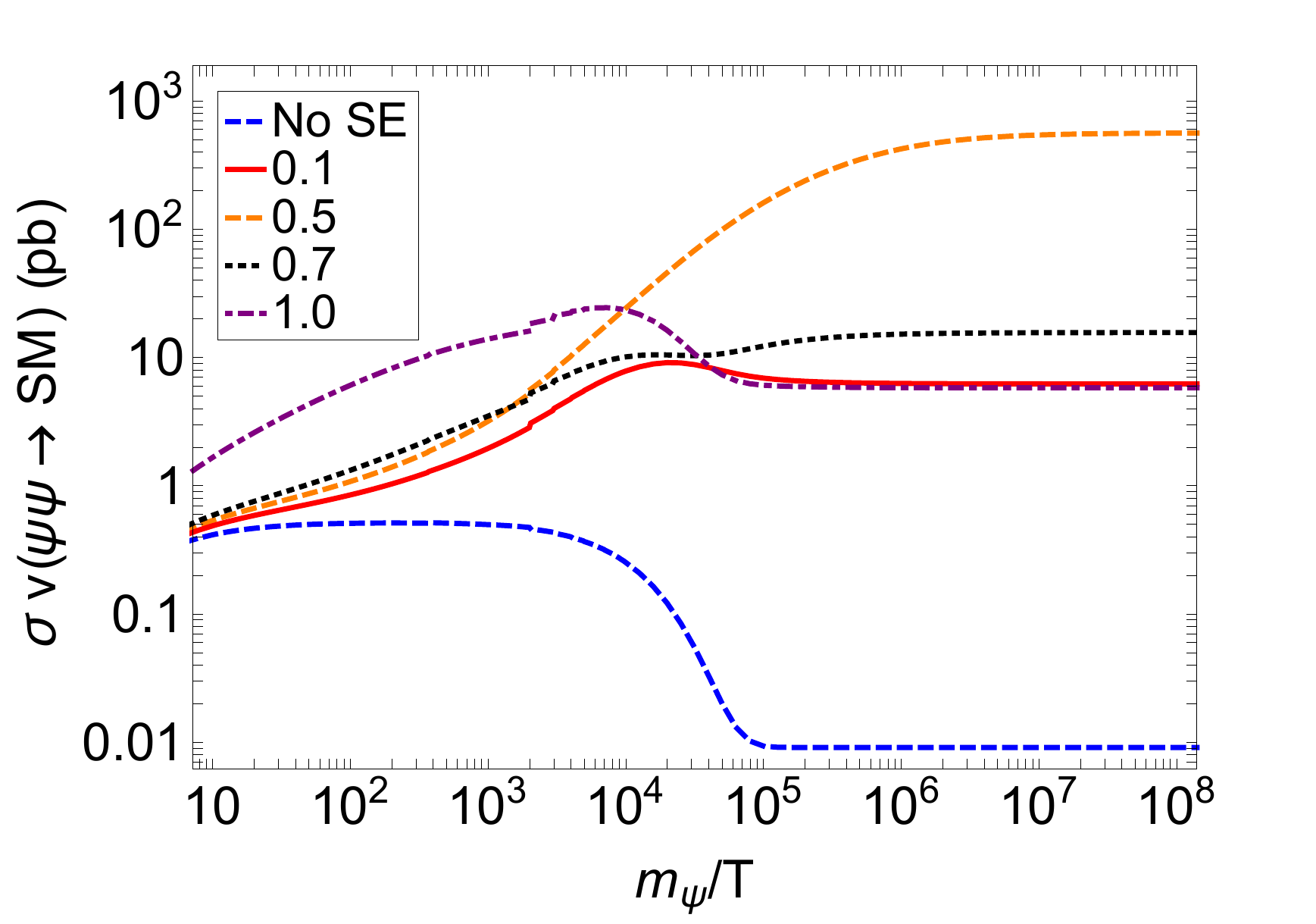}
  \includegraphics[width=0.49\textwidth]{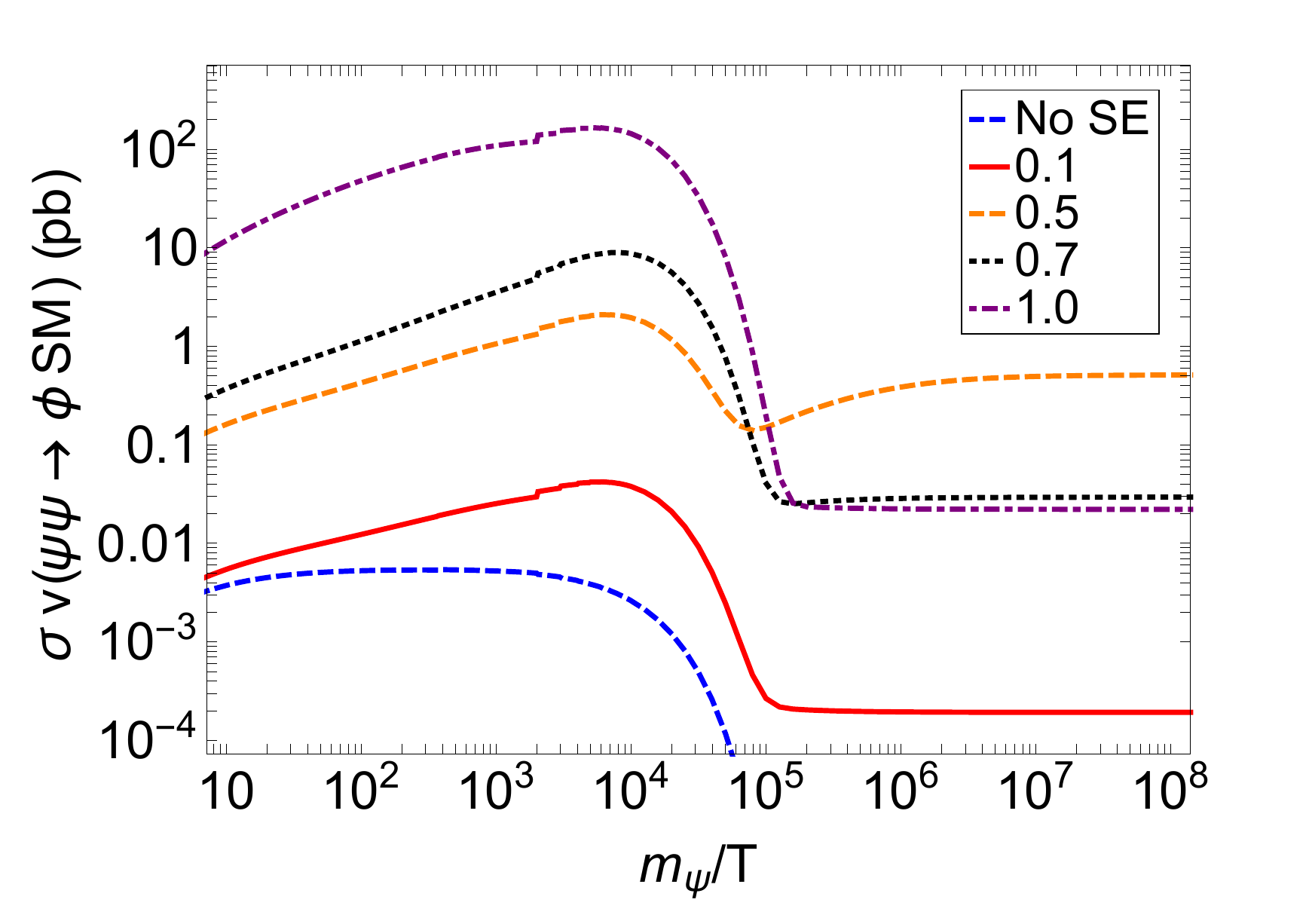}
  \caption{Thermally-averaged cross sections for annihilation (left) and SA (right).  These results are for $m_\psi = 2$~TeV, $m_\phi = 158$~GeV and values of $y$ as labelled, except for the blue dashed line which shows the cross sections without including the SE.}\label{fig:SE}
\end{figure}

We approximate the SE as only applying to the $s$-wave piece of the annihilation cross section, so that
\begin{equation}
  \sigv = \langle \mathcal{S} \rangle \, \sigma_0 + ( \sigv_0 - \sigma_0) \,,
\label{eq:SEapprox}\end{equation}
with $\sigv_0$ ($\sigma_0$) the unenhanced thermally averaged ($s$-wave) cross section and $\langle\mathcal{S}\rangle$ the thermally averaged SE factor.  We plot the thermally averaged annihilation and SA cross sections for a sample of points in \figref{fig:SE}.  Note that at late times, the annihilation cross section is always greater than the SA one; this is for the reasons discussed below \modeqref{eq:S0Q0SA}.

From \figref{fig:SE}, we see that in the absence of the SE, the SA channel effectively vanishes at low temperatures.  This is because when $T < \delta m_\psi$, the charged states are no longer thermally populated.  Scattering processes $\psi^+ Z \to \psi^0 W^+$ convert charged fermions to neutral ones, and the $\psi^0\psi^0$ initial state does not interact.  This is in contrast with the annihilation channel, where $\psi^0 \bar{\psi}^0 \to W^+ W^-$ gives a non-zero contribution even in the absence of the SE.  It follows that the SE is relatively even more important to the SA channel than the annihilation channel.

\begin{figure}
  \centering
  \includegraphics[width=0.6\textwidth]{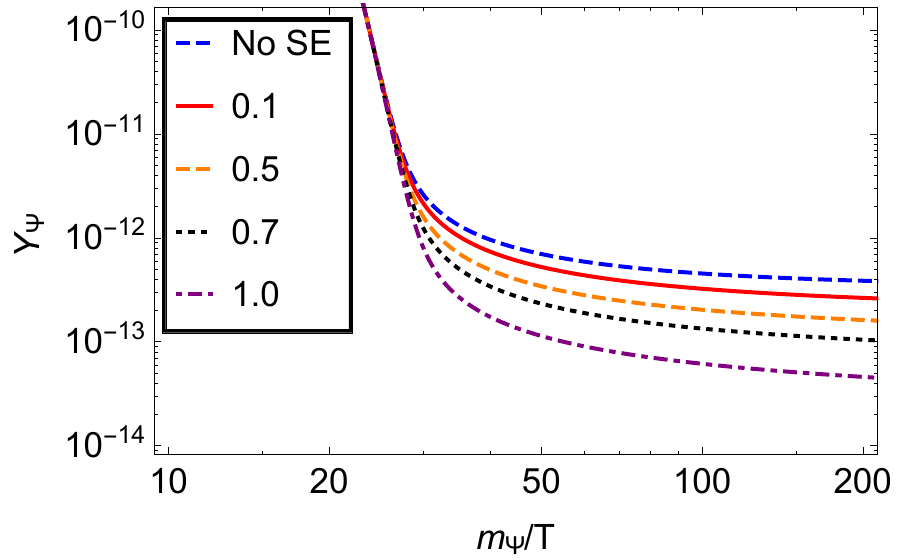}
  \caption{The fermion abundance $Y_\Psi$ for $m_\psi = 2$~TeV, $m_\phi = 158$~GeV, $\lhp = 0.1$ and $y$ as labelled ($y = 0.1$ for the no SE line).  Note that the effects of the SE shown in \figref{fig:SE} only modify the evolution at late times, $m_\Psi/T \gtrsim 25$.}\label{fig:rdev}
\end{figure}

\figref{fig:SE} also illustrates that the SE always \emph{increases} the total cross section, with the largest effect happening at late times, $m_\Psi/T \gtrsim 10^3$.  This motivates the approximation of \modeqref{eq:SEapprox}: it implies that the SE will only modify solutions to the Boltzmann equation at late times when $p$-wave and higher terms are suppressed by $\langle v^2\rangle = 6T/m_\Psi$.  We illustrate this by plotting $Y_\Psi$ as a function of $m_\Psi/T$ in \figref{fig:rdev} for the same parameter choices as in \figref{fig:SE}.   The SE only modifies the evolution for $m_\Psi/T \gtrsim 25$ or $\langle v^2\rangle \lesssim 25\%$.  However, we expect the final uncertainty in the relic density to be less than this for several reasons.  First, we are not completely neglecting the non-$s$-wave pieces, only the SE piece.  Using the $s$-wave results as a guide, the SE is probably only an $\mathcal{O}(1)$ effect at $m_\Psi/T \gtrsim 25$ (except for $y = 1$).  Second, the typical error in neglecting the $p$-wave term is suppressed by one-half; if the DM annihilation cross section can be approximated as $\sigma v = a + b v^2$, then
\begin{equation}
  \Omega \approx \frac{\Omega_0}{a + 3b/x_f} \,, \qquad x_f \approx 25 \,.
\end{equation}
Third, since the mean separation is smallest in the $s$-wave, we can reasonably expect the SE to be smaller for the neglected terms.  These factors suggest a theoretical uncertainty associated with \modeqref{eq:SEapprox} of approximately 10\%.  \figsref{fig:SE}{fig:rdev} also determine the error associated with neglecting relativistic corrections to \modeqref{eq:SomEffSchEq}.  These scale as $\langle v^2\rangle$ and will also have their largest effect around $m_\Psi/T \approx 25$.  Again, because the uncertainty decreases at later times we can expect a theoretical uncertainty of 10\%.  However, these estimates may overstate matters; a previous study~\cite{Cirelli:2015bda} gave the total effect of both approximations  in a similar context to be only $\mathcal{O}(5\%)$.

\begin{figure}
  \centering
  \includegraphics[width=0.7\textwidth]{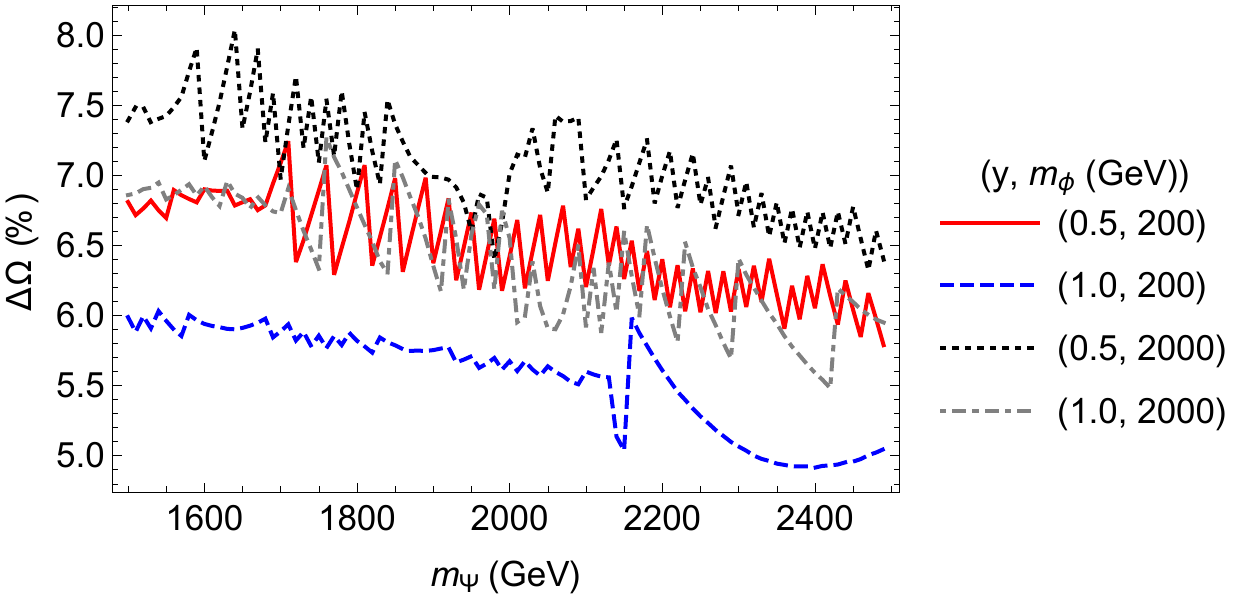}
  \caption{Differences between the relic density computed in our code and those computed using micrOMEGAs~4.1, for $\lhp = 0.1$ and other parameters as labelled.}\label{fig:diffMO}
\end{figure}

In computing the relic densities including SA and the SE, we numerically integrate the Boltzmann equations~\eqref{eq:fermdY} and~\eqref{eq:scadY} using the SE cross sections computed as discussed above.  For scalar annihilations to the SM, we follow \refcite{Cline:2013gha} in computing the cross section for $m_\phi \leq 150$~GeV, writing
\begin{equation}
  \sigma v (s) = \frac{2\lhp^2 \vew^2}{\sqrt{s}} \, \frac{\Gamma_h (\sqrt{s})}{(s - m_h^2)^2 + m_h^2 \Gamma_h^2 (m_h)} 
\end{equation}
with $\Gamma_h (m)$ the Higgs partial width as a function of mass taken from~\refcite{Dittmaier:2011ti}.  For $m_\phi > 150$~GeV, we use the perturbative cross sections (also taken from~\cite{Cline:2013gha}).  We cross-checked our relic density calculations by comparing the output in the absence of the SE with the output of micrOMEGAs~4.1, and found good agreement in the relic densities.  We show in \figref{fig:diffMO} the difference for a selection of points, which is in the 5--10\% region.  Using this as an estimate of the theoretical uncertainty associated with solving the Boltzmann equation, and combining with the other errors discussed above, we estimate a total theoretical uncertainty in our calculation of the relic density in the 10--20\% range.

\section{Collider Constraints}\label{sec:LHC}
Searching for dark matter has been among the primary motivations for recent collider experiments.  The limits have the important feature for us that they are insensitive to SA processes; the initial state has zero dark sector charge.  It follows that the constraints we derive for our models are independent of $y$.   Additionally, we can place limits separately on scalar and fermion production.

\begin{figure}
  \centering
  \includegraphics[width=0.5\textwidth]{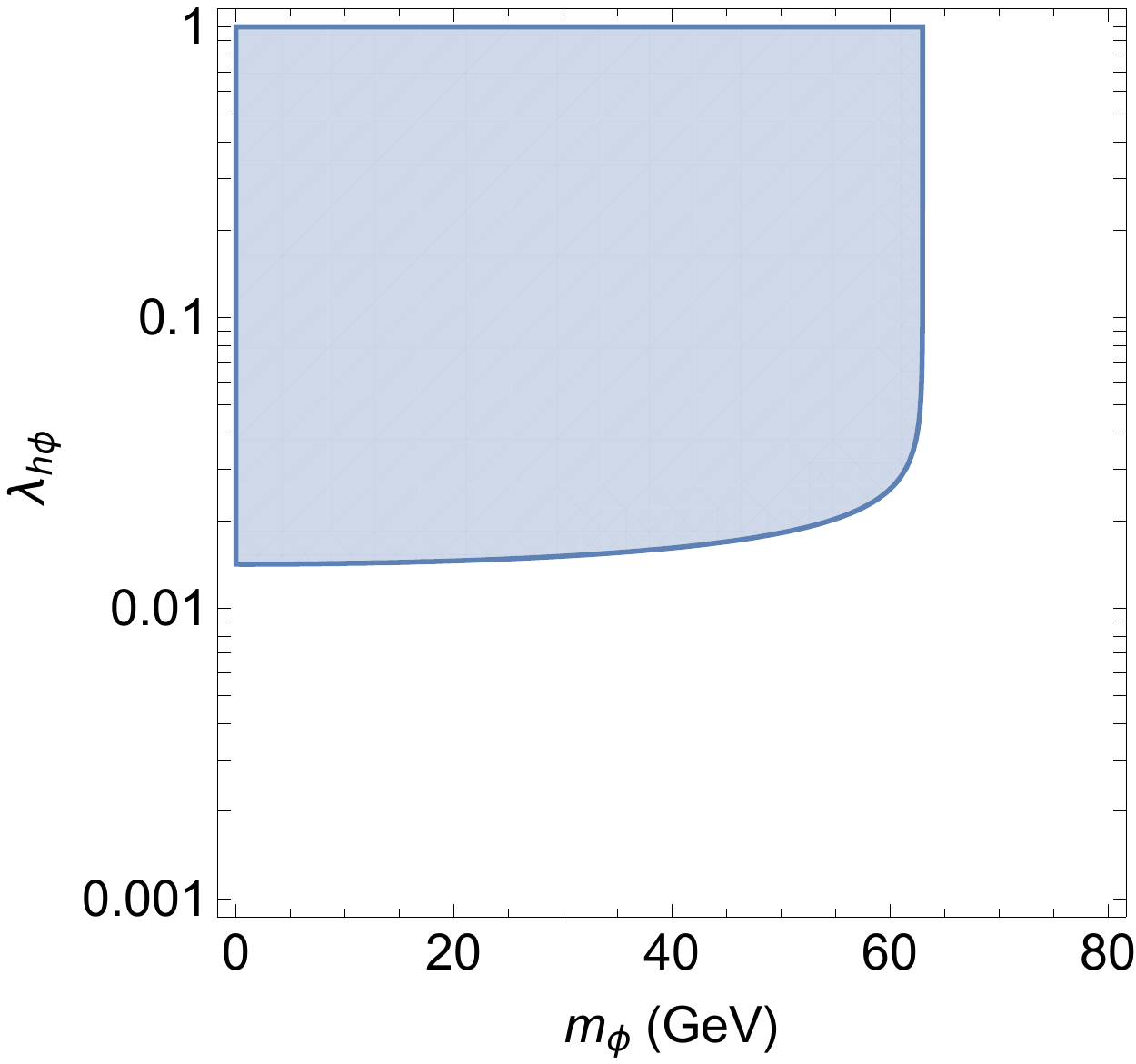}
  \caption{Excluded parameter space from the invisible branching fraction of the Higgs.}\label{fig:HInv}
\end{figure}

The scalar couples to the SM through the Higgs portal.  This opens two production channels.  At low masses, $2m_\phi < m_h$, it contributes to the invisible decay of the Higgs.  We use the limits from \refcite{Cline:2013gha}.  The contribution to the invisible decay width is
\begin{equation}
  \Gamma (h \to \phi\phi) = \frac{\lhp^2 \vew^2}{32\pi m_h} \, \sqrt{1 - \frac{4m_{\phi}^2}{m_h^2}} \,.
\end{equation}
\refcite{Belanger:2013xza} obtained a bound $Br (h \to\,$invisible$)<0.19$ at 95\% confidence level.  We show the limits in \figref{fig:HInv}.  When the decay is kinematically allowed, we have the constraint $\lhp \lesssim 0.01$--0.02.

For heavy scalars, $\phi$ will be pair produced via the Higgs portal.  Limits from the 8~TeV LHC were studied in \refcite{Endo:2014cca}, and future constraints at 14 and 100~TeV in \refcite{Craig:2014lda}.  Both found that the strongest limits come from jets plus missing energy searches via vector boson fusion.  However, for $\lhp \lesssim 1$, the limits are weaker than those from direct detection searches and so we do not include them in our results.

\begin{table}
  \centering
  \begin{tabular}{|c|c|}
    \hline
    Signal Region & $p_T^{track}$ Lower Bound \\
    \hline
    A & 75~GeV \\
    B & 100~GeV \\
    C & 150~GeV \\
    D & 200~GeV \\
    \hline
  \end{tabular}
  \caption{Definitions of the four signal regions for the ATLAS disappearing tracks search.}\label{tab:SR}
\end{table}

For the fermion singlet model, the production of $\psi$ will be heavily suppressed.  The lack of a direct fermion-SM coupling means the fermion can only be produced via scalar intermediate states, and we expect the limits on scalar production to be stronger.  For the fermion triplet model, in contrast, we have production through electroweak interactions:
\begin{equation}
pp\rightarrow W^\pm \rightarrow \psi^\pm \psi^0 
\qquad
pp\rightarrow \gamma/Z \rightarrow \psi^\pm \bar{\psi}^\pm 
\end{equation}
The subsequent decay of the charged fermions are
\begin{align}
&\psi^\pm \rightarrow \psi^0 \pi^\pm   &  &Br_\pi = 97.7\% \\
&\psi^\pm \rightarrow \psi^0 l^\pm \nu  & & Br_l = 2.3 \%  ,
\end{align} 
where $l$ denotes electron and muon.  Since the decay length of $\psi^\pm$ is of $\mathcal{O}(10)\; \rm{cm}$ and the pion is soft, the charged fermions appear as disappearing tracks in the inner detectors of LHC experiments. Both ATLAS~\cite{ATLAS_DT} and CMS~\cite{CMS_DT} have performed corresponding searches and excluded chargino masses below 270 GeV and 260 GeV, respectively, at $95 \%$ confidence level.  We expect limits on our model to be larger, as our fermions are Dirac, leading to a factor of 2 increase in the production cross sections. 

\begin{figure}[tp]
  \centering
  \includegraphics[width=0.5\textwidth]{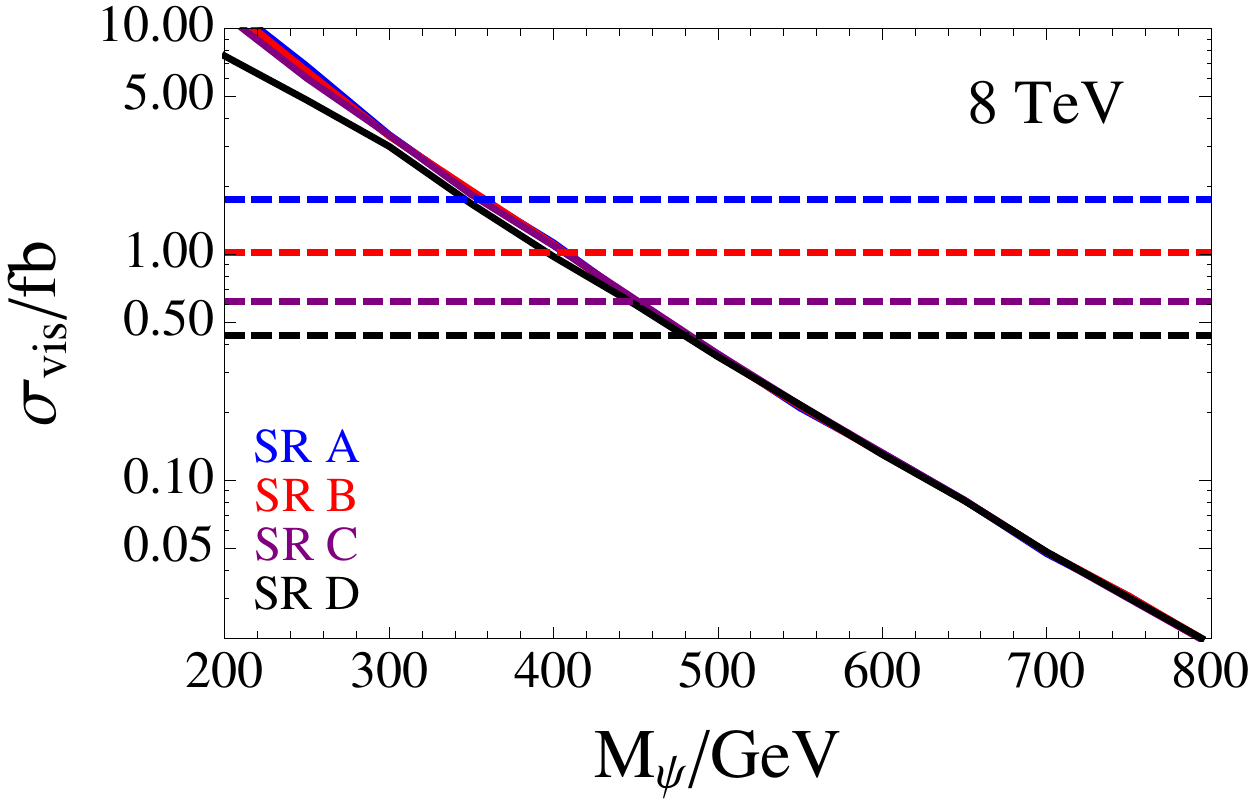}
  \caption{Results for the four signal regions for the ATLAS disappearing tracks search.  Solid (dashed) lines correspond to production (exclusion) cross sections.}\label{fig:lhclimit}
\end{figure}

We use the ATLAS model-independent limits provided in four signal regions, defined by different visible track $p_T$ as given in \tabref{tab:SR}.  We implement our model in FeynRules~2.0~\cite{Alloul:2013bka}, and generate event samples using MadGraph~5~\cite{Alwall:2011uj} and PYTHIA~8.2~\cite{Sjostrand:2014zea}.  Detector effects are simulated with Delphes~3~\cite{deFavereau:2013fsa} with the standard ATLAS card.  The decay lengths of the charged fermions are calculated using the two-loop mass splittings from \refcite{Ibe:2012sx}.  To match the experimental cuts, we require the hardest jet with $p_T > 90 \, \rm{GeV}$, a large missing transverse energy $E_T^{miss}>90\, \rm{GeV}$, and a separation $\Delta \phi_{min}^{jet-E_T^{miss}} > 1.5$ between the missing momentum vector and both the hardest jet and the second hardest jet with $p_T>45\, \rm{GeV}$ (if it exists). We also apply a lepton veto.  To mimic the ATLAS disappearing-track selection, we reject all events where the $\psi^\pm$ decays before the silicon microstrip (SCT) detectors or after the straw-tube radiation tracker (TRT).  We validate our analysis by checking our selection efficiency agrees with that of  the benchmark point given in \refcite{ATLAS_DT} to within $10\%$.  The final visible cross sections including all cuts are shown in \figref{fig:lhclimit} together with the model independent limits from ATLAS, from which we can draw the lower limit for the fermion triplet mass $m_\psi \gtrsim 480 \, \rm{GeV}$. 

The disappearing track search at higher energies, both in the 14 TeV LHC run and at a 100 TeV proton-proton collider, has been investigated in~\cite{Low:2014cba} for a wino.  Assuming the major background at the signal region after the selection comes from $Z+jets$, the background at higher energies is estimated by scaling the normalization according to the $Z+jets$ cross section.  With $100\%$ background normalization, the $5\sigma$ discovery limit at 100 TeV is set to a wino mass of 2.2 TeV.  Given our fermionic triplet is a Dirac particle, we expect the 100 TeV search will be able to discover our model to at least the same fermion mass.

\section{Direct Detection}\label{sec:DD}
DM direct detection experiments can impose stringent constraints on the model parameter space.  Like the collider constraints discussed above, SA processes are generally not relevant.  In most models (including ours) the strongest constraints come from elastic scattering cross sections of the form $p\chi \to p\chi$, for $\chi$ a dark sector state.  However, limits in our model are \emph{indirectly} sensitive to $y$ through the individual relic densities of $\phi$ and $\psi$.

The scalar $\phi$ interacts with nucleons via the Higgs portal.  The spin-independent scattering cross section is
\begin{equation}
  \sigma_{SI} (\phi N \to \phi N) = \frac{\lhp^2 f_N^2}{4\pi} \, \frac{m_N^4}{m_H^4 (m_N + m_\phi)^2} \,,
\end{equation}
where $f_N$ is the Higgs-nucleon coupling
\begin{equation}
  f_N = \sum_q f_q = \sum_q \frac{m_q}{m_N} \, \langle N \lvert \bar{q} q \rvert N \rangle \,,
\end{equation}
and $m_N = 0.946$~GeV.  We follow \refcite{Cline:2013gha} and take $f_N = 0.345$ in placing our limits.  See \refscite{Cline:2013gha,Alarcon:2011zs,Alarcon:2012kn,Junnarkar:2013ac,Ellis:2008hf,Akrami:2010dn,Bertone:2011nj,Crivellin:2013ipa,Hoferichter:2015dsa,Alarcon:2012nr} for more details.

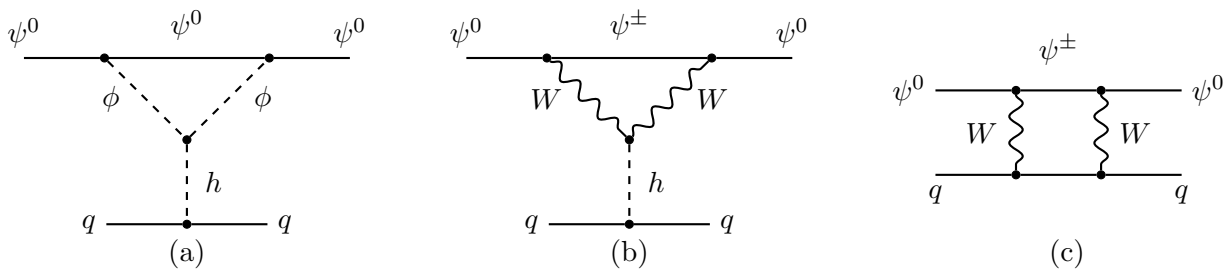
\begin{figure}[ht]
\centering
\begin{subfigure}
\centering
\begin{tikzpicture}[node distance=1cm and 1cm]
\coordinate[vertex] (v1);
\coordinate[vertex, below right = of v1 ] (v2);
\coordinate[vertex, above right = of v2 ] (v3);
\coordinate[vertex, below = of v2,label=below:(a)] (v4);

\coordinate[left = of v1, label=above:$\psi^0$] (f1);
\coordinate[left = of v4, label=left:$q$] (f3);
\coordinate[right = of v3, label=above:$\psi^0$] (f2);
\coordinate[right = of v4, label=right:$q$] (f4);

\draw[fermionnoarrow] (f1) -- (v1);
\draw[fermionnoarrow] (v1) -- (v3) node[midway,above=0.1cm]{$\psi^0$} ;
\draw[fermionnoarrow] (v3) -- (f2);
\draw[scalar] (v1) -- (v2) node[midway,left=0.2cm] {$\phi$};
\draw[scalar] (v2) -- (v3) node[midway,right=0.2cm] {$\phi$};
\draw[scalar] (v2) -- (v4) node[midway,right=0.1cm] {$h$};
\draw[fermionnoarrow] (f3) -- (v4);
\draw[fermionnoarrow] (v4) -- (f4);
\end{tikzpicture}
\end{subfigure}
\hfill
\begin{subfigure}
\centering
\begin{tikzpicture}[node distance=1cm and 1cm]
\coordinate[vertex] (v1);
\coordinate[vertex, below right = of v1 ] (v2);
\coordinate[vertex, above right = of v2 ] (v3);
\coordinate[vertex, below = of v2,label=below:(b)] (v4);

\coordinate[left = of v1, label=above:$\psi^0$] (f1);
\coordinate[left = of v4, label=left:$q$] (f3);
\coordinate[right = of v3, label=above:$\psi^0$] (f2);
\coordinate[right = of v4, label=right:$q$] (f4);

\draw[fermionnoarrow] (f1) -- (v1);
\draw[fermionnoarrow] (v1) -- (v3) node[midway,above=0.1cm] {$\psi^\pm$};
\draw[fermionnoarrow] (v3) -- (f2);
\draw[photon] (v1) -- (v2) node[midway,left=0.2cm] {$W$};
\draw[photon] (v2) -- (v3) node[midway,right=0.2cm] {$W$};
\draw[scalar] (v2) -- (v4) node[midway,right=0.1cm] {$h$};
\draw[fermionnoarrow] (f3) -- (v4);
\draw[fermionnoarrow] (v4) -- (f4);
\end{tikzpicture}
\end{subfigure}
\hfill
\begin{subfigure}
\centering
\begin{tikzpicture}[node distance=1cm and 1cm]
\coordinate[vertex] (v1);
\coordinate[vertex, right = of v1] (v2);
\coordinate[vertex, below = of v1] (v3);
\coordinate[vertex, below = of v2] (v4);
\coordinate[below right = of v3, label=left: (c)](v5);

\coordinate[left = of v1, label=left:$\psi^0$] (f1);
\coordinate[right = of v2, label=right:$\psi^0$] (f2);
\coordinate[left = of v3, label=below:$q$] (f3);
\coordinate[right = of v4, label=below:$q$] (f4);

\draw[fermionnoarrow] (f1) -- (v1);
\draw[fermionnoarrow] (v1) -- (v2) node[midway,above=0.2cm] {$\psi^\pm$} ;
\draw[fermionnoarrow] (v2) -- (f2);
\draw[photon] (v1) -- (v3) node[midway, left=0.1cm] {$W$};
\draw[photon] (v2) -- (v4) node[midway, right=0.1cm] {$W$};
\draw[fermionnoarrow] (f3) -- (v3);
\draw[fermionnoarrow] (v3) -- (v4);
\draw[fermionnoarrow] (v4) -- (f4);
\end{tikzpicture}
\end{subfigure}
\caption{ Feynman diagrams for interactions between $\psi^0$ and quarks at one-loop level.}
\label{fig:DDloop}
\end{figure}

In the singlet model, the fermion $\psi$ can only interact with nucleons through a Higgs penguin with a $\phi$ loop, as shown in \figref{fig:DDloop}\,(a).  In the triplet model, $\psi^0$ has two additional loop contributions, a Higgs penguin with $W$ loop as in \figref{fig:DDloop}\,b and the box diagram of \figref{fig:DDloop}\,(c).\footnote{The triplet model also has inelastic couplings through a $t$-channel $W$.  However, these are negligible at direct detection experiments due to the relatively large mass splitting, $\delta m_\psi \gg K$ where $K$ is the $\psi$ kinetic energy.}  The contribution from the box diagram is subdominant for $m_\psi \gtrsim m_W$, which is enforced by the LHC constraints of \secref{sec:LHC}.  The low energy effective Lagrangian for the $\psi^0$-quark interaction is 
\begin{align}
\mathcal{L}_{\psi^0 q} & = \sum_i \lambda_q^i \bar{\psi}^0 \psi^0 \bar{q}_i q_i\\
\nonumber
\lambda_q^i & =  \frac{y^2 \lhp}{32 \pi^2} \frac{m_{q_i}}{ m_\psi m_h^2}
      \frac{5-8\eta_\phi+3 \eta_\phi^2 +2(2 - \eta_\phi)\log\eta_\phi}{(1-\eta_\phi)^3} \\
\nonumber
& \quad - \alpha_2^2 \frac{m_{q_i}}{m_\psi m_h^2} \frac{1-4\eta_W+3\eta_W^2 + (2-4\eta_W) \log\eta_W}{ (1-\eta_W)^3}\\
& \quad +  \alpha_2^2  \frac{m_{q_i}}{m_\psi m_W^2} 
   \frac{(2 -3 \eta_W +6\eta_W^2-5\eta_W^3+3\eta_W (1+\eta_W^2)\log\eta_W) }{6 (1-\eta_W)^4} \; , 
\label{eqn:qqnn}
\end{align} 
where $q_i$ runs through all quarks and $\eta_{\phi,W} \equiv \frac{m_{\phi,W}^2}{m_\psi^2}$.  The three terms in \modeqref{eqn:qqnn} come from three diagrams in \figref{fig:DDloop} in the same order.  The last two terms only appear in the fermionic triplet model.  Note there can be cancellation among the contributions depending on the sign of the quartic coupling $\lambda_{h\phi}$. 

The strongest spin-independent limits for DM masses $\gtrsim 5$~GeV are from the preliminary run at LUX~\cite{Akerib:2013tjd}.  We include the contributions from both fermion and scalar scattering, computed by implementing the effective Lagrangian of \modeqref{eqn:qqnn} into FeynRules~2.0~\cite{Alloul:2013bka} and computing event rates with micrOMEGAs~3~\cite{Belanger:2013oya}.  We compute the total number of events expected with the fiducial target mass of $118\, \rm{kg}$ in $85.3$ days.  The relative detection efficiency is extracted directly from the LUX results~\cite{Akerib:2013tjd}, and the contributions from $\phi,\psi$ are scaled by $f_{\phi,\psi} \equiv \Omega_{\phi,\psi}/\Omega_{cdm}$. $\Omega_{cdm}$ takes the central value of the Planck~2015 results~\cite{Planck2015}, {\it i.e.} $\Omega_{cdm} h^2 = 0.1186$.  The LUX collaboration observed zero signal event with $0.64$ background events from the electron recoil leakage, which translates with the Feldman-Cousins method~\cite{1998PhRvD..57.3873F} into an upper limit of the number of signal events $N_s < 1.82$ at $90\%$ C.L.. 

\begin{figure}
  \centering
  \includegraphics[width=0.5\textwidth]{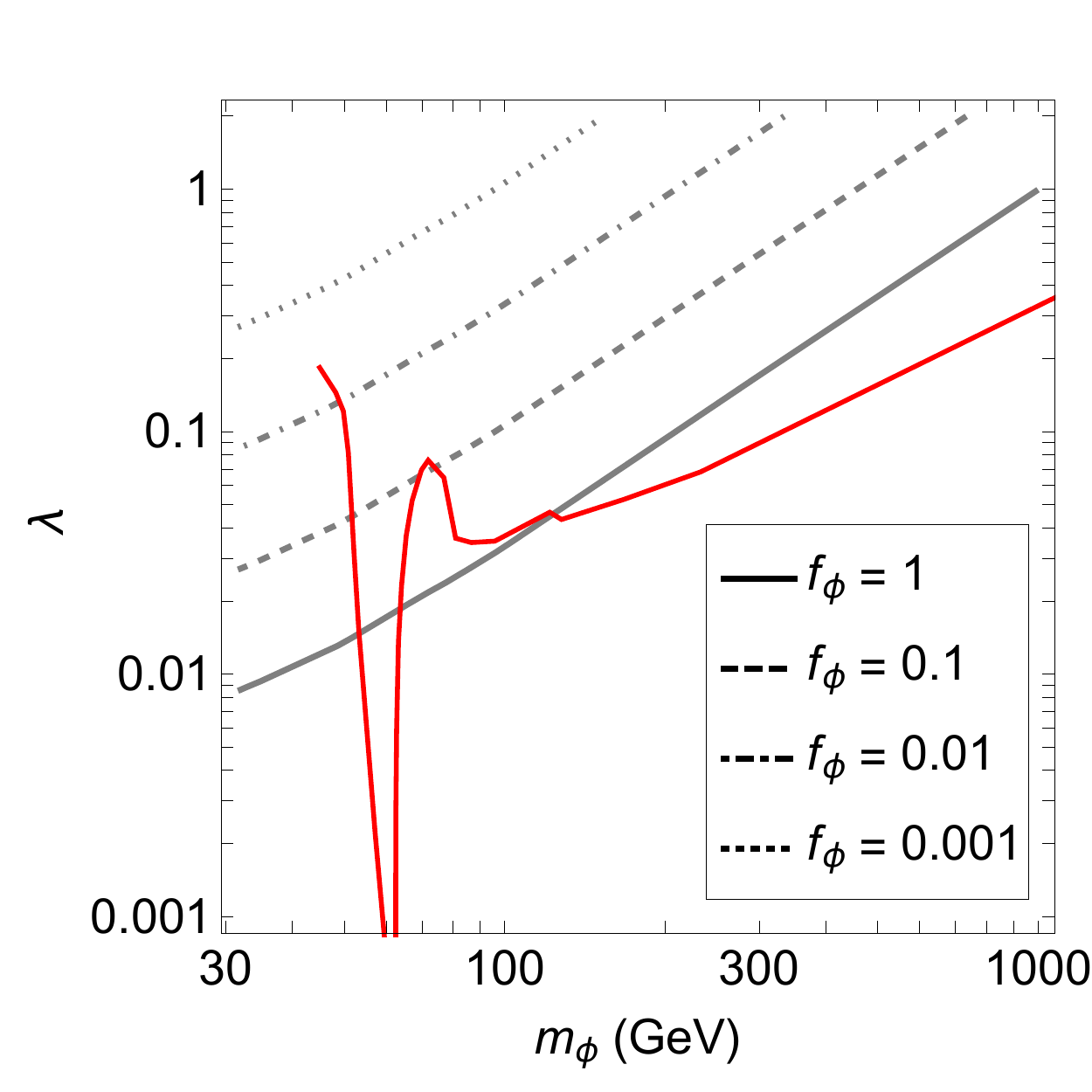}
  \caption{Exclusion contours from LUX using \modeqref{eq:appLUX}.  The regions above each contour are excluded for a scalar DM fraction of at least the appropriate amount.  The red line shows the contour for which scalar annihilation alone gives $f_\phi = 1$.}\label{fig:LUXexcl}
\end{figure}

We find that the contribution from fermion scattering is always negligible.  The exclusion contours are equivalent to applying the bound
\begin{equation}
  f_\phi \, \sigma_{SI} (\phi N \to \phi N) < \sigma_{LUX}^{90\%} (m_\phi) \,,
\label{eq:appLUX}\end{equation}
where $\sigma_{LUX}^{90\%} (m)$ is the bound given by the LUX collaboration on the scattering cross section for DM mass $m$.  We show these bounds in \figref{fig:LUXexcl} for differing values of $f_\phi$, together with the contour where scalar annihilations give $f_\phi = 1$.  When no other processes are relevant to the scalar relic density, there is an approximate cancellation between $f_\phi \propto \lhp^{-2}$ and $\sigma_{SI} \propto \lhp^2$, giving a $\lhp$-independent bound $m_\phi \gtrsim 130$~GeV or $m_\phi \approx m_h/2$.  This applies to our models in the limit $m_\psi \gg m_\phi$.  We use the scalar-only approximation to derive prospective limits from the full LUX run~\cite{Szydagis:2014xog}, Xenon~1T~\cite{2012arXiv1206.6288A} and LZ~\cite{2011arXiv1110.0103M}.  

\section{Indirect Detection}\label{sec:ID}
DM indirect detection experiments search for 
its annihilation or decay in regions of high DM density. They are the only one of the three search channels that we discuss here that are \emph{directly} sensitive to the fermion-scalar coupling $y$, in that SA processes can contribute to experimental signals.  Searches exist for several possible SM final states, including positrons~\cite{Adriani:2008zr, Adriani:2013uda, FermiLAT:2011ab, Aguilar:2013qda}, anti-protons~\cite{Adriani:2010rc}, and neutrinos~\cite{Aartsen:2013dxa}.  However, both our dark sector states dominantly annihilate into gauge bosons, and in such cases the strongest limits are known to come from searches for high-energy $\gamma$-rays~\cite{Cirelli:2007xd,Cirelli:2015bda,Masiero:2004ft,Ibe:2015tma}.  We therefore focus on this channel.

The differential photon flux from dark matter annihilation is given by
\begin{equation}
  \frac{d^2\Phi}{d\Omega d E_\gamma} = \frac{1}{8\pi} \sum_i \frac{f_a f_b}{m_a m_b} \frac{dN_i}{d E_\gamma} \,\langle\sigma_i v\rangle \int_{l.o.s.} \rho^2 dl \,.
\label{eq:defID}\end{equation} 
This factors into a product of astrophysics and particle physics.  The former is given by the integral, taken along the line-of-sight, with $\rho$ the dark matter density.  This is conventionally parameterised by a $J$-factor, defined as
\begin{equation}
  \bar{J} (\Delta \Omega) = \frac{1}{\Delta\Omega} \int_{\Delta \Omega} d\Omega \int_{l.o.s.} \rho^2 dl \,,
\end{equation}
where $\Delta\Omega$ is the relevant field of view.  The particle physics component of \modeqref{eq:defID} is given by everything else, where the sum runs over all annihilation and SA processes $ab \to SM$.  The fraction of the total dark matter made up by each component is $f_{a,b} = \Omega_{a,b}/\Omega_{cdm}$, which we assume to be spatially constant.  $dN/dE_\gamma$ is the differential energy spectrum of photons for each process, normalized per (semi-)annihilation, and with cross section $\langle\sigma_i v\rangle$.  We first discuss the particle physics details specific to our model, which we can divide into three classes depending on the initial state: two scalars, two fermions, or scalar plus fermion.

For the two-scalar initial state, we note that for a pure scalar singlet, there are no current limits from scalar annihilation~\cite{Cline:2013gha}.  This cannot change with the presence of the fermion, since $f_\phi < 1$.  We are instead interested in two signals here.  First, we wish to consider future limits from Cherenkov Telescope Array (CTA)~\cite{Consortium:2010bc}.  Second, as we discuss below, for $m_\phi \sim m_\psi \lesssim 200$~GeV, it is possible to explain the Fermi excess using a combination of annihilation and SA.  In both cases, we compute the photon spectra using PPPC~4~DM~ID~\cite{Cirelli:2010xx}, combining the results for different final states proportionally.  Note that while gauge boson final states are generally most important, at low masses $m_\phi \lesssim 1$~TeV annihilations to $t\bar{t}$ make a significant contribution; while for $m_\phi \lesssim m_W$, annihilations to $b\bar{b}$ will dominate.

For CTA limits, we demand that the signal flux always be lower than the sensitivities of \refcite{Silverwood:2014yza}, using the $J$-factors they compute based on Einasto~\cite{Navarro:2003ew} profile.  For the regions of the galactic sky that they use to set limits, these are $J_{ON} = 7.4\times10^{21}$~GeV$^2$~cm$^{-5}$ and $J_{OFF} = 1.2\times10^{22}$~GeV$^2$~cm$^{-5}$, where ON (OFF) are the signal (background) regions.  These bounds are conservative in that for a continuous spectrum, a $\chi$-squared analysis would improve sensitivity by a factor of $\sim 2$; however, we find that this does not significantly affect the excluded regions of parameter space.  These limits are also weaker than those of ~\cite{Cline:2013gha} due to the inclusion of systematic uncertainties.

For fermion-scalar initial states, we can further subdivide into two processes: $\psi\phi \to \bar{\psi}h$ and $\psi\phi \to \bar{\psi}V$, with $V$ a SM gauge boson.  The former process is present in both models, while the latter only exists in the fermion triplet case.  For the Higgs final state, we again use PPPC~4~DM~ID to construct the final state spectra.  Note that the Higgs is produced with energy 
\begin{equation}
  E^\ast (m_h) = \frac{m_\phi^2 + 2 m_\psi m_\phi + m_h^2}{2(m_\psi + m_\phi)} \,.
\label{eq:estar}\end{equation}
This channel is most important in the fermion singlet model, where we combine the flux with that from scalar annihilation according to \modeqref{eq:defID}.  This is discussed further in \secref{sec:FXS} below.

For the process $\psi\phi \to \bar{\psi}V$, we make use of the results of \refcite{Cirelli:2015bda}, which gives limits directly on the cross sections (see also \refcite{Garcia-Cely:2015dda}).  Note that since the neutral fermion does not directly couple to $\gamma/Z$, we need only consider $\psi^0\phi \to \bar{\psi}^\pm W^\mp$.  Given a bound $\sigma^{exc} (m)$ from \refcite{Cirelli:2015bda}, we impose the constraint
\begin{equation}
  \frac{f_\psi f_\phi}{m_\psi m_\phi} \, \sigma (\psi^0\phi \to \bar{\psi}^+ W^-) \leq \frac{1}{(E^\ast (m_W))^2} \, \sigma^{exc} \bigl(E^\ast (m_W) \bigr) \,,
\label{eq:FSexc}\end{equation}
with $E^\ast$ as given in \modeqref{eq:estar}, and a factor of 2 from summing over final state charges cancels a factor of one-half due to producing only a single gauge boson.  However, we find that limits from this channel are generally suppressed compared to SE processes with two fermions in the initial state.

\begin{figure}
  \centering
  \includegraphics[width=0.48\textwidth]{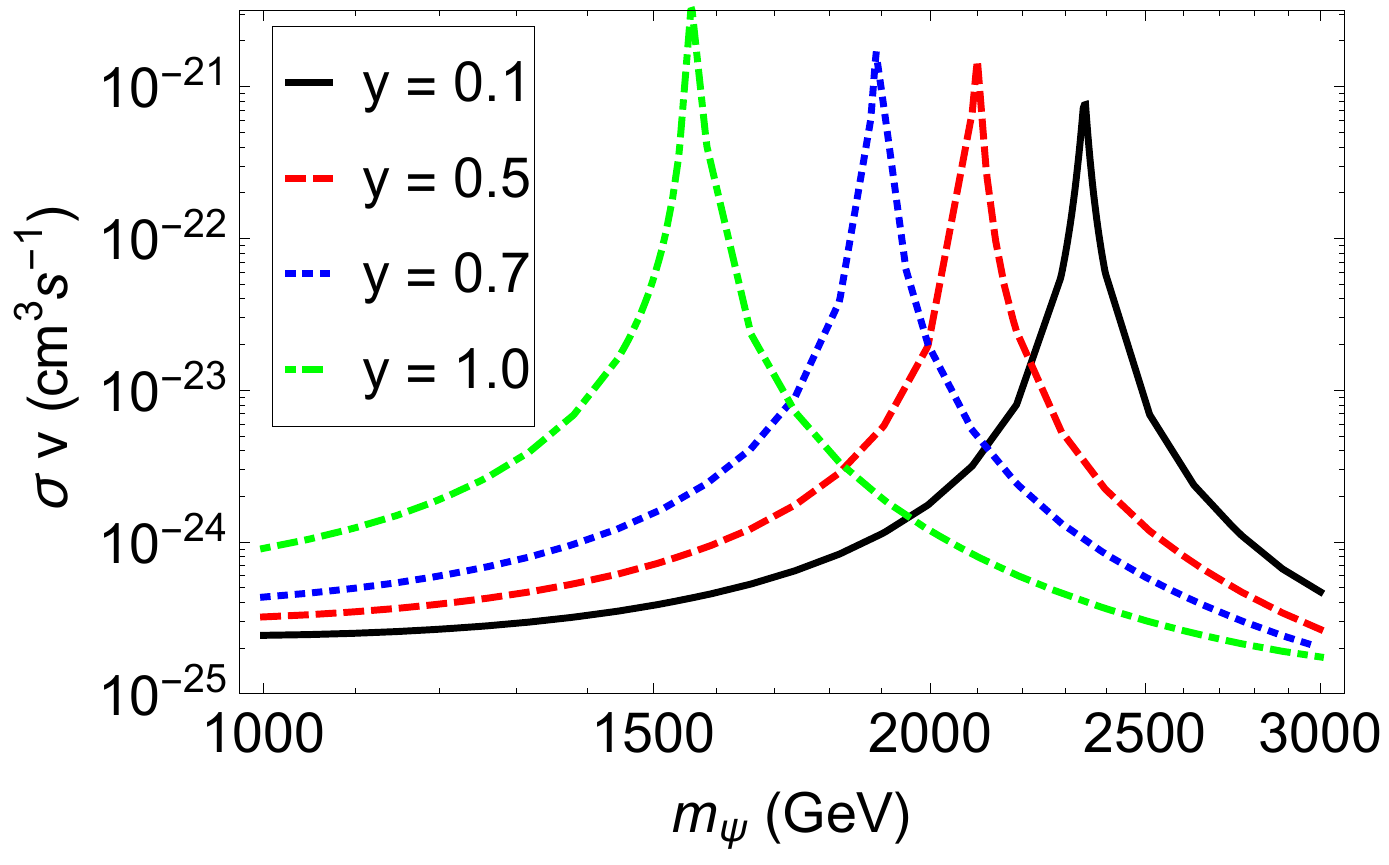}
  \includegraphics[width=0.48\textwidth]{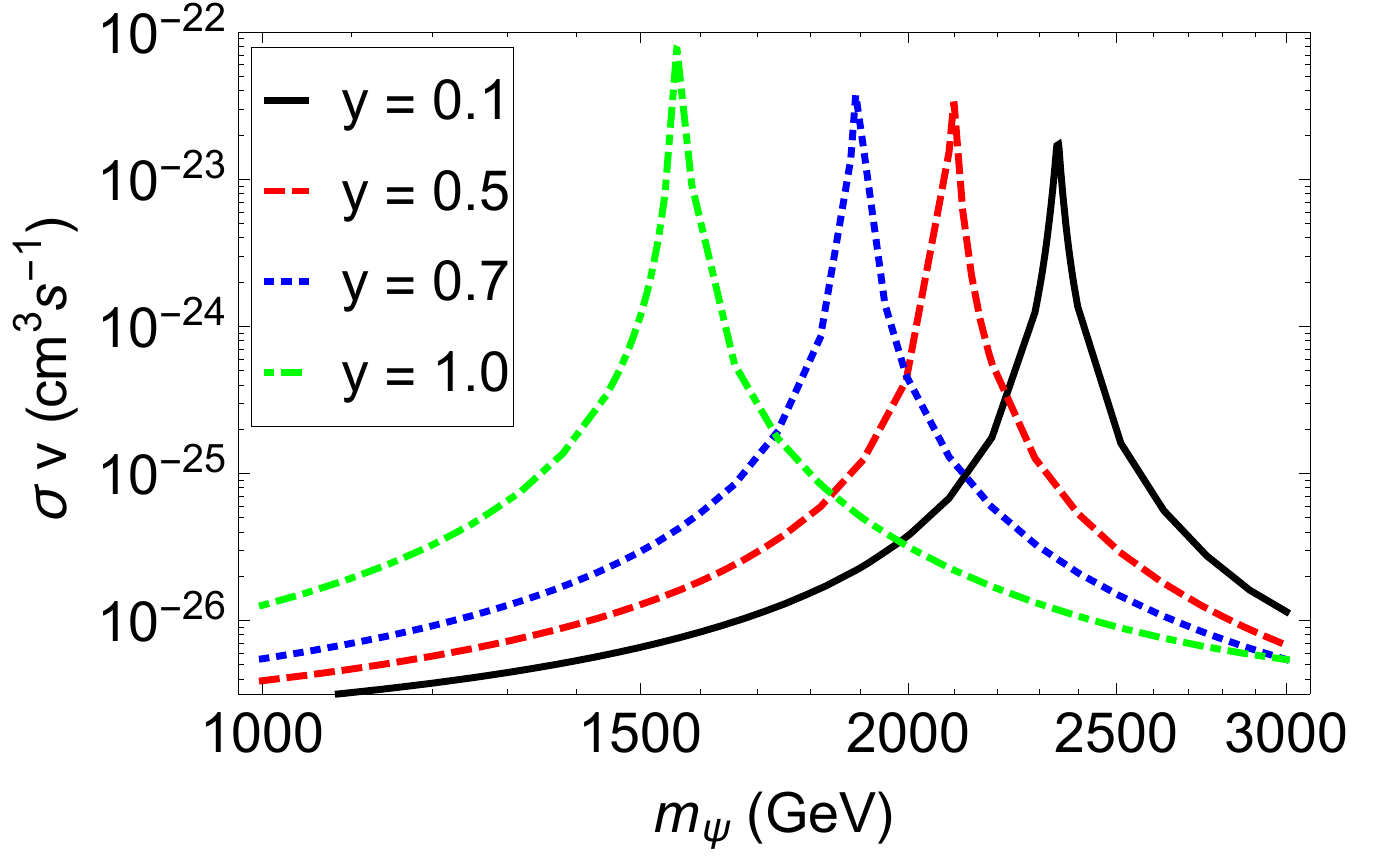}
  \caption{Combined annihilation cross sections to $WW$, $ZZ$ and $Z\gamma$ (left) and $\gamma\gamma$ and $Z\gamma$ (right), for $m_\phi = 200$~GeV and different values of $y$.  Note that the resonant peak occurs at smaller values of $m_\psi$ as $y$ increases.}\label{fig:IDXSec}
\end{figure}

We similarly use \refcite{Cirelli:2015bda} to set bounds on processes with two fermions in the initial state.  Since the LHC bounds on the fermion triplet model enforce $m_\psi \gg m_{W,Z}$ we may neglect the $W$--$Z$ mass difference, and sum over $W$ and $Z$ final states.  Our computation of the signal cross sections must again include the SE, but there are some important differences with the relic density calculation of \secref{sec:RD}.  Most importantly, we have only neutral fermions at late times, so we focus on the $Q=0$ subspaces.  Further, we can reasonably approximate that only the $S = 0$ subspace contributes, as it is the only case with a non-vanishing tree-level $\psi^0\bar{\psi}^0$ annihilation.  We thus focus on the potential matrix from \modeqref{eq:q0pot} and the annihilation from \modeqref{eq:s0q0ann}.  Finally, we impose different boundary conditions at infinity to \modeqref{eq:SEBC}:
\begin{equation}
  \begin{split}
    \lim_{r\to\infty} \Phi_{ij} (r) & = 0 \quad \text{if } i = 1, 3 \,, \\
    \lim_{r\to\infty} \frac{\Phi'_{2j} (r)}{\Phi_{2j} (r)} & = i \sqrt{M K} \,.
  \end{split}
\end{equation}

In principle, a similar consideration would apply for the SA process $\psi\psi\to\phi Z$.  However, this process turns out to be negligible, as already discussed in \secref{sec:RD} and observable in \figref{fig:SE}.  Specifically, two neutral fermions can only exist in the $S=0$ state, but SAs are suppressed compared to annihilations both by $m_Z^2/m_\psi^2$, and the absence of a tree-level term.  We have confirmed this numerically for a selection of points.  For limits from continuous spectra, this gives us the effective cross sections for 
\begin{equation}
  \sigma^{eff} (\psi\bar{\psi} \to VV) \equiv f_\psi^2 \biggl( \sigma (\psi^0\bar{\psi}^0 \to W^+W^-) + \sigma (\psi^0\bar{\psi}^0 \to ZZ) + \frac{1}{2} \, \sigma (\psi^0\bar{\psi}^0 \to Z\gamma) \biggr) \,.
\end{equation} 
Of these, the first is the only term that is non-vanishing at tree-level, and in practice dominates.

\begin{figure}
  \centering
  \includegraphics[height=0.4\textwidth]{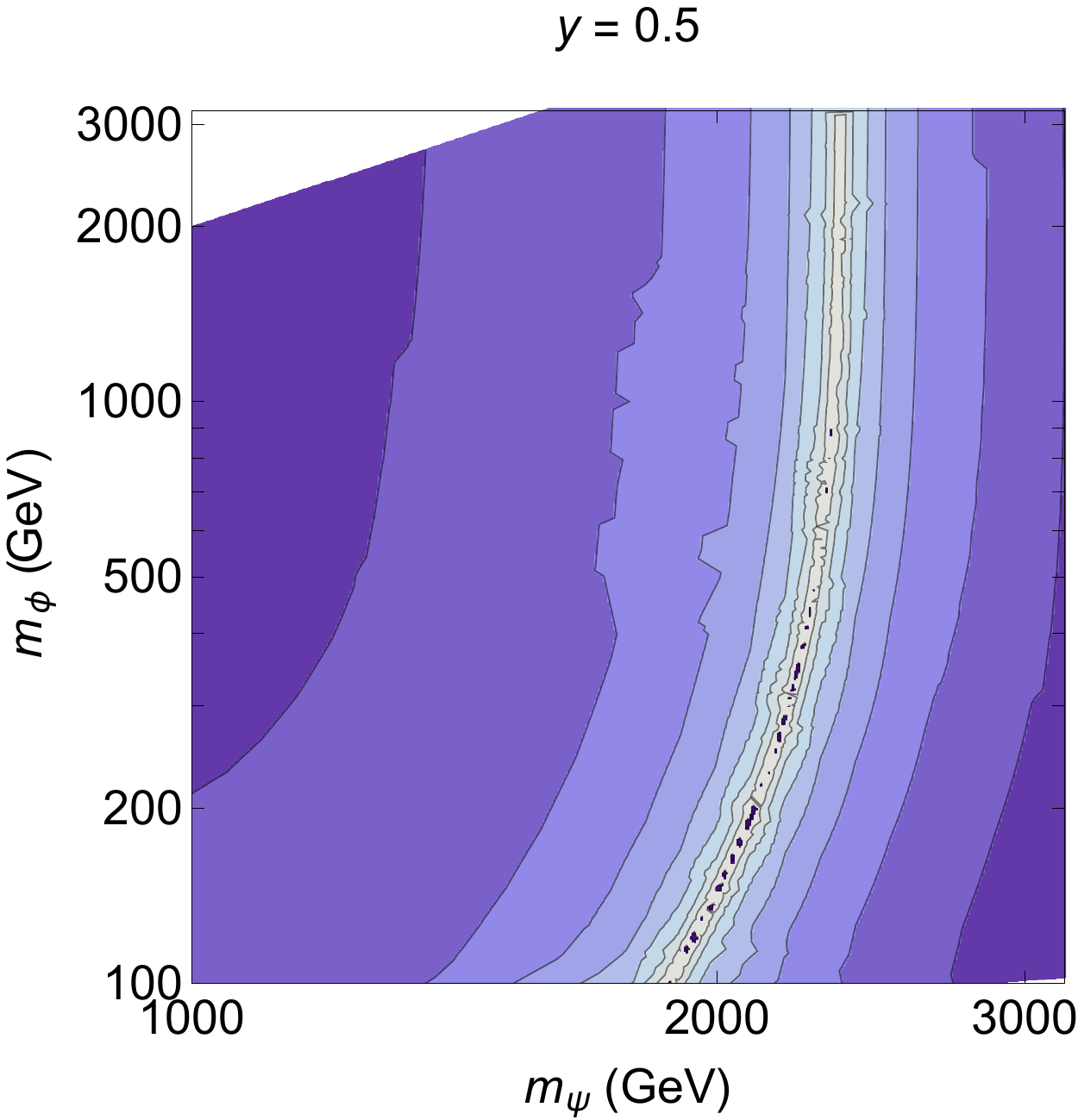}
  \includegraphics[height=0.4\textwidth]{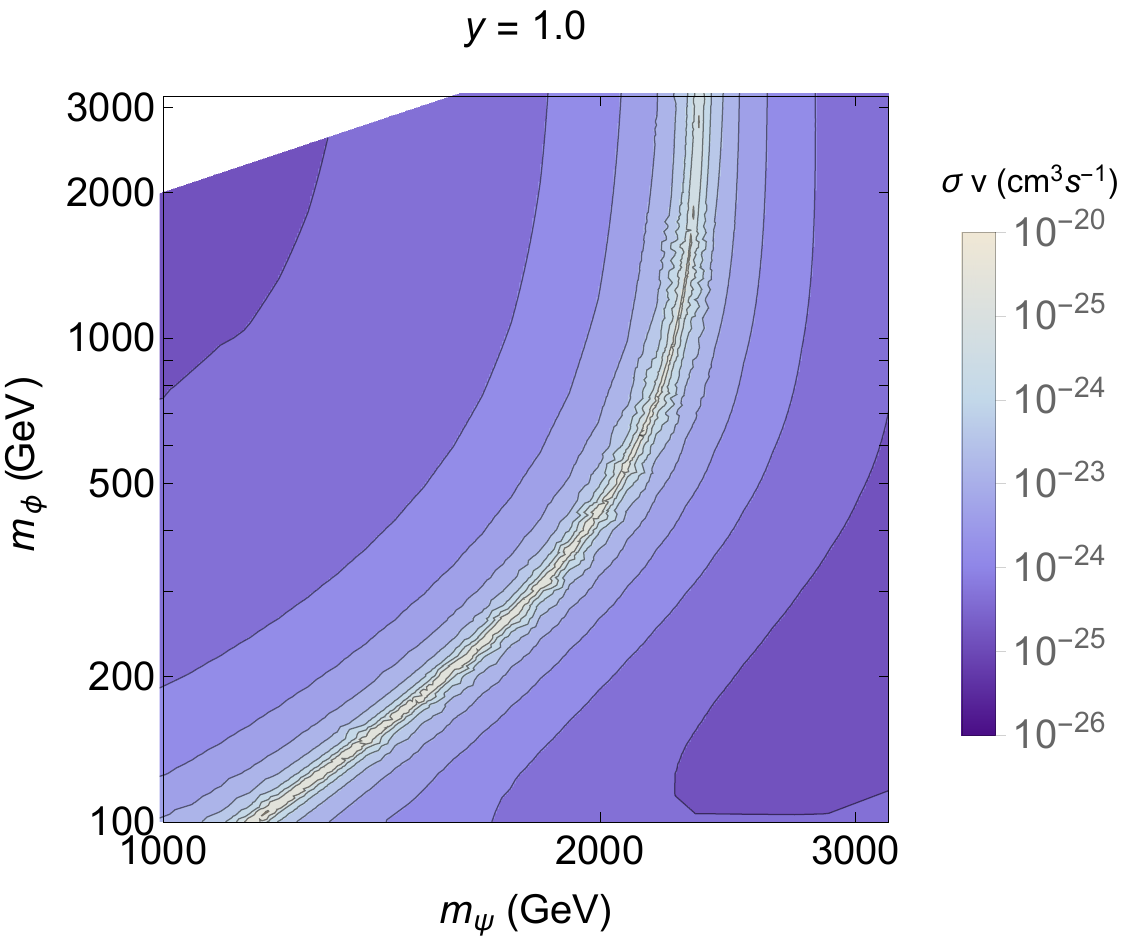}
  \caption{Annihilation cross sections for fermions to massive gauge bosons for $y = 0.5$ (1.0) in the left (right) figure.  Note that the resonant peak occurs at smaller values of $m_\psi$ as $m_\phi$ decreases, and this effect is stronger for larger $y$.}\label{fig:respos}
\end{figure}

We plot annihilation cross sections (without the $f_\psi$ factor) in \figref{fig:IDXSec} (left).  Despite the tree-level cross sections being a function of SM parameters only, we see that increasing $y$ has the effect of making the resonance stronger and shifting to lower masses.  Both of these facts are consistent with the scalar-mediated force being always attractive, combining with the electroweak potential.  The case for $y = 0.1$, when the scalar-mediated potential is much weaker than the vector-mediated one, can be compared to previous calculations for fermion triplets~\cite{Hryczuk:2011vi,Cirelli:2007xd}.  We show how the SE effect varies with the scalar and fermion masses in \figref{fig:respos}.  We can see that for $m_\phi\gtrsim m_\psi$, the position of the resonance is set by the electroweak gauge bosons independent of $y$.  As $m_\phi$ decreases, the resonance moves to lower values of $m_\psi$, with the strength of the effect increasing with $y$.

We also consider monochromatic $\gamma$-rays from $\gamma\gamma$ and $Z\gamma$ final states.  The SA contribution is again subleading, though for a slightly different reason: the lack of a tree-level $\psi^+\psi^-\to \phi \gamma$ process in the spin-0 case.  For fermion annihilation, the leading contribution comes from the SE process $\psi^0 \bar{\psi}^0 \to \psi^+ \bar{\psi}^- \to \gamma+\gamma/Z$, with the direct loop process being higher order in the weak coupling.  Consistently including the loop contribution would require calculating higher-order terms in the fermion potential, so we follow \refcite{Hryczuk:2011vi} and omit them.  Since these higher-order terms are the leading contributions for the SA channel, we expect them to be negligible.

The energy splitting between the $\gamma\gamma$ and $Z\gamma$ lines is $\delta E = E_{\gamma\gamma} - E_{Z\gamma} < 0.01 \, E_{\gamma\gamma}$ for $m_\psi > 500$~GeV, and is unlikely to be observable at current and near-future experiments.  As such we place limits on
\begin{equation}
  \sigma^{an,\gamma} \equiv f_\psi^2 \biggl( \sigma (\psi^0\bar{\psi}^0 \to \gamma\gamma) + \frac{1}{2} \sigma(\psi^0\bar{\psi}^0 \to Z\gamma) \biggr) \,.
\end{equation}
We plot this cross-section (without the $f_\psi$ factors) in \figref{fig:IDXSec} (right).  As with the production of massive gauge bosons, the position of the resonance decreases with increasing $y$.

\begin{figure}
  \centering
  \includegraphics[width=0.48\textwidth]{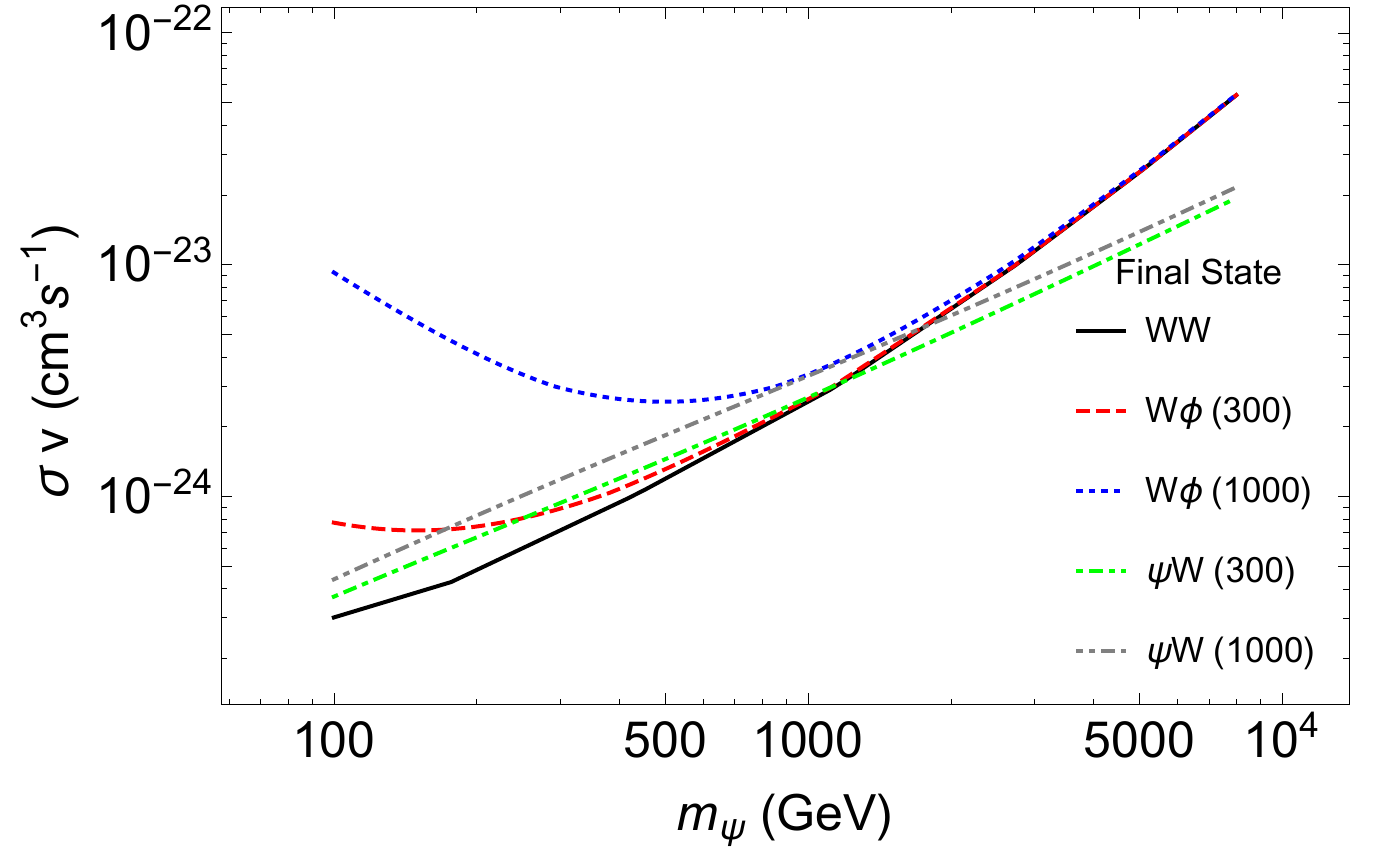}
  \includegraphics[width=0.48\textwidth]{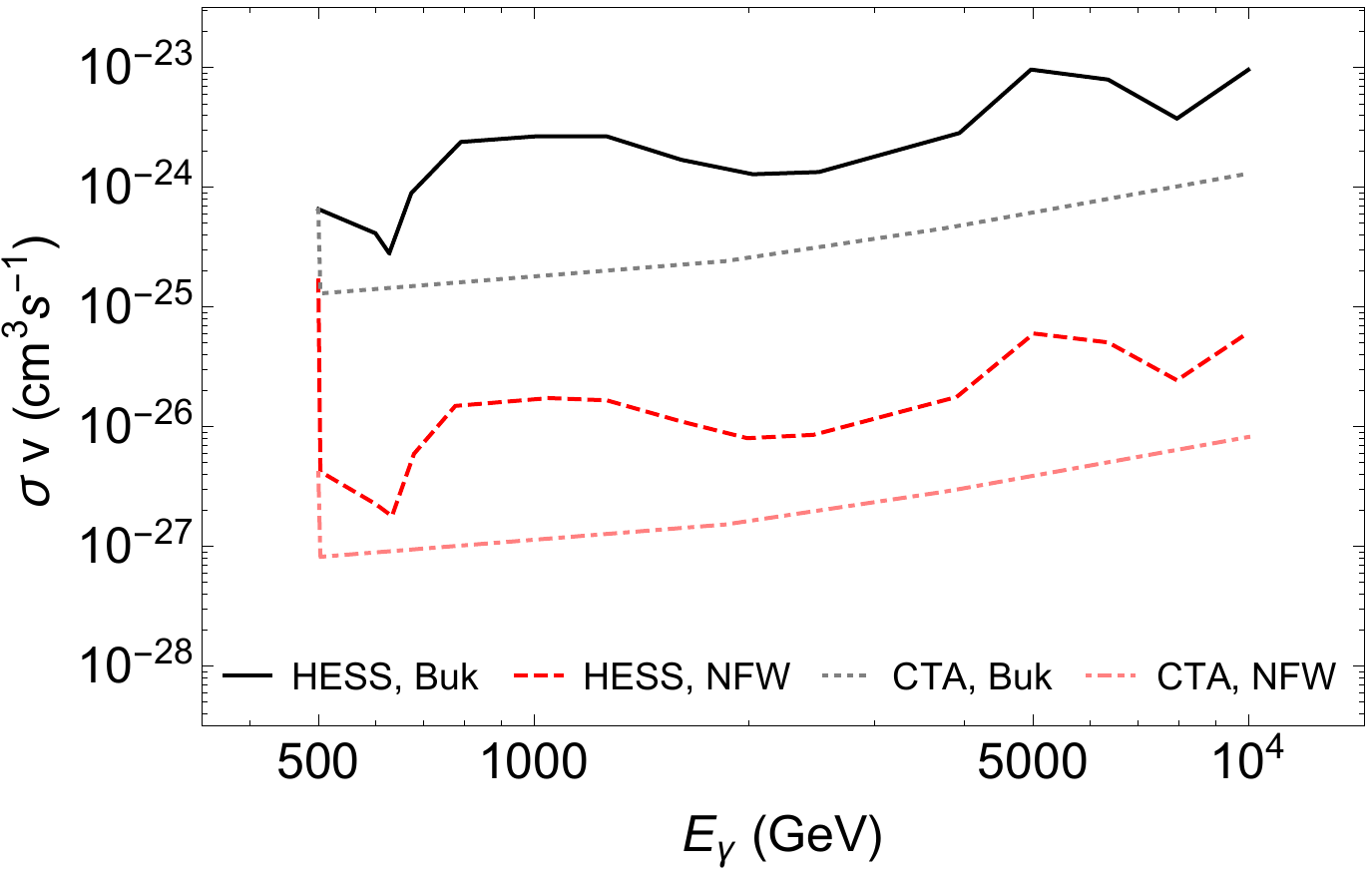}
  \caption{(Left): Excluded cross sections for continuous gamma-ray spectra, taken from Fermi observations of dwarf spheroidals, for different final states as labelled.  For the $W\phi$ and $\psi W$ final states, the number labels the scalar mass.  (Right): Excluded and projected cross sections for gamma-ray lines from HESS and CTA observations of the galactic center. }\label{fig:IDVexcl}
\end{figure}

\refscite{Cirelli:2015bda,Garcia-Cely:2015dda} performed a thorough study of different limits on DM annihilations to gauge bosons.  To avoid cluttering our results, we only use constraints that are the most stringent in some region of parameter space.  However, indirect limits depend on the poorly-known DM galactic distribution.  We choose to take two limits, an ``optimistic'' and a ``conservative'' one, corresponding to DM profiles that are respectively peaked or cored at the galactic centre.  For the former, we find that the strongest limit in all cases is from the HESS search for $\gamma$-ray lines~\cite{Abramowski:2013ax} for an Einasto~\cite{Navarro:2003ew} or NFW~\cite{Navarro:1995iw} profile (which are roughly equal).\footnote{We do not use an even stronger constraint given in~\cite{Cirelli:2015bda} based on a variant of the Einasto profile, which we consider too optimistic.}  For the conservative constraints, we use limits based on Fermi observations of the continuum spectrum from dwarf spheroidals~\cite{Ackermann:2015zua}; indeed, we use the ``Fermi$-$'' limits of \refcite{Cirelli:2015bda}, which weaken the limits by a factor of 10 to account for uncertainties in the dwarf galaxy $\bar{J}$-factors.  We also consider prospective limits from a CTA $\gamma$-line search, considering NFW/Einasto (optimistic) and Bukert~\cite{Burkert:1995yz} (conservative) profiles.  We plot the limits on annihilation and SA cross sections from these searches in \figref{fig:IDVexcl}.

\section{The Fermi Excess}\label{sec:FXS}

One of the most exciting results in recent dark matter phenomenology has been the identification of a gamma ray excess from the galactic center in data measured by the Fermi Large Area Telescope.  This Galactic Centre Excess (GCE) has been identified by several theoretical groups~\cite{Hooper:2010mq,Boyarsky:2010dr,Hooper:2011ti,deBoer:2011zz,Abazajian:2012pn,Daylan:2014rsa,Abazajian:2014fta}, peaks at energies of several GeV, and has the expected morphology in the sky for a dark matter signal.  While possible alternative explantations, including unresolved millisecond Pulsars~\cite{Abazajian:2014fta,Yuan:2014rca,Petrovic:2014xra} or cosmic rays at the galactic centre~\cite{Gordon:2014gya,Petrovic:2014uda,Gaggero:2015nsa}, have been advanced, and the Fermi collaboration has yet to release an official publication on the GCE, we will investigate whether either of our models can explain the anomaly.  In particular, we make use of recent analyses~\cite{Calore:2014xka,Agrawal:2014oha} showing that the excess can be explained by self-annihilation of dark matter with a mass of $\mathcal{O}(10^2)$~GeV.  For another interpretation using semi-annihilating dark matter, see \refcite{Fonseca:2015rwa}.

\begin{figure}[ht]
\centering
\includegraphics[width=0.5\textwidth]{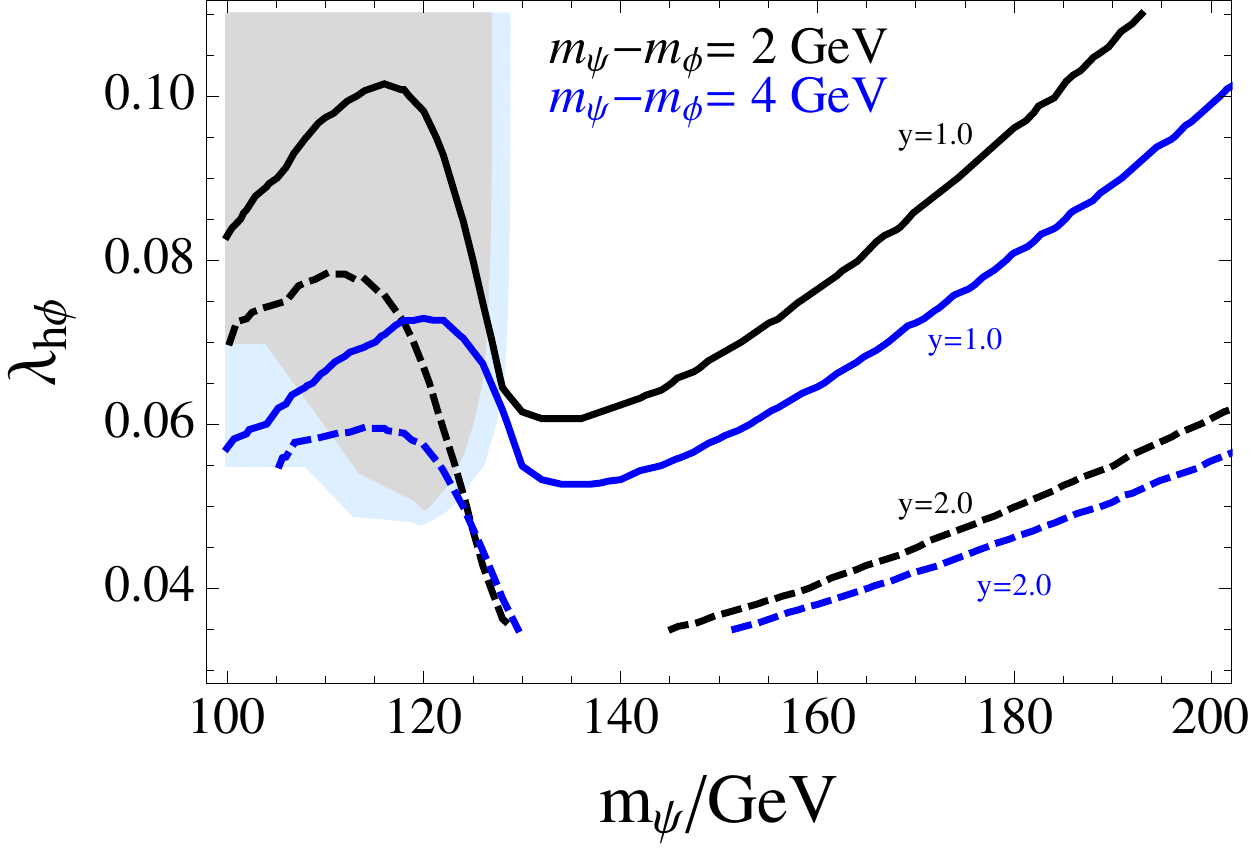}
\caption{The contour plot for the Yukawa coupling $y$ that gives the right relic density in black and blue for $\Delta m = 2 / 4 \; \rm{GeV}$. The LUX exclusion regions are shown in light gray and light blue respectively. 
}
\label{fig:ycontour}
\end{figure}

\begin{figure}[ht]
  \includegraphics[width=0.45\textwidth]{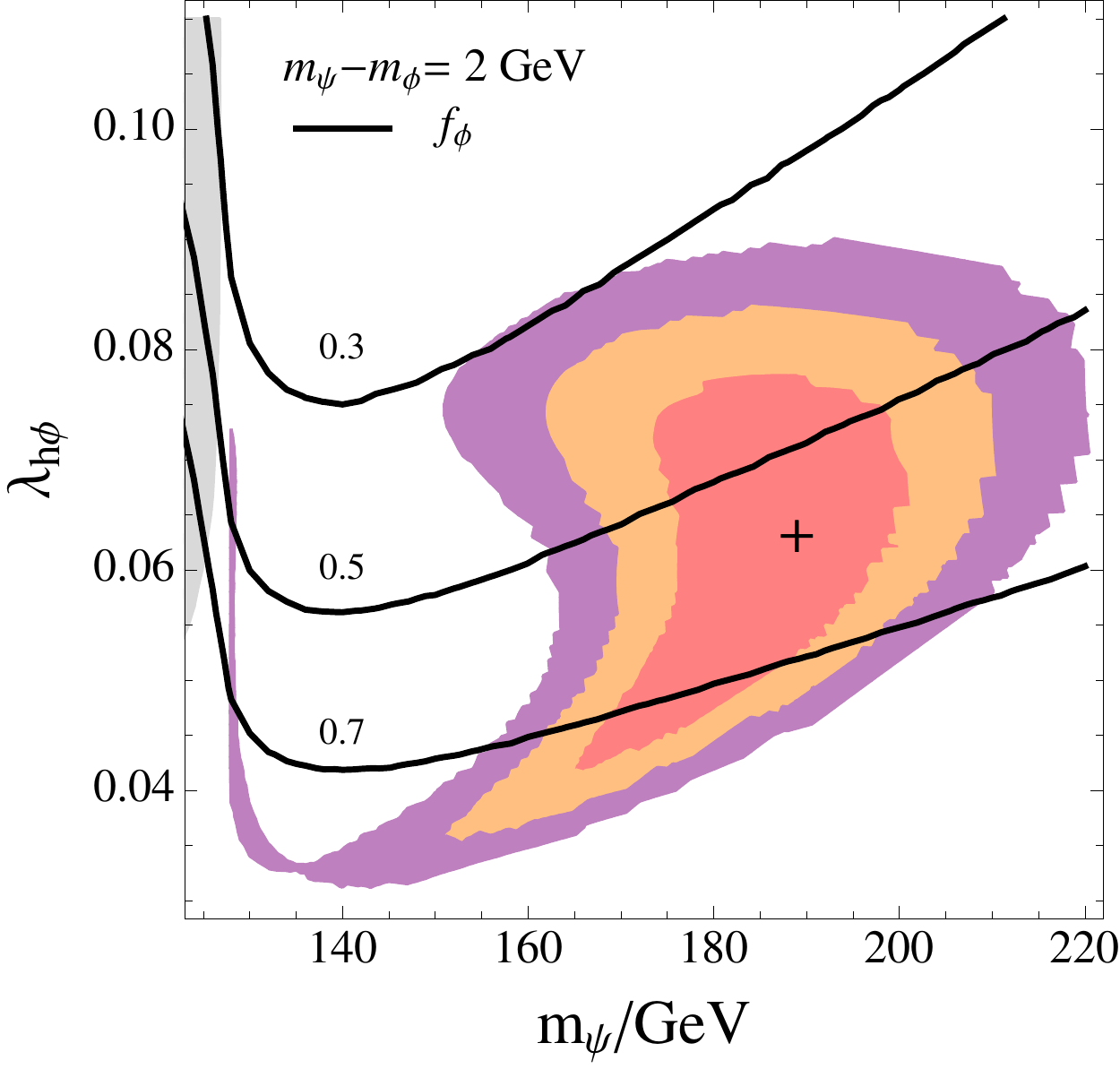}
  \hfill
  \includegraphics[width=0.45\textwidth]{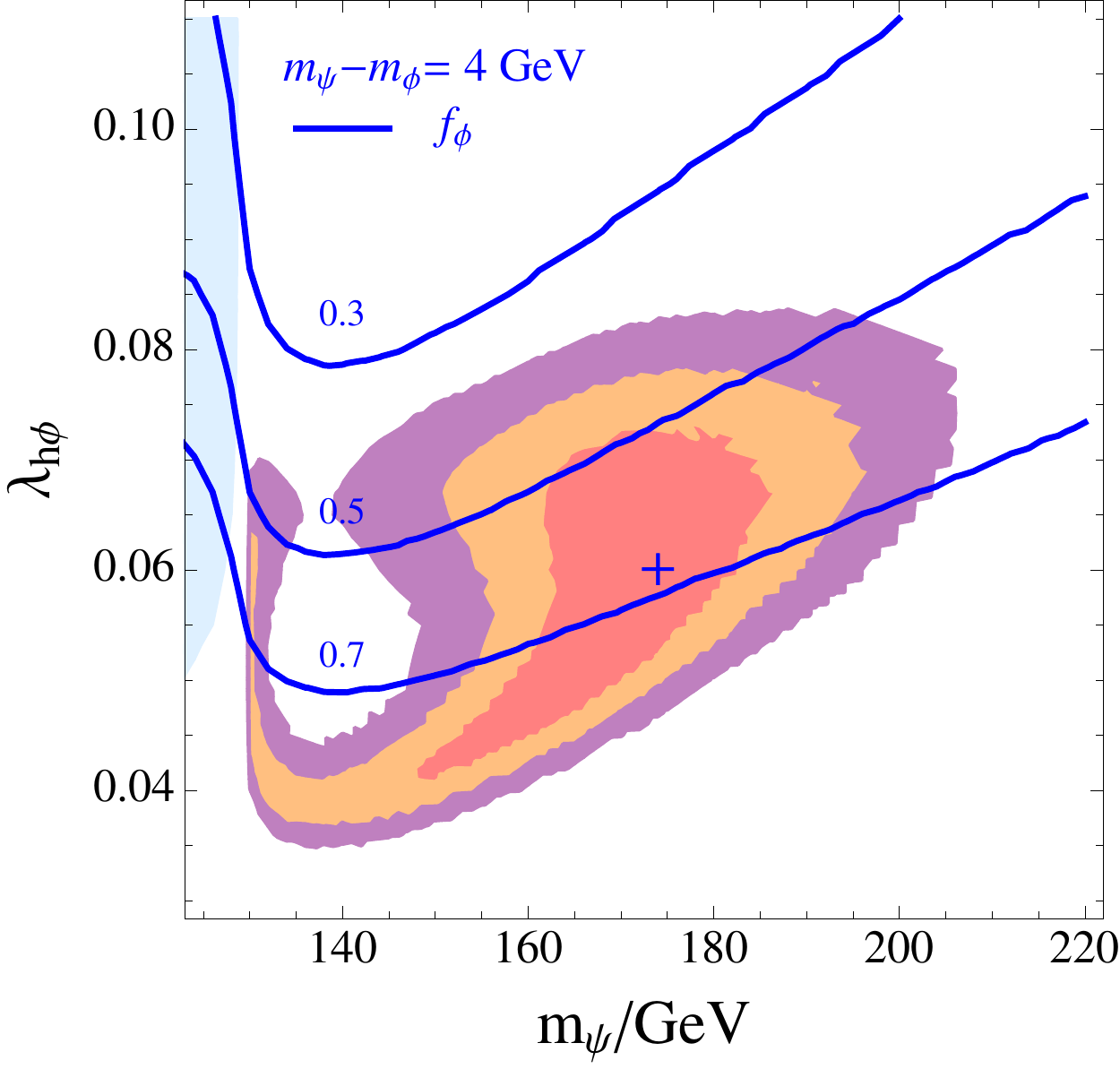}
  \caption{The best fit points and the 1, 2 and 3 $\sigma$ shown in the shaded regions. The fraction of the scalar relic density is shown in the solid contours. }
  \label{fig:chiscontour}
\end{figure}

For the fermion singlet model, only two processes, $\phi\phi \rightarrow SM$ and $\psi\phi\rightarrow \psi h$, have non-vanishing $s$-wave terms that would contribute to the gamma ray excess.  Of these, the annihilation channel was already considered in \refcite{Calore:2014xka}, so we focus on regions of parameter space where both contribute.  We perform a scan on the parameter space in the following region,
\begin{equation}
m_\psi /\rm{GeV} \in [80,220] \, , \qquad \lambda_{h\phi} \in [0.01,0.2] \, , 
\end{equation}   
with a fixed mass difference between the fermionic  and the scalar dark matter $\Delta m\equiv m_\psi - m_\phi = 2 (4) \; \rm{GeV}$.  The Yukawa coupling $y$ is chosen between 0 and $\sqrt{4\pi}$  to generate the correct relic abundance given $m_\psi$, $m_\phi$ and $\lambda_{h\phi}$.  In \figref{fig:ycontour} we plot contours of the Yukawa couplings in black (blue) together with the LUX exclusions in light gray (light blue) for $\Delta m = 2 (4) \; \rm{GeV}$.  We generate and combine photon spectra using results from PPPC~4~DM~ID~\cite{Cirelli:2010xx} and as described in \secref{sec:ID}.  For the astrophysical $\bar{J}$ factor, we follow \refcite{Calore:2014xka} in writing
\begin{equation}
  \bar{J} = \mathcal{J} \, \bar{J}_{canonical} \,,
\end{equation}
where  $\bar{J}_{canonical}=1.58\times 10^{24}\, \rm{GeV}^2/\rm{cm}^5$ and $\mathcal{J}\in [0.14, 4.0]$.  We perform a $\chi^2$ analysis on the computed photon spectrum in 20 bins, with the observed Fermi spectrum taken from \refcite{Cline:2015qha}.  The goodness of fit is found to be optimized for larger $\mathcal{J}$ values.  The best fit points and the 1, 2 and 3 $\sigma$ regions for $\Delta m = 2 (4) \; \rm{GeV}$ and $\mathcal{J}=4$ are depicted in \figref{fig:chiscontour}.  The best fit point is found at $m_\psi = 189 (174) \; \rm{GeV}$ and $\lambda_{h\phi} = 0.063 (0.060)$ for $\Delta m = 2 (4) \; \rm {GeV}$, with $\chi^2=12.4 (11.5) $ for 20 data points, which shows that our model describes the GCE very well.

\begin{figure}[ht]
\centering
\includegraphics[width=0.47\textwidth]{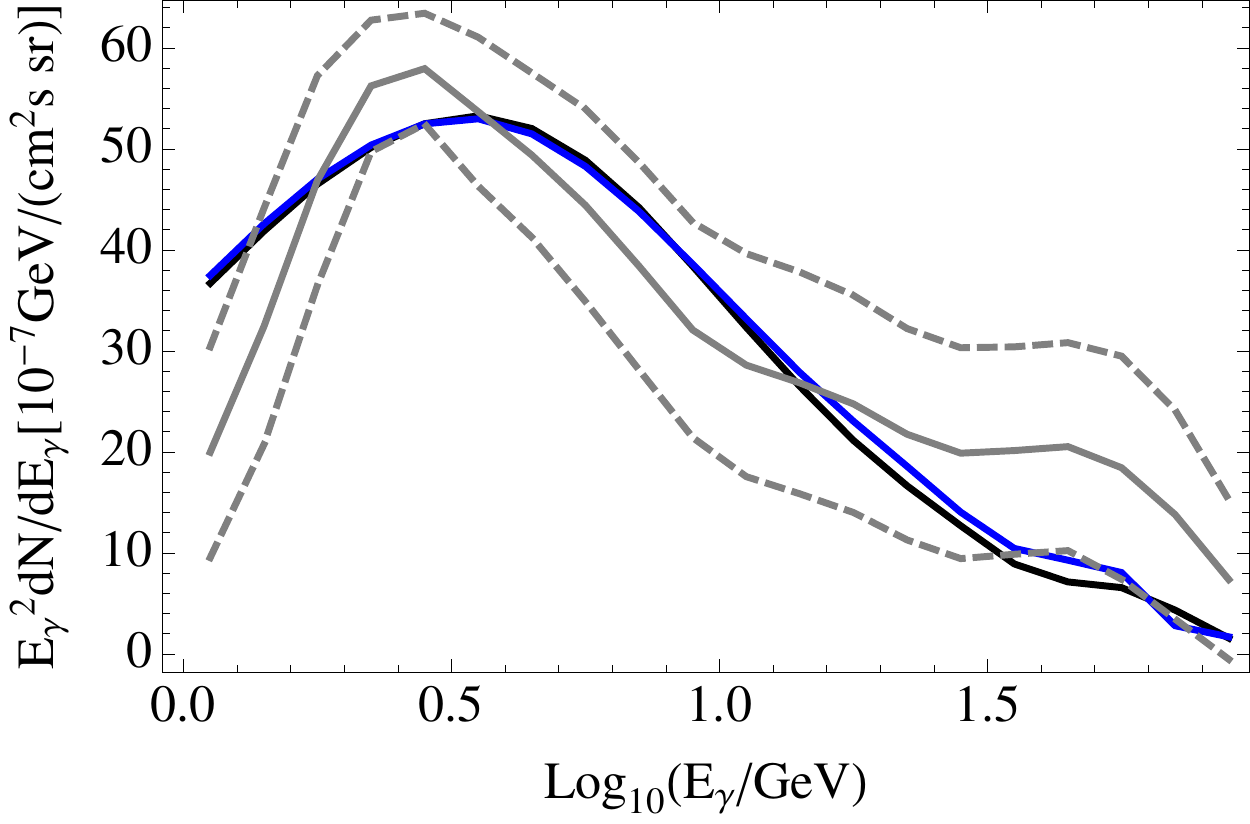}
\hfill
\includegraphics[width=0.47\textwidth]{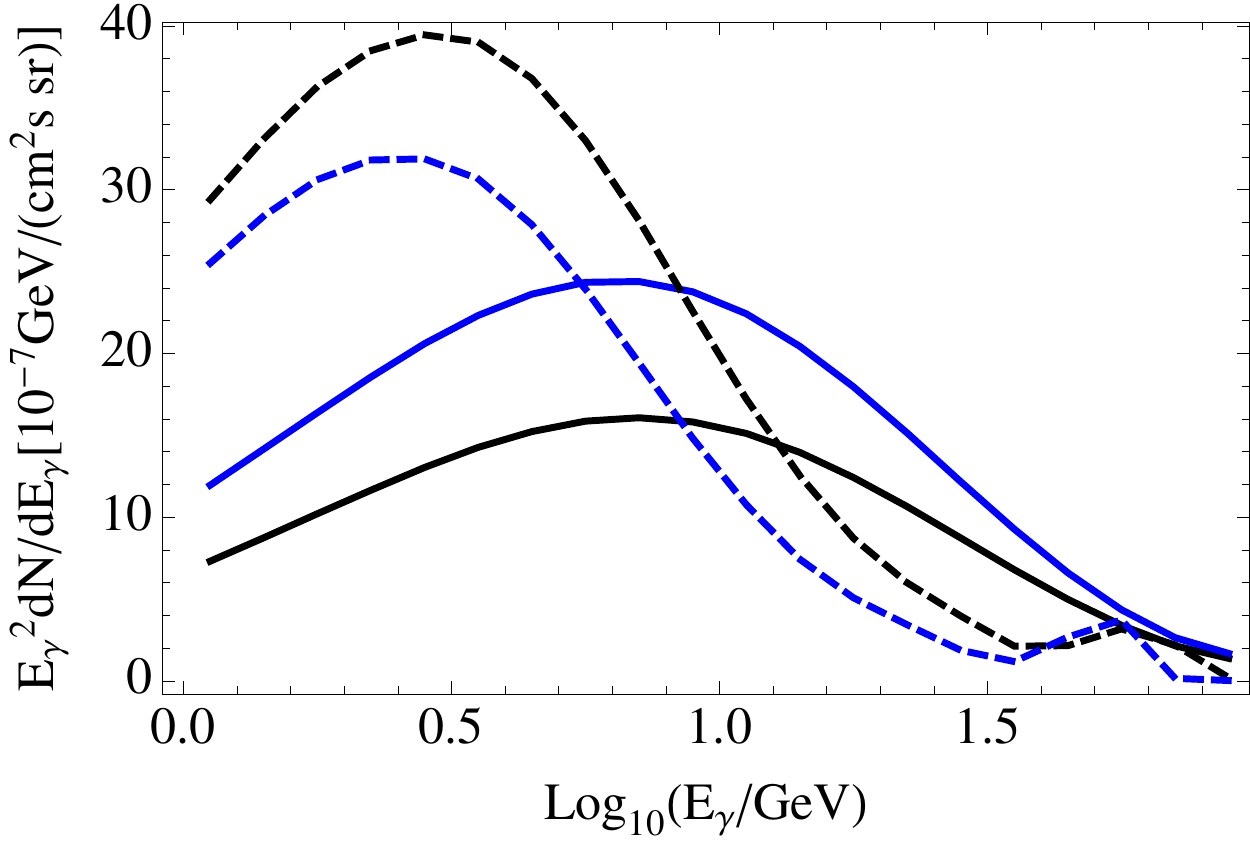}
\caption{
Left: the predicted gamma spectrum at the best fit points shown in Fig.~\ref{fig:chiscontour}.
The gray solid and dashed lines are the Fermi observation and errors.
Right: the contributions from the scalar self-annihilation  and the SA to the gamma spectra  
in solid and dashed black (blue) lines for $\Delta m = 2 (4) \; \rm{GeV}$. 
} 
\label{fig:fermifit}
\end{figure}

At the best fit points, the Yukawa coupling $y$ takes large values, $1.69$ and $1.30$ for $\Delta m=2$ and 4~GeV.  This enhances the scalar-fermion SA channel relative to scalar annihilation.  The fractions of the scalar dark matter and the fermion dark matter are of comparable size, which also helps to relax the direct detection constraints from LUX (as shown in \modeqref{eq:appLUX}).  The predicted photon spectra at the best fit points for $\Delta m = 2 (4) \; \rm{GeV}$ are plotted in black (blue) in the left panel of \figref{fig:fermifit}, together with the GCE spectrum and uncertainties from \refcite{Cline:2015qha}.  The contributions from the scalar self-annihilation and the SA to the gamma spectra are also shown in the right panel of Fig.~\ref{fig:fermifit}.  Since the SA contribution is a $t$-channel process with a heavy fermion final state, the photon spectrum is shifted to lower energies.  The small peaks at higher energies in the SA channel are due to the Higgs to di-photon decay and broadened because of the boosted Higgs.   

SA plays an essential role in two aspects in this explanation of the GCE.  Firstly smaller Higgs portal couplings are allowed compared to the standard scalar DM model, as the DM relic density can be significantly modified by SA. This alleviates the stringent constraints from LUX; as seen in \figsref{fig:ycontour}{fig:chiscontour}, there are no exclusions for $m_\phi > m_h$. Furthermore, the SA channel in the singlet model has a large cross section and thus substantially enhances the production of Higgses. In contrast, DM annihilation in the standard Higgs portal scalar DM model preferentially produces gauge bosons over Higgses.  The sizeable contribution from SA reshapes the total gamma spectrum and greatly improves the fitting to the GCE, as shown in \figref{fig:fermifit}.

For the fermion triplet model, the LHC  excludes the fermion masses considered for the singlet model.  Additionally, fermion annihilation through their gauge couplings suppresses the fermion relic density.  These two factors make the fermion-scalar SA channel negligible in the region of parameter space relevant to the GCE, and without it we cannot generate a good fit to the GCE spectrum.

\section{Fermion Triplet Phenomenology}\label{sec:triplet}

\begin{figure}
  \centering
  \includegraphics[width=0.5\textwidth]{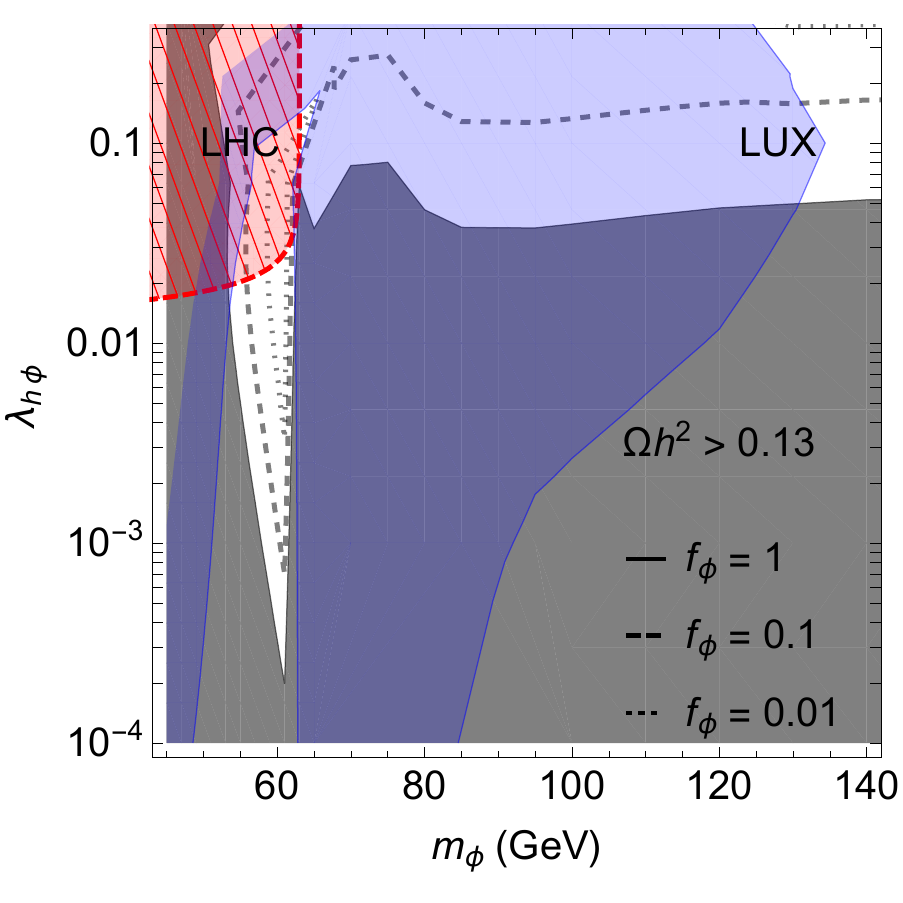}
  \caption{Scalar relic density for $m_\psi = 500$~GeV and $y = 1.0$, together with constraints from the Higgs invisible width (red) and from LUX (blue).  The relic density in this parameter range is set by scalar annihilation only.}\label{fig:lowmass}
\end{figure}

We now combine the results of \threesecsref{sec:LHC}{sec:DD}{sec:ID} for the fermion triplet model.  First we infer lower bounds on the fermion and scalar masses.  As discussed in \secref{sec:LHC}, we have a strict bound $m_\psi > 480$~GeV from the LHC, independent of all other parameters.  This then implies either 53~GeV$\,< m_\phi < 63$~GeV or $m_\phi > 130$~GeV, because the relic density for scalars in this mass range is set by scalar annihilation to the SM; SA and DME processes are irrelevant. We show the scalar relic density in the $m_\phi$--$\lhp$ plane for $m_\psi = 500$~GeV and $y = 1.0$ in \figref{fig:lowmass}, which is essentially identical to previous results for the scalar singlet model, \emph{e.g.} Fig.~1 of \refcite{Cline:2013gha}.  Including constraints from the Higgs invisible width (see \figref{fig:HInv}) and LUX (see \figref{fig:LUXexcl}) lead to the allowed regions given above, in the limit of vanishing fermion relic density.  Non-zero $\Omega_\psi$ would reduce the allowed region near the Higgs resonance.

To explore how semi-annihilation modifies the phenomenology of the model, we scanned the $m_\phi$--$m_\psi$ plane of parameter space for fixed $\lhp = 0.1$ and for increasing values of the Yukawa, $y = 0.1$, 0.5, 0.7 and 1.0.  For $y = 0.1$, we expect semi-annihilation to be subdominant to fermion and scalar annihilation.  As discussed in \secref{sec:RD}, the fermion annihilation and semi-annihilation channels have roughly equal cross sections for $y \approx 0.7$, which motivated our other choices above.  The lower bounds on our scan are set by the limits discussed above.  There are $y$-dependent upper bounds on $m_\psi$ from demanding $\Omega_{\phi+\psi} < \Omega_{cdm}$.  For small $y$, this bound is set by fermion annihilation and is $m_\psi \lesssim 2.4$~TeV.  Larger Yukawa couplings allow heavier fermion masses due to DME processes for $m_\psi \gg m_\phi$.  While there is no absolute upper bound on the scalar mass $m_\phi$, for $m_\phi > 2m_\psi$ the model largely reduces to the well-studied case of a fermion triplet; the scalar decays before the present epoch and all SA and DME processes are negligible

\begin{figure}
  \centering
  \includegraphics[width=0.48\textwidth]{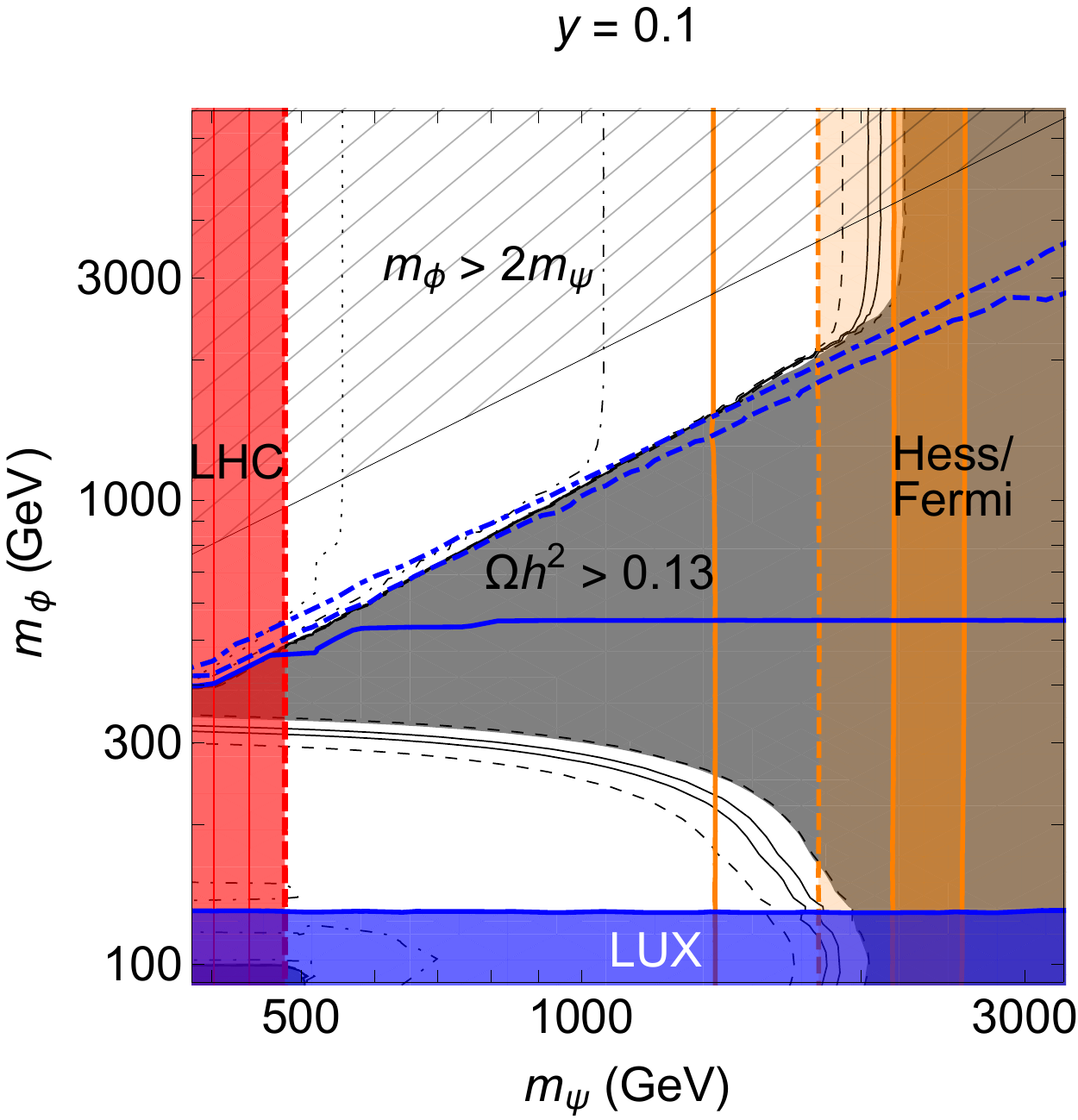}
  \includegraphics[width=0.48\textwidth]{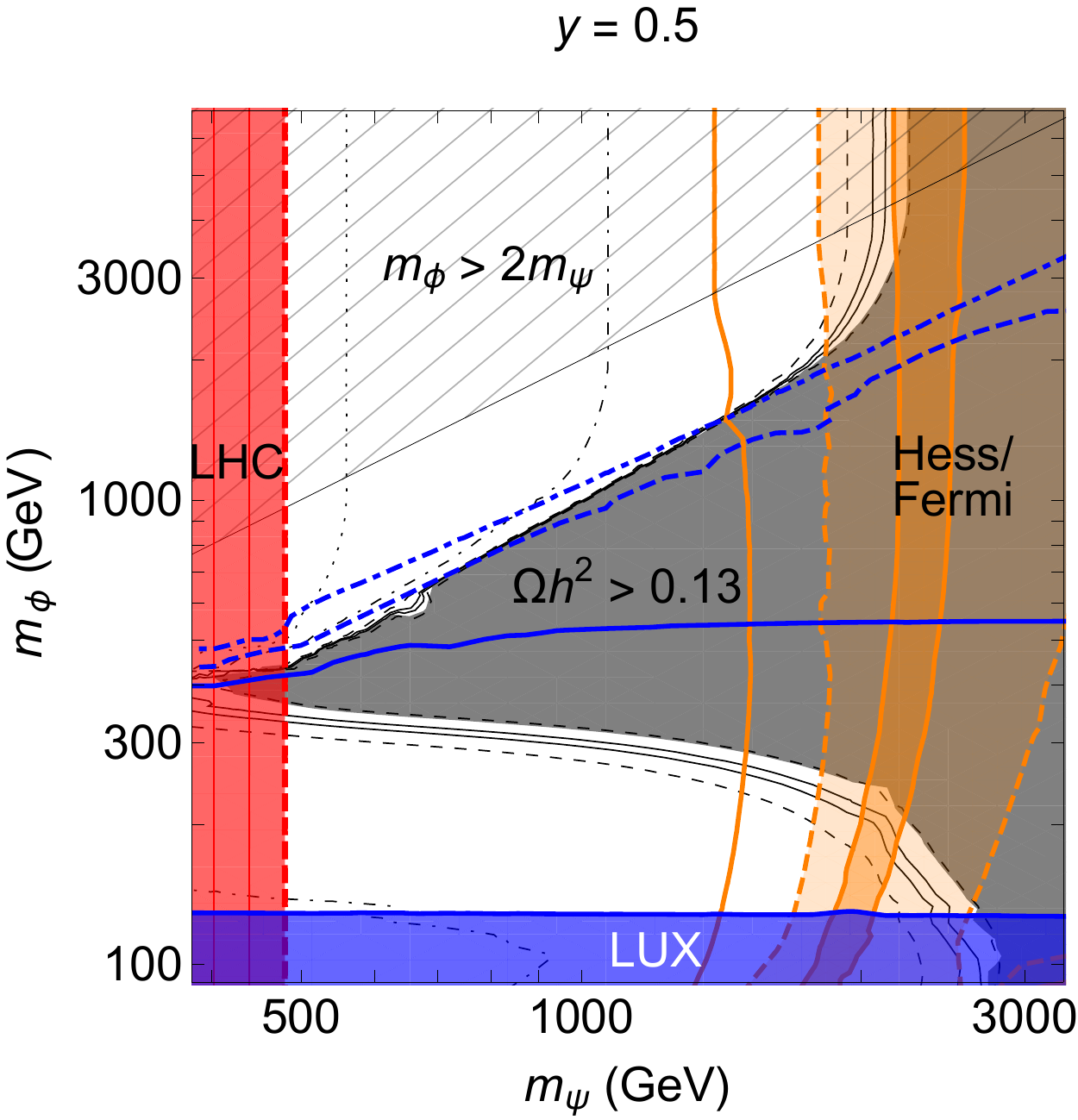}\\
  \includegraphics[width=0.48\textwidth]{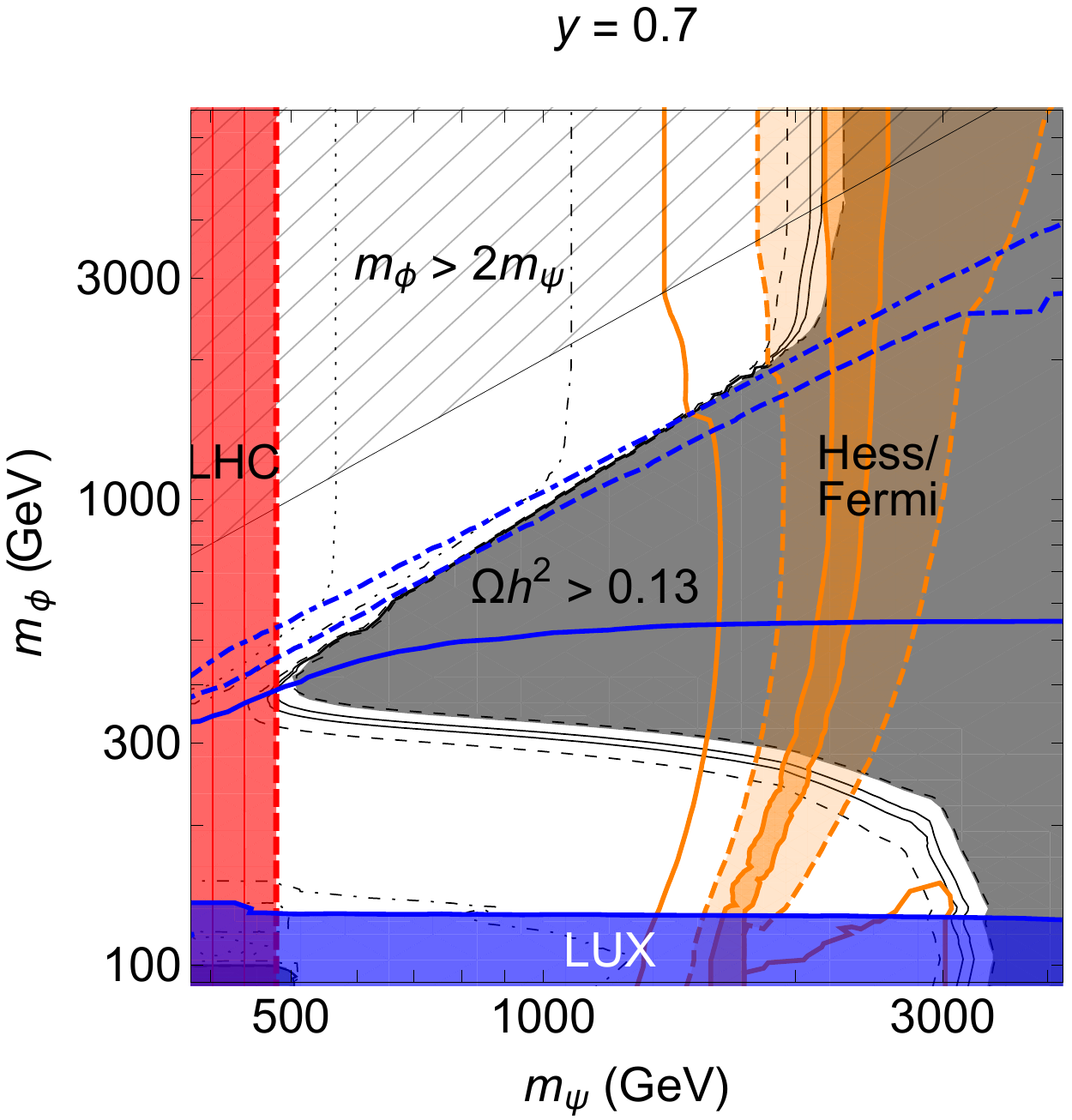}
  \includegraphics[width=0.48\textwidth]{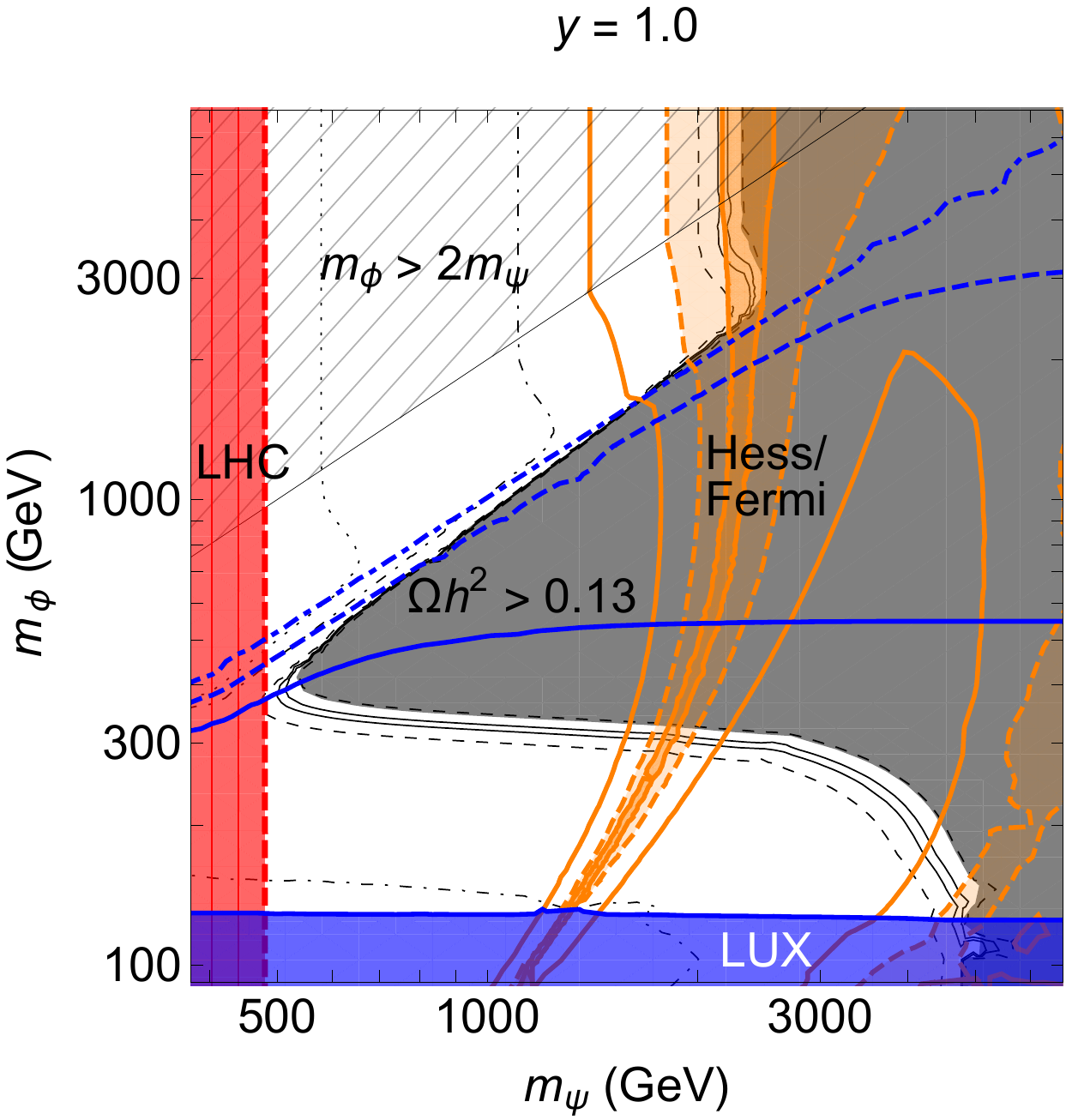}
  \caption{Slices of the fermion triplet parameter space in the $m_\phi$--$m_\psi$ plane, for $\lhp = 0.1$ and $y$ as labelled.  The grey shaded regions show regions where the total relic density is larger than observations.  In the white hatched region, the scalar decays to two fermions.  The red (blue, orange) shaded regions and contours show current and future bounds from the LHC (direct detection, indirect searches).  See the text for more details.}\label{fig:tripletresults}
\end{figure}

We show our scan results in in \figref{fig:tripletresults}.  The grey shaded region is excluded by a too large total DM relic density: $\Omega_{\phi + \psi} > 1.1 \,\Omega_{cdm}$, where we have included a 10\% theoretical uncertainty as discussed in \secref{sec:RD}.  The black solid (dashed) contours mark the regions where we find the correct relic density within the Planck 3$\sigma$ result with (without) our 10\% theoretical uncertainty.  Within the remainder of the white unshaded regions, our model does not explain the full observed DM density, but might still be consistent in the presence of additional sources of dark matter, \emph{e.g.} axions.  The dotted (dash-dotted) contours show $\Omega_{\phi + \psi} = 0.1 \,(0.3) \,\Omega_{cdm}$.  In the white region with diagonal hatching, the scalar is unstable to the decay $\phi \to \psi\psi$.  The red regions with vertical hatching denote the LHC bounds of \secref{sec:LHC}.  The blue unhatched region the LUX constraints from \secref{sec:DD}; additionally, we show prospective limits from LUX (Xenon1T, LZ) by blue solid (dashed, dot-dashed) contours.  The orange regions give the indirect detection bounds and prospects.  The lighter (darker) shaded regions with a dashed (solid) boundaries shows the bound for an optimistic (conservative) profile, as discussed in \secref{sec:ID}.  We also show prospective limits from CTA for an NFW profile by an orange contour.  The CTA limits for a cored profile are slightly weaker than the Fermi dwarf spheroidal limits (our current conservative exclusions) so we do not show them.

\begin{figure}
  \centering
  \includegraphics[width=0.48\textwidth]{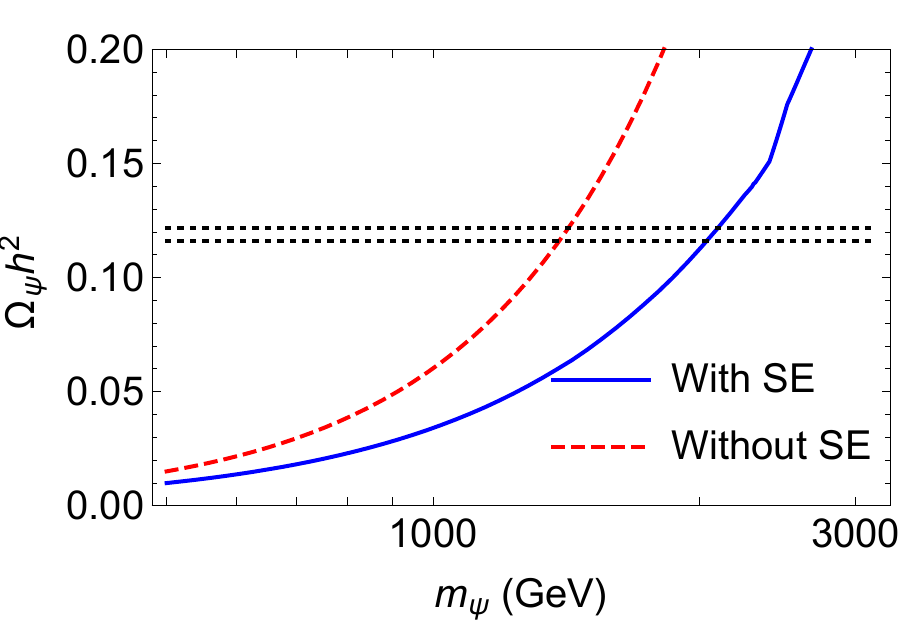}\hfill
  \includegraphics[width=0.48\textwidth]{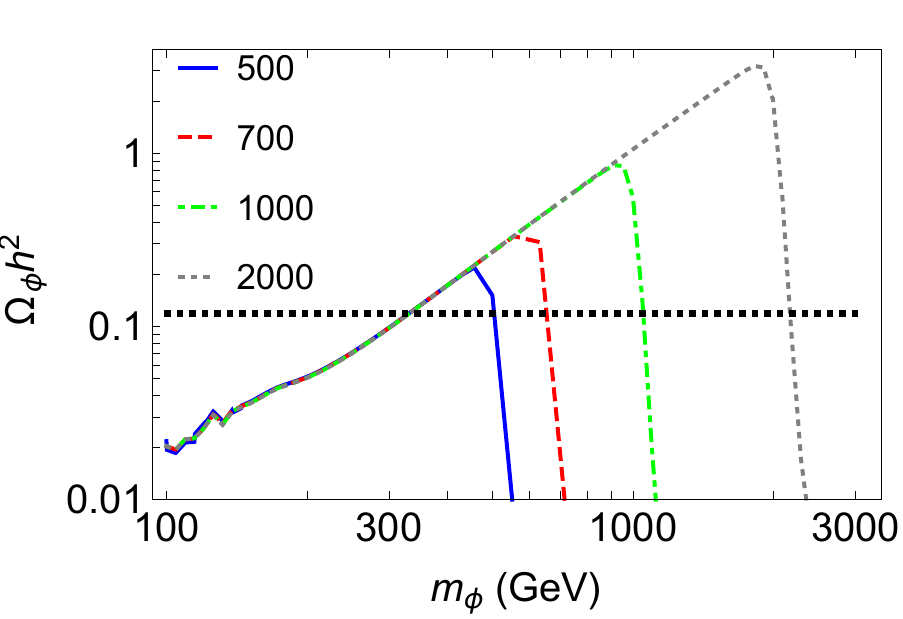}
  \caption{Fermion (left) and scalar (right) relic densities for $y = 0.1$.  The dotted bands denote the Planck 3\,$\sigma$ measurement.  For the fermion, we show the results with and without the SE; for the scalar, we show the results for different fermion masses as labelled.}\label{fig:y01RD}
\end{figure}

To better interpret these results, let us first consider $y = 0.1$.  The fermion and scalar relic densities are shown in \figref{fig:y01RD}.  The salient feature is that $\Omega_\psi$ is a function only of $m_\psi$; while $\Omega_\phi$ is a function only of $m_\phi$, unless $m_\phi > m_\psi$.  This is the expected situation in the absence of semi-annihilation: each DM particle freezes out separately.  The only effect of non-zero $y$ is that, for $m_\phi > m_\psi$, the DM exchange process $\phi\phi \to \psi\bar{\psi}$ becomes important.  We can then divide the parameter space into two regions, according to which of the two particles is more massive.

When $m_\phi < m_\psi$, the two DM particles freeze out essentially independently, and the correct relic density is produced when these two processes coincidentally sum to the observed value.  In this region, $f_\phi \sim f_\psi$ due to the lower bounds on $m_{\phi,\psi}$ giving lower bounds on $\Omega_{\phi,\psi}$.  For $y = 0.1$, SA and DME are irrelevant in this region of parameter space.  This region is currently weakly constrained, but can be completely ruled out by the full LUX data set.

When $m_\phi > m_\psi$, DME leads to a rapid decrease in the scalar relic density as can be seen from \figref{fig:y01RD}.  This results in the correct relic density only being reproduced for $m_\phi \approx m_\psi$ (when $f_\phi \sim f_\psi$); or $m_\phi > m_\psi \approx 2.1$~TeV, when $f_\psi \sim 1 \gg f_\phi$.  The presence of DME makes this region more interesting, as it allows points in the $m_\phi$--$\lhp$ plane that are excluded for the pure scalar singlet model.  While not all of the parameter space in this region is accessible to current and near-future searches, the band where our model produces the full DM relic density can be excluded for optimistic DM galactic profiles.  For $m_\psi \lesssim 1.5$~TeV, direct searches at LZ are most sensitive due to the moderate values of $f_\phi$.  In the complementary region $m_\psi \gtrsim 1.5$~TeV, indirect searches for NFW/Einasto DM profiles are strongest.  As with a pure Wino, HESS already excludes the region around $m_\psi \approx 2.1$~TeV where $f_\psi \sim 1$, and CTA can exclude the rest.  If the DM profile is more conservative, then exploring this part of parameter space would require a 100~TeV collider as discussed in \secref{sec:LHC}.

As we increase $y$, we see several effects on our results.  In the low scalar mass region, the SA and DME processes $\psi\psi \to \phi V$ and $\psi\bar{\psi} \to \phi\phi$ become important to the relic density, while the SE resonance moves to lower masses as shown in \figref{fig:IDXSec}.  This results in the fermion relic density acquiring a dependence on $m_\phi$.  We also see that these additional channels increase the maximal allowed $m_\psi$, in particular allowing masses forbidden in the pure fermion triplet model.  As $\Omega_\psi$ is suppressed, exclusions from indirect searches are weakened and more strongly focused on the resonance.  The scalar relic density is only weakly affected by SA and DME; because the fermion freezes out first, $Y_\psi \ll Y_\phi$ during scalar freeze out.  On resonance, where the late-time SA channels are strongly enhanced, $Y_\psi$ is suppressed even further.  We see at most a $\sim 5\%$ variation in $\Omega_\phi$ with $m_\psi$.  Since for large $y$, $f_\psi \ll f_\phi$ in much of this region, contours of total relic density are nearly lines of constant $m_\phi$.  The direct detection constraints remain the same, and in particular the whole region can be excluded by LUX.

\begin{figure}
  \centering
  \includegraphics[height=0.32\textwidth]{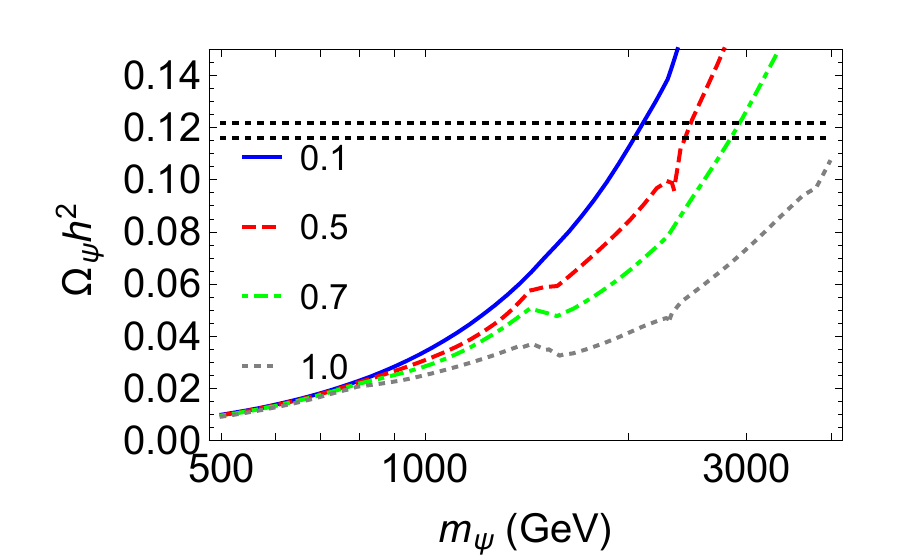}\hfill
  \includegraphics[height=0.32\textwidth]{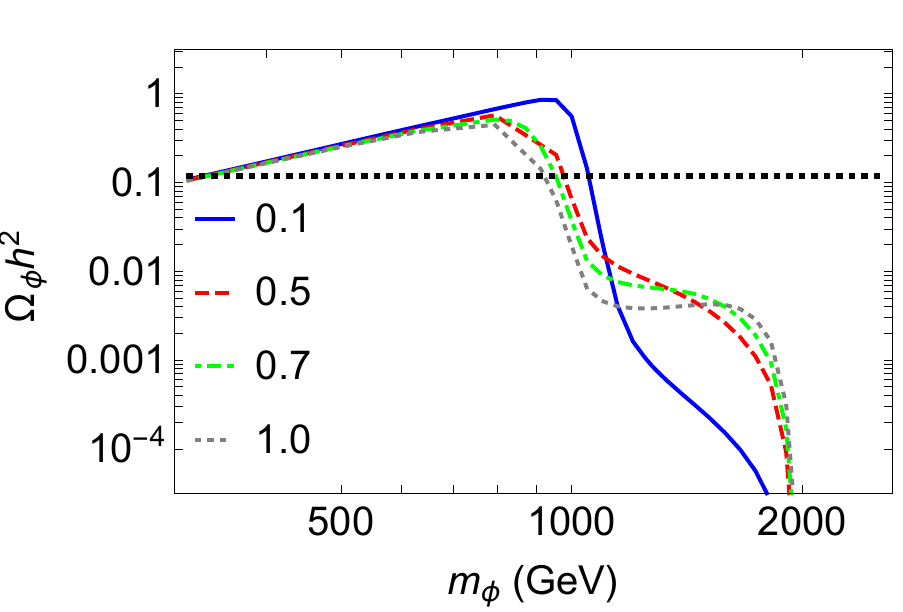}
  \caption{Fermion (left) and scalar (right) relic densities for different values of $y$ as labelled, and $m_\phi = 1.5$~TeV (left) or $m_\psi = 1$~TeV (right).  The dotted bands denote the Planck 3$\sigma$ measurement.}\label{fig:RDydep}
\end{figure}

More interesting is the high scalar mass case.  From \figref{fig:tripletresults}, we can see that this region extends to $m_\phi \lesssim m_\psi$ for large $y$.  This can easily be understood from \figref{fig:RDydep} (right), where we show the scalar relic densities for different $y$ and $m_\psi  = 1$~TeV.  As $y$ increases, the semi-annihilation process $\psi\phi \to \bar{\psi} + SM$ becomes increasingly relevant at $m_\phi < m_\psi$, lowering $\Omega_\phi$.  We also see that, for $m_\phi > m_\psi$, the scalar relic density increases with SA; the scalars are replenished through the process $\psi\psi \to \phi V$.  The effect of SA on the fermion relic density is more modest, and can be seen in \figref{fig:RDydep} (left): $\Omega_\psi$ decreases monotonically with $y$.   It also increases when $m_\phi \gtrsim m_\psi$, as DME now converts scalars to fermions instead of the reverse.  The net effect on the correct relic density band is that it is shifted to lower scalar masses for larger $y$; the size of the shift weakens near the resonance at $m_\psi \approx 2.1$~TeV, as $f_\psi$ increases towards 1.  From an observational perspective, it remains the case that the band with $\Omega_{\phi + \psi} = \Omega_{cdm}$ can be probed up to $m_\psi \sim 1.5$~TeV through direct detection at LZ; and at higher masses by CTA (for optimistic DM density profiles), and eventually by a 100~TeV collider.

Finally, let us briefly discuss the prospects for identifying this model if a discovery is made at a future DM search.  The best possible situation involves inconsistent discoveries at both LUX and CTA, which would clearly point to a two-component DM sector.  Such a situation is possible in our model only for $m_\phi \lesssim 300$~GeV and $m_\psi \sim 1$--2~TeV.  Even in this ideal case, measuring the SA coupling $y$ would be difficult due to the degeneracy between the $\bar{J}$-factor and the annihilation cross section.  However, a signal for $m_\psi \sim 1$~TeV would point towards a large $y$, as even with optimistic DM profiles this would normally be unobservable.

In other situations, we would likely have evidence for only one DM particle for some time.  In such cases, the only evidence for the existence of a second state would be difficulty in reconciling the measured DM properties with the observed relic abundance.  However, without direct evidence of both DM states it would be difficult to rule out alternative explanations, such as non-thermal production.  It would likely require a 100~TeV collider to provide the necessary evidence in this case.

\section{Conclusion}\label{sec:conc}

With the absence of any unequivocal evidence of dark matter, it is important to explore a wide range of possible DM signals to ensure no stone is left unturned.  In particular, we should consider models with non-standard phenomenology, and see to what extent they can be probed by current and planned experiments.  In this paper, we have applied this philosophy to semi-annihilating fermionic dark matter.  We have constructed two simple models with two-component dark sectors stabilised by a $\set{Z}_4$ symmetry.  In addition to the fermionic DM $\psi$, the additional dark sector state is a scalar singlet $\phi$.  The difference between our two models lies only in whether the fermion is an electroweak singlet or triplet.  The former case is the minimal fermionic example of SADM; the latter the minimal model with a non-singlet fermion with zero hypercharge.  Both our models demonstrate how semi-annihilation can play an important role modifying both the relic density and indirect detection signals even in models with fermionic dark matter.  The triplet model also features the first example of semi-annihilation subject to the Sommerfeld enhancement.

For the fermion singlet model, the only coupling between the dark sector and the visible sector is a Higgs portal involving $\phi$.  It follows that in much of the parameter space, the phenomenology reduces to that of the scalar singlet model, except that there is an DM particle so it is natural to have $\Omega_\phi < \Omega_{cdm}$.  However, when $m_\phi \sim m_\psi$, interesting semi-annihilation phenomenology can arise.  We have found that when the DM masses are close to the Higgs mass, this model can provide a good fit to the Galactic Centre excess in $\gamma$-rays.  The semi-annihilation channel $\psi\phi\to\bar{\psi} h$ is essential in both providing this fit, and allowing the correct relic density while avoiding direct detection constraints.

For the fermion triplet model, the gauge couplings have two important effects.  First, it opens a second channel connecting the visible and dark sectors, allowing the fermion to directly annihilate to visible states.  It also leads to a significant Sommerfeld enhancement of both the annihilation channel $\psi\bar{\psi} \to SM$ and the semi-annihilation channels $\psi\psi \to \phi + SM$.  For large fermion-scalar coupling $y$ and $m_\psi \gg m_\phi$, the scalar can also significantly contribute to the Sommerfeld enhancement.  These factors have several effects on the phenomenology, compared to that of a fermion triplet without semi-annihilation:
\begin{itemize}
  \item The parameter space where the DM relic density can be explained is substantially enlarged.  In particular, regions of heavy fermions $m_\psi \gtrsim 3$~TeV are allowed that would produce too much DM without the scalar.
  \item For large $y$ and small $m_\phi$, the Sommerfeld resonance is moved to smaller fermion masses $m_\psi$.  As a consequence, limits from indirect searches are shifted to lower masses and are much weaker.
  \item Fermion masses 500~GeV$\,< m_\psi < 2$~TeV can reproduce the observed total DM relic density due to the presence of the scalar.  In particular, semi-annihilation and dark matter exchange result in the correct abundance being produced for $m_\phi \sim m_\psi$ and a Higgs portal coupling $\lhp$ that is not too small.
\end{itemize}
This last point is also relevant for comparisons with the scalar singlet model.  This region is generically excluded in that case as it overproduces DM.  

Despite the expanded parameter space, we find excellent prospects for probing this model at current and near-future searches.  In particular, the entire region that produces the full observed relic abundance can be excluded by a combination of direct (LZ) and indirect (CTA) searches for the NFW DM density profile.  When indirect searches are weakened by fermion semi-annihilation, direct detection searches for the scalar DM state will be very effective.  If there is an additional source of DM (\emph{e.g.} axions), then regions of parameter space where our model is only part of the total DM abundance exist that are difficult to probe.  However, in this sense our model is no worse than pure Wino DM; and even this section of parameter space can eventually be excluded by a 100~TeV collider.

\section*{Acknowlegements}
We thank F. Gao and M. A. Schmidt for valuable discussions about the galactic center excess. We also thank an anonymous referee for their input.  YC and AS were supported by the Australian Research Council.  This work was supported by IBS under the project code, IBS-R018-D1.

\bibliography{ref}{}
\bibliographystyle{JHEP}

\end{document}